\documentclass[prc,nofootinbib,showpacs,eqsecnum,byrevtex,floatfix,superscriptaddress,twocolumn]{revtex4}
\usepackage{graphics}
\usepackage{pstcol}
\usepackage{afterpage}
\usepackage[dutch,english]{babel}

\begin{document}

\title{Soft-core meson-baryon interactions. II. $\pi N$ and $K^+ N$ scattering}
\author{H.\ Polinder}
\affiliation{Institute for Theoretical Physics, Radboud University Nijmegen,
         Nijmegen, The Netherlands }
\affiliation{Forschungszentrum J\"{u}lich, Institut f\"{u}r Kernphysik (Theorie), D-52425 J\"{u}lich, Germany}
\author{Th.A.\ Rijken}
\affiliation{Institute for Theoretical Physics, Radboud University Nijmegen,
         Nijmegen, The Netherlands } 
 
\date{version of: \today}

\begin{abstract}
The $\pi N$ potential includes the $t$-channel exchanges of the scalar-mesons $\sigma$ and $f_0$, vector-meson $\rho$, tensor-mesons $f_2$ and $f_2'$ and the Pomeron as well as the $s$- and $u$-channel exchanges of the nucleon $N$ and the resonances $\Delta$, Roper and $S_{11}$. These resonances are not generated dynamically. We consider them as, at least partially, genuine three-quark states and we treat them in the same way as the nucleon. The latter two resonances were needed to find the proper behavior of the phase shifts at higher energies in the corresponding partial waves. The soft-core $\pi N$-model gives an excellent fit to the empirical $\pi N$ $S$- and $P$-wave phase shifts up to $T_{{\rm lab}}=600$ MeV. Also the scattering lengths have been reproduced well and the soft-pion theorems for low-energy $\pi N$ scattering are satisfied.
The soft-core model for the $K^+ N$ interaction is an $SU_f(3)$-extension of the soft-core $\pi N$-model.
The $K^+ N$ potential includes the $t$-channel exchanges of the scalar-mesons $a_0$, $\sigma$ and $f_0$, vector-mesons $\rho$, $\omega$ and $\varphi$, tensor-mesons $a_2$, $f_2$ and $f_2'$ and the Pomeron as well as $u$-channel exchanges of the hyperons $\Lambda$ and $\Sigma$. The fit to the empirical $K^+ N$ $S$-, $P$- and $D$-wave phase shifts up to $T_{{\rm lab}}=600$ MeV is reasonable and certainly reflects the present state of the art. Since the various $K^+ N$ phase shift analyses are not very consistent, also scattering observables are compared with the soft-core $K^+ N$-model. A good agreement for the total and differential cross sections as well as the  polarizations is found.
\end{abstract}

\maketitle


 \hyphenation{Po-lin-der}
 \hyphenation{Nij-me-gen}
\hyphenation{iso-scalar}

\section{Introduction}
\label{sec:1}

In the previous paper (paper I) \cite{Pol05} the Nijmegen soft-core model for the pseudoscalar-meson baryon interaction in general (NSC model) is derived. In this paper (paper II) we apply the NSC model to the $\pi N$ and $K^+ N$ interactions.

The interaction between a pion and a nucleon has been investigated experimentally as well as theoretically for many years. For the early literature we would like to refer to Chew and Low \cite{Che56}, who presented one of the best early models that described the low energy $P$-wave scattering successfully, Hamilton \cite{Ham67}, 
Bransden and Moorhouse \cite{Bra73} and H{\"o}hler \cite{Hol79}.

Although the underlying dynamics of strong hadron interactions in general and the $\pi N$ interaction specifically are believed to be given by quark-gluon interactions, it is in principle not possible to use ab initio these degrees of freedom to describe the low and intermediate energy strong interactions. This problem is related to the phase transition between low energy and high energy strong interactions and the nonperturbative nature of confinement. Instead an effective theory with meson and baryon degrees of freedom must be used to describe strong interaction phenomena at low and intermediate energies, at these energies the detailed quark-gluon structure of hadrons is expected to be unimportant.

In particular meson-exchange models have proven to be very successful in describing the low and intermediate energy baryon-baryon interactions for the $NN$ and $YN$ channels \cite{Rij02a, Rij02b, Rij99, Mac87, Hol89, Reu94}. Similarly it is expected that this approach can also successfully be applied to the meson-baryon sector, i.e. $\pi N$, $K^+ N$, $K^- N$, etc...

The last decade the low and intermediate energy $\pi N$ interaction has been studied theoretically, analogous to the $NN$ interaction, in the framework of meson-exchange by several authors \cite{Pea90, Gro93, Sch94, Sch95, Sat96, Lah99, Pas00, Gas03}. The $K^+ N$ interaction has been investigated in this framework only by the J\"{u}lich group \cite{BHM90, HDH95} and in this work. In the same way as the Nijmegen soft-core $YN$ model was derived in the past as an $SU_f(3)$ extension of the Nijmegen soft-core $NN$ model, we present the NSC $K^+ N$-model as an $SU_f(3)$ extension of the NSC $\pi N$-model.

The above $\pi N$ meson-exchange models have in common that besides the nucleon pole terms also the $\Delta_{33}(1232)$ ($\Delta$) pole terms are included explicitly, i.e. the $\Delta$ is not considered to be purely dynamically generated as a quasi-bound $\pi N$ state, which might be possible if the $\pi N$ potential is sufficiently attractive in the $P_{33}$ wave. This possibility was investigated in the past by \cite{Nie68, Nie71}. From the quark model point of view the $\Delta$ resonance and other resonances are fundamental three-quark states and should be treated on the same footing as the nucleons.

We remark that the exact treatment of the propagator of the $\Delta$ and its coupling to $\pi N$ is different in each model. The NSC $\pi N$-model uses the same coupling and propagator for the $\Delta$ as Sch\"{u}tz et al. \cite{Sch94}.

The above $\pi N$ models differ, however, in the treatment of the other resonances, $P_{11}(1440)$ (Roper or $N^*$), $S_{11}(1535)$, etc... Gross and Surya \cite{Gro93} include the Roper resonance explicitly but the $S_{11}(1535)$ resonance is generated dynamically in their model, which gives a good description of the experimental data up to $T_{{\rm lab}}=600$ MeV. Sch\"{u}tz et al. \cite{Sch94} do not include the Roper resonance explicitly but generate it dynamically. However their model describes the $\pi N$ data only up to $T_{{\rm lab}}=380$ MeV, and in this energy region the Roper is not expected to contribute much. Pascalutsa and Tjon \cite{Pas00} include the above resonances explicitly in their model in order to find a proper description of the experimental data up to $T_{{\rm lab}}=600$ MeV.
The resonances that are relevant in the energy region we consider, the $\Delta$, Roper and $S_{11}(1535)$, are included explicitly in the NSC $\pi N$-model.

Several other approaches to the $\pi N$ interaction can be found in the literature, quark models have been used to describe $\pi N$ scattering \cite{LPL95}. Also models in the framework of chiral perturbation theory exist \cite{Gas88,Ber97,Dat97,Fet98,Mei00,Fet00,Fet01,Lut02}, however, heavier degrees of freedom, such as vector-mesons, are integrated out in this framework. We do not integrate out these degrees of freedom, but include them explicitly in the NSC model. 

For the $\pi N$ interaction accurate experimental data exist over a wide range of energy and both energy-dependent and energy-independent phase shift analyses of that data have been made, e.g. \cite{Arn95,Koch80,Car73}. Several partial wave analyses for the $\pi N$ interaction as well as for other interactions are available at http://gwdac.phys.gwu.edu/ (SAID).

Contrary to pions, the kaon ($K$) and antikaon ($\bar{K}$) interaction with the nucleons is completely different. This is due to the difference in strangeness, which is conserved in strong interactions. Kaons have strangeness $S=1$, meaning that they contain an $\bar{s}$-quark and a $u$- or $d$-quark in case of $K^+$ and $K^0$ respectively. Antikaons have strangeness $S=-1$, meaning that they contain an $s$-quark and a $\bar{u}$- or $\bar{d}$-quark in case of $K^-$ and $\bar{K}^0$ respectively. Since the $\bar{u}$- or $\bar{d}$-quark of the antikaon can annihilate with a $u$- or $d$-quark of the nucleon, the $\bar{K}N$ interaction is strong because low-lying resonances can be produced, giving a large cross section. This situation can be compared with the $\Delta$-resonance in $\pi N$ interactions.

The $\bar{s}$-quark of the kaon can not annihilate with one of the quarks of the nucleon in strong interactions, therefore three-quark resonances can not be produced, only heavy exotic five-quark ($qqqq\bar{q}$) resonances (referred to as $Z^*$ in the old literature or the pentaquark $\Theta^+$ in the new literature) can be formed, so the $K^+ N$ interaction is weak at energies below the energy of $Z^*$. The cross sections are not large and the $S$-wave phase shifts are repulsive.

However, in four recent photo-production experiments \cite{Nak03, Bar03, Ste03, Bart03} indications are found for the existence of a narrow exotic $S=1$ light resonance in the $I=0$ $K^+ N$ system with $\sqrt{s}\simeq 1540$ MeV and $\Gamma \leq$ 25 MeV. The existence of such an exotic resonance was predicted by Diakonov et al. \cite{Dia97}, they predicted the exotic resonance to have a mass of about 1530 MeV and a width of less than 15 MeV and spin-parity $J^P=\frac{1}{2}^+$.

The existing $K^+ N$ scattering data, which we use to fit the NSC $K^+ N$-model, does, however, not show this low-lying exotic resonance. On the other hand, this exotic resonance has not been searched for at low energies in the scattering experiments. At these energies not much scattering data exists and a narrow resonance could have escaped detection.

For the early literature on the $K^+N$ interaction we would like to refer to the review article by Dover and Walker \cite{Dov82}. The $K^+ N$ interaction has been studied by the J\"{u}lich group, they presented a model in the meson-exchange framework, B\"{u}tgen et al. \cite{BHM90} and Hoffmann et al. \cite{HDH95}, in analogy to the Bonn $NN$ model \cite{Mac87}.

In \cite{BHM90} a reasonable description of the empirical phase shifts is obtained, here not only single particle exchanges, ($\sigma$,$\rho$,$\omega$,$\Lambda$,$\Sigma$,$Y^*$), are included in the $K^+ N$ model, also fourth-order processes with $N,\Delta,K$ and $K^*$ intermediate states are included in analogy to the Bonn NN model, in which $\sigma$-exchange effectively represents correlated two-pion-exchange. Coupling constants involving strange particles are obtained from the known $NN\pi$ and $\pi\pi\rho$ coupling constants assuming $SU(6)$ symmetry.

However an exception had to be made for the $\omega$-coupling, which had to be increased by 60$\%$ in order to find enough short-range repulsion and to obtain a reasonable description of the $S$-wave phase shifts, model A. But this also caused too much repulsion in the higher partial waves and it was concluded that the necessary repulsion had to be of much shorter range. In model B the $\omega$ coupling was kept at its symmetry value and a phenomenological short-ranged repulsive $\sigma _0$ with a mass of 1200 MeV was introduced, which led to a more satisfactory description of the empirical phase shifts.

In \cite{HDH95} the model of \cite{BHM90} is extended by replacing the $\sigma$- and $\rho$-exchange by the correlated two-pion-exchange. A satisfactory description of the experimental observables up to $T_{{\rm lab}}$=600 MeV, having the same quality as in \cite{BHM90}, is achieved. Just as in \cite{BHM90} the phenomenological short ranged $\sigma_0$ was needed in this model in order to keep the $\omega$ coupling at its symmetry value. B\"{u}tgen et al. suggest that this short ranged $\sigma_0$ might be seen as a real scalar-meson or perhaps as a real quark-gluon effect.

The most recent quark models for the $K^+ N$ interaction are from Barnes and Swanson \cite{Bar94}, Silvestre-Brac et al. \cite{Sil95, Sil97} and Lemaire et al. \cite{Lem01, Lem02}. The agreement of these quark models with the experimental data is not good. The results of \cite{Sil95}- \cite{Lem02} show that there is enough repulsion in the $S$-waves, but the other waves can not be described well.

Recently a hybrid model for the $K^+ N$ interaction was published by Hadjimichef et al. \cite{Had02}. They used the J\"{u}lich model extended by the inclusion of the isovector scalar-meson $a_0(980)$exchange, which was taken into account in the Bonn $NN$ model \cite{Mac87}, but not in the J\"{u}lich $K^+ N$ models \cite{BHM90, HDH95}. The short ranged phenomenological $\sigma_0$-exchange was replaced by quark-gluon exchange. A nonrelativistic quark model, in which one-gluon-exchange and the interchange of the quarks is considered, was used. This quark-gluon exchange is, contrary to the $\sigma_0$-exchange, isospin dependent. A satisfactory description of the empirical phase shifts, having the same quality as \cite{HDH95}, was obtained. However Hadjimichef et al. conjecture that the short ranged quark-gluon dynamics they include could perhaps be replaced by the exchange of heavier vector-mesons.

Another approach for the $K^+ N$ interaction is given by Lutz and Kolomeitsev \cite{Lut02}. Meson-baryon interactions in general and $K^+ N$ interactions specifically are studied by means of chiral Lagrangians in this work. A reasonable description of the $K^+ N$ differential cross sections and phases was achieved, but only up to $T_{{\rm lab}}=360$ MeV.


The major differences between the existing $\pi N$ and $K^+ N$ models and the NSC model presented in this work are briefly discussed below. Form factors of the Gaussian type are used in the soft-core approach in this work, while monopole type form factors and other form factors are used for the $\pi N$-model by Pascalutsa and Tjon \cite{Pas00} and the $K^+ N$-model by Hoffmann et al. \cite{HDH95}. The Roper resonance in the $\pi N$ system is, at least partially, considered as a three-quark state and treated in the same way as the nucleon and is included explicitly in the potential. However, we renormalize the Roper contribution at its pole, while Pascalutsa and Tjon \cite{Pas00} renormalize it at the nucleon pole.

An other difference is our treatment of the scalar-mesons $\sigma$ etc., we consider them as belonging to an $SU_f(3)$ nonet, while in all other models they are considered to represent correlated two-pion-exchange effectively. Also we include Pomeron-exchange, where the physical nature of the Pomeron can be seen in the light of QCD as (partly) a two-gluon-exchange effect \cite{LOW75, NUS75}, in order to comply with the soft-pion theorems for low-energy $\pi N$ scattering \cite{Wei66,Adl65,Swa89}.
Furthermore, the exchange of tensor-mesons is included in the NSC model mainly to find a good description of the $K^+ N$ scattering data. We use only one-particle exchanges to find this description while Hoffmann et al. \cite{HDH95} need to consider two-particle exchanges in their $K^+ N$-model.

The contents of this paper are as follows. In Sec. \ref{chap:5} the $SU_f(3)$ relations between the coupling constants used in the $\pi N$ and $K^+ N$ interactions are shown.
The $\pi N$ total cross section shows several resonances
in the considered energy range. 
The renormalization procedure we use to include the $s$-channel Feynman diagrams for the resonances in the $\pi N$ potential is described in Sec. \ref{chap:4}.
In Sec. \ref{chap:6} the NSC $\pi N$-model is discussed and the results of the fit to the empirical phase shifts of the lower partial waves are presented.
The NSC $\pi N$-model is, via $SU_f(3)$-symmetry, extended to the NSC $K^+ N$-model in Sec. \ref{chap:7}. The results of the fit to the empirical phase shifts are given, since the different phase shift analyses are not always consistent, also the model calculation of some scattering observables is given. The NSC $K^+ N$-model is used to give a theoretical estimate for the upper limit of the decay width of the recently discovered exotic resonance in the isospin zero $K^+ N$ system.

Finally the summary gives an overview of the research in this work and its main results. Also, some suggestions for improvement and extension of the present NSC model are given.  In Appendix \ref{app:AA} details are given on the calculation of the isospin factors for $\pi N$ and $K^+ N$ interactions.

\section{Meson-baryon channels and $SU_f(3)$}
\label{chap:5}

We consider in this work the $\pi N$ and $K^+ N$ interactions, they make up only a subset of all meson-baryon interactions. Because the NSC $K^+ N$-model is derived from the NSC $\pi N$-model, using $SU_f(3)$ symmetry, we define an $SU_f(3)$ invariant interaction Hamiltonian describing the baryon-baryon-meson and meson-meson-meson vertices. The Lorentz structure of the baryon-baryon-meson interaction is discussed in paper I
, here we deal with its $SU_f(3)$ structure. In order to describe the interaction Hamiltonian we define the octet irreducible representation (irrep) of $SU_f(3)$ for the $J^P=\frac{1}{2}^+$ baryons and the octet and singlet irreducible representations of $SU_f(3)$ for the mesons. Using the phase convention of \cite{Swa63}, the $J^P=\frac{1}{2}^+$ baryon octet irrep can be written as traceless $3\times 3$ matrix
\begin{eqnarray}
{\mathcal B}&=&
\left(
\begin{array}{ccc}
\frac{\Sigma^0}{\sqrt{2}}+\frac{\Lambda}{\sqrt{6}} & \Sigma^+ & p \\
\Sigma^- & -\frac{\Sigma^0}{\sqrt{2}}+\frac{\Lambda}{\sqrt{6}} & n \\
-\Xi^- & \Xi^0 & -\frac{2\Lambda}{\sqrt{6}}
\end{array}
\right) \ ,
\end{eqnarray}
similarly the pseudoscalar-meson octet irrep can be written as
\begin{eqnarray}
{\mathcal P}_8&=&
\left(
\begin{array}{ccc}
\frac{\pi^0}{\sqrt{2}}+\frac{\eta_8}{\sqrt{6}} & \pi^+ & K^+ \\
\pi^- & -\frac{\pi^0}{\sqrt{2}}+\frac{\eta_8}{\sqrt{6}} & K^0 \\
K^- & K^0 & -\frac{2\eta_8}{\sqrt{6}} \ 
\end{array}
\right) \ ,
\end{eqnarray}
while the pseudoscalar-meson singlet irrep is the $3\times 3$ diagonal matrix ${\mathcal P}_1$ with the elements $\eta_1/\sqrt{3}$ on the diagonal. The pseudoscalar-meson nonet, having a nonzero trace, is given by
\begin{eqnarray}
{\mathcal P}&=&{\mathcal P}_8+{\mathcal P}_1 \ .
\end{eqnarray}
The physical mesons $\eta$ and $\eta '$ are superpositions of the octet and singlet mesons $\eta_8$ and $\eta_1$, usually written as
\begin{eqnarray}
\eta '&=&\sin\theta\ \eta_8+\cos\theta\ \eta_1\ , \nonumber \\
\eta &=&\cos\theta\ \eta_8-\sin\theta\ \eta_1\ .
\end{eqnarray}
Similar expressions hold for the physical coupling constant of the $\eta$ and $\eta '$. The octets and singlets for the scalar- and vector-mesons are defined in the same way and the expressions for the physical ($\omega$,$\varphi$) and ($\sigma$,$f_0$) are analogous to ($\eta '$,$\eta$). From these octets and nonets, $SU_f(3)$-invariant baryon-baryon-meson interaction Hamiltonians can be constructed, using the invariants ${\rm Tr}\left({\bar{\mathcal B}}{\mathcal P}{\mathcal B}\right)$, ${\rm Tr}\left({\bar{\mathcal B}}{\mathcal B}{\mathcal P}\right)$ and ${\rm Tr}\left({\bar{\mathcal B}}{\mathcal B}\right){\rm Tr}\left({\mathcal P}\right) $. We take the antisymmetric ($F$) and symmetric ($D$) octet couplings and the singlet ($S$) coupling\\
\begin{eqnarray}
\left[{\bar{\mathcal B}}{\mathcal B}{\mathcal P}\right]_F&=&{\rm Tr}\left({\bar{\mathcal B}}{\mathcal P}{\mathcal B}\right)-{\rm Tr}\left({\bar{\mathcal B}}{\mathcal B}{\mathcal P}\right)
\nonumber \\ &=&
{\rm Tr}\left({\bar{\mathcal B}}{\mathcal P}_8{\mathcal B}\right)-{\rm Tr}\left({\bar{\mathcal B}}{\mathcal B}{\mathcal P}_8\right)\ ,
\nonumber \\
\left[{\bar{\mathcal B}}{\mathcal B}{\mathcal P}\right]_D
&=&
{\rm Tr}\left({\bar{\mathcal B}}{\mathcal P}{\mathcal B}\right)+{\rm Tr}\left({\bar{\mathcal B}}{\mathcal B}{\mathcal P}\right)-\frac{2}{3} {\rm Tr}\left({\bar{\mathcal B}}{\mathcal B}\right){\rm Tr}\left({\mathcal P}\right)
\nonumber \\
&=&{\rm Tr}\left({\bar{\mathcal B}}{\mathcal P}_8{\mathcal B}\right)+{\rm Tr}\left({\bar{\mathcal B}}{\mathcal B}{\mathcal P}_8\right)\ ,
\nonumber \\
\left[{\bar{\mathcal B}}{\mathcal B}{\mathcal P}\right]_S&=&{\rm Tr}\left({\bar{\mathcal B}}{\mathcal B}\right){\rm Tr}\left({\mathcal P}\right)={\rm Tr}\left({\bar{\mathcal B}}{\mathcal B}\right){\rm Tr}\left({\mathcal P}_1\right)\ .
\end{eqnarray}
The $SU_f(3)$-invariant baryon-baryon-meson interaction Hamiltonian is a linear combination of these quantities and defined according to \cite{Swa63}
\begin{eqnarray}
m_{\pi^+}{\mathcal H}&=&f_8\sqrt{2}\left(\alpha \left[{\bar{\mathcal B}}{\mathcal B}{\mathcal P}\right]_F+(1-\alpha)\left[{\bar{\mathcal B}}{\mathcal B}{\mathcal P}\right]_D\right)+
\nonumber \\ &&
f_1\sqrt{\frac{1}{3}}\left[{\bar{\mathcal B}}{\mathcal B}{\mathcal P}\right]_S\ .
\end{eqnarray}
Here, $\alpha$ is the $F/(F+D)$-ratio. The most general interaction Hamiltonian that is invariant under isospin transformations is given by
\begin{eqnarray}
m_{\pi^+}{\mathcal H}_1&=&\left[f_{NN\eta_1}\left(\bar{N}N\right)+f_{\Lambda\Lambda\eta_1}\left(\bar{\Lambda}\Lambda\right)+f_{\Sigma\Sigma\eta_1}\left(\bar{\mathbf{ \Sigma}}\cdot\mathbf{ \Sigma}\right)
\right. \nonumber \\ && \left.
+f_{\Xi\Xi\eta_1}\left(\bar{\Xi}\Xi\right)\right]\eta_1\ ,\nonumber \\
m_{\pi^+}{\mathcal H}_8&=&f_{NN\pi}\left(\bar{N}\mbox{\boldmath $\tau$}N\right)\cdot\mbox{\boldmath $\pi$} -if_{\Sigma\Sigma\pi}\left(\bar{\mathbf{ \Sigma}}\times\mathbf{ \Sigma}\right)\cdot\mbox{\boldmath $\pi$} \nonumber \\
&&+f_{\Lambda\Sigma\pi}\left(\bar{\Lambda}\mathbf{ \Sigma}+\bar{\mathbf{ \Sigma}}\Lambda\right)\cdot\mbox{\boldmath $\pi$}+f_{\Xi\Xi\pi}\left(\bar{\Xi}\mbox{\boldmath $\tau$}\Xi\right)\cdot\mbox{\boldmath $\pi$} \nonumber \\
&&+f_{\Lambda NK}\left[\left(\bar{N}K\right)\Lambda+\bar{\Lambda}\left(K^\dagger N\right)\right] 
\nonumber \\&&
+f_{\Xi\Lambda K}\left[\left(\bar{\Xi}K_c\right)\Lambda+\bar{\Lambda}\left(K^\dagger_c\Xi\right)\right]
\nonumber \\&&
+f_{\Sigma NK}\left[\bar{\mathbf{ \Sigma}}\cdot\left(K^\dagger\mbox{\boldmath $\tau$}N\right)+\left(\bar{N}\mbox{\boldmath $\tau$}K\right)\cdot\mathbf{ \Sigma}\right]
\nonumber \\&&
+f_{\Sigma \Xi K}\left[\bar{\mathbf{ \Sigma}}\cdot\left(K^\dagger_c\mbox{\boldmath $\tau$}\Xi\right)+\left(\bar{\Xi}\mbox{\boldmath $\tau$}K_c\right)\cdot\mathbf{ \Sigma}\right]
\nonumber \\&&
+f_{NN\eta_8}\left(\bar{N}N\right)\eta_8
+f_{\Lambda\Lambda\eta_8}\left(\bar{\Lambda}\Lambda\right)\eta_8
\nonumber \\&&
+f_{\Sigma\Sigma\eta_8}\left(\bar{\mathbf{ \Sigma}}\cdot\mathbf{ \Sigma}\right)\eta_8
+f_{\Xi\Xi\eta_8}\left(\bar{\Xi}\Xi\right)\eta_8 \ ,
\label{eq:40}
\end{eqnarray}
for the singlet and octet coupling respectively, and $f_{NN\pi}=f_8$ and $f_{NN\eta_1}=f_{\Lambda\Lambda\eta_1}=f_{\Sigma\Sigma\eta_1}=f_{\Xi\Xi\eta_1}=f_1$.  We have introduced the isospin doublets\\
\begin{equation}
N=\left(\begin{array}{c}\! p\! \\ \! n\! \end{array}\right),\ \Xi=\left(\begin{array}{c}\! \Xi^0\! \\\! \Xi^-\! \end{array}\right),\ K=\left(\begin{array}{c}\! K^+\! \\\! K^0\! \end{array}\right),\ K_c=\left(\begin{array}{c}\! \bar{K}^0\! \\\! -K^-\! \end{array}\right)\ ,
\end{equation}
the phases have been chosen according to \cite{Swa63}, such that the inner product of the isovectors $\mathbf{ \Sigma}$ and $\mbox{\boldmath $\pi$}$ is
\begin{equation}
\mathbf{ \Sigma}\cdot\mbox{\boldmath $\pi$}=\Sigma^+\pi^-+\Sigma^0\pi^0+\Sigma^-\pi^+\ .
\end{equation}
The interaction Hamiltonians in Eq. (\ref{eq:40}) are invariant under $SU_f(3)$ transformations if the coupling constants are expressed in terms of the octet coupling $f_8\equiv f$ and $\alpha$ as, \cite{Swa63},
\begin{equation}
\begin{array}{rlrl}
f_{NN\pi}  = & f & f_{NN\eta_8}  = & \frac{1}{\sqrt{3}}(4\alpha -1)f\\
f_{\Xi\Xi\pi}  = & -(1-2\alpha)f &  f_{\Xi\Xi\eta_8}  = & -\frac{1}{\sqrt{3}}(1+2\alpha )f\\
f_{\Lambda\Sigma\pi}  = & \frac{2}{\sqrt{3}}(1-\alpha)f & f_{\Sigma\Sigma\eta_8}  = & \frac{2}{\sqrt{3}}(1-\alpha )f\\
f_{\Sigma\Sigma\pi}  = & 2\alpha f &  f_{\Lambda\Lambda\eta_8}  = & -\frac{2}{\sqrt{3}}(1-\alpha )f\\ 
f_{\Lambda NK}  = & -\frac{1}{\sqrt{3}}(1+2\alpha)f  & f_{\Xi\Lambda K}  = & \frac{1}{\sqrt{3}}(4\alpha-1)f  \\
f_{\Sigma NK}  = & (1-2\alpha)f  & f_{\Xi\Sigma K} = & -f  \ ,
\end{array} \\
\end{equation}
and the singlet coupling $f_1$ as
\begin{equation}
f_{NN\eta_1}=f_{\Lambda\Lambda\eta_1}=f_{\Sigma\Sigma\eta_1}=f_{\Xi\Xi\eta_1}=f_1\ .
\end{equation}
The baryon-baryon-meson vertices are thus characterized by only four parameters if $SU_f(3)$-symmetry is assumed, the octet coupling constant $f_8$, the singlet coupling constant $f_1$, the $F/(F+D)$-ratio $\alpha$ and the mixing angle, which gives the relation between the physical and octet and singlet isoscalar mesons. 
The $SU_f(3)$ invariant local interaction densities we use for the triple-meson (MMM) vertices are given below.
\begin{itemize}
\item[(i)] \underline{$J^{PC}=1^{--}$ Vector-mesons}:
\begin{eqnarray}
{\cal H}_{PPV} &=& g_{PPV}\ f_{abc}\ V_\mu^a\ P^b \stackrel{\leftrightarrow}{\partial}^\mu P^c \nonumber \\ 
&=& -i\sqrt{2}\ g_{PPV}\ Tr\ {\cal P}_8 \left(\partial_\mu{\cal P}_8\cdot{\cal V}^\mu_8 -{\cal V}^\mu_8\ \partial_\mu{\cal P}_8\right) \nonumber \\
 &=& g_{PPV}\left[ \mbox{\boldmath $\rho$}_\mu\cdot\left( \mbox{\boldmath $\pi$}
 \times\stackrel{\leftrightarrow}{\partial}^\mu \mbox{\boldmath $\pi$} +
 iK^\dagger \mbox{\boldmath $\tau$} \stackrel{\leftrightarrow}{\partial}^\mu K\right) +                  
 \right. \nonumber \\ && 
 \left( i K^{*\dagger}_\mu\ \mbox{\boldmath $\tau$} K\cdot
 \stackrel{\leftrightarrow}{\partial}^\mu \mbox{\boldmath $\pi$} + H.c. \right) +
\nonumber \\ && 
\sqrt{3}\left( i K^{*\dag}_\mu K
 \stackrel{\leftrightarrow}{\partial}^\mu \eta + H.c. \right) +
\nonumber \\ &&
\left.
 \sqrt{3}\ i \varphi_{8,\mu}\ K^\dag
 \stackrel{\leftrightarrow}{\partial}^\mu K \right]\ ,
\label{eq:7.6} \end{eqnarray}
where $H.c.$ stands for the Hermitian conjugate of the preceding term, and we use the usual notation for the derivative $\stackrel{\leftrightarrow}{\partial}^\mu$ acting on the pseudoscalar-mesons, $P^b \stackrel{\leftrightarrow}{\partial}^\mu P^c \equiv  P^b\left(\partial^\mu P^c\right) - \left(\partial^\mu P^b\right)\cdot P^c$. The coupling of the vector-mesons to the pseudoscalar-mesons is $SU_f(3)$ antisymmetric, the symmetric coupling can be excluded by invoking a generalized Bose symmetry for the pseudoscalar-mesons, interchanging the two pseudoscalar-mesons leaves ${\cal H}_{PPV}$ invariant.
The coupling constant for the decay of a $\rho$-meson into two pions is defined as $g_{\pi\pi\rho}=2\,g_{PPV}$, which can be estimated using the decay width of the $\rho$-meson, see Eq. (\ref{eq:dec1}).

\item[(ii)] \underline{$J^{PC}=0^{++}$ Scalar-mesons}:
\begin{eqnarray}
{\cal H}_{PPS}& = &\frac{\sqrt{3}}{2}\ g_{PPS} \ d_{abc}\ S^a\ P^b\ P^c  \nonumber \\
&=& \frac{\sqrt{3}}{2\sqrt{2}}\ g_{PPS}\ 
 Tr\ {\cal P}_8\left({\cal P}_8\cdot{\cal S}_8 + {\cal S}_8\cdot{\cal P}_8\right)\ \nonumber \\
&=&g_{PPS} 
\left[\mbox{\boldmath{$a$}}_0\cdot\left(\mbox{\boldmath{$\pi$}}\eta+\frac{\sqrt{3}}{2}K^\dag\mbox{\boldmath{$\tau$}}K\right) 
+
\right. \nonumber \\ &&
\frac{\sqrt{3}}{2}\left(K_0^\dag\mbox{\boldmath{$\tau$}}K\cdot\mbox{\boldmath{$\pi$}}+H.c.\right)
-\frac{1}{2}\left(K_0^\dag K\eta+H.c.\right)
\nonumber \\
&&\left.
+\frac{1}{2}f_0\left(\mbox{\boldmath{$\pi$}}\cdot\mbox{\boldmath{$\pi$}}-K^\dag K-\eta\eta\right) \vphantom{\frac{\sqrt{3}}{2}}\right]\ .
\label{eq:7.7} 
\end{eqnarray}
For the scalar-mesons we have a symmetric coupling. The dimensionless coupling constant for the decay of the $\sigma$-meson into two pions is defined as $g_{\pi\pi\sigma}=g_{PPS}/ m_{\pi^+}$, which can be estimated using the decay width of the $\sigma$-meson, see Eq. (\ref{eq:dec1}).
    
\item[(iii)] \underline{$J^{PC}=2^{++}$ Tensor-mesons}:
\begin{eqnarray}
{\cal H}_{PPT} 
&=&\frac{2g_{PPT}}{m_{\pi^+}} \left[\mbox{\boldmath{$a$}}_2^{\mu\nu}\cdot\left(\partial_\mu\mbox{\boldmath{$\pi$}}\partial_\nu\eta+\frac{\sqrt{3}}{2}\partial_\mu K^\dag\mbox{\boldmath{$\tau$}}\partial_\nu K\right) \right. \nonumber \\ 
&&+\frac{\sqrt{3}}{2}\left(K_2^{\mu\nu\dag}\mbox{\boldmath{$\tau$}}\partial_\mu K\cdot\partial_\nu\mbox{\boldmath{$\pi$}}+H.c.\right)-
\nonumber \\ &&
\frac{1}{2}\left(K_2^{\mu\nu\dag} \partial_\mu K\partial_\nu\eta+H.c.\right) \nonumber \\
&&\left.+\frac{1}{2}f_2^{\mu\nu}\left(\partial_\mu\mbox{\boldmath{$\pi$}}\cdot\partial_\nu\mbox{\boldmath{$\pi$}}-\partial_\mu K^\dag \partial_\nu K-\partial_\mu\eta\partial_\nu\eta\right) \vphantom{\frac{\sqrt{3}}{2}}\right]\ .\nonumber \\
\label{eq:7.8} 
\end{eqnarray}
The coupling constant for the decay of the $f_2$-meson into two pions is given by $g_{\pi\pi f_2}=g_{PPT}$, which is estimated in Eq. (\ref{eq:dec1}).
\end{itemize}
Some numerical values for the previous coupling constants are given by Nagels et al. \cite{NRS79}.
The isospin factors resulting from the previous interactions are discussed in Appendix \ref{app:AA} and listed in Tables \ref{tab:6.0} and \ref{tab:7.0} for $\pi N$ and $K^+ N$ interactions respectively.
We remark that in the NSC model the $SU_f(3)$-symmetry is broken dynamically, since we use the physical masses for the baryons and mesons. The $SU_f(3)$-symmetry for the coupling constants is not necessarily exact, in fact, we allow for a breaking in the NSC $K^+ N$-model, Sec. \ref{chap:7}.


\section{Renormalization}
\label{chap:4}
\label{sec:19}
The Lagrangians used are effective Lagrangians, expressed in terms of the physical coupling constants and masses. Then, in principle, counter-terms should be added to the Lagrangian and fixed by renormalization conditions. This is particularly to the point in channels where bound-states and
resonances occur. For example, the famous $\Delta$ resonance 
at $M_\Delta=1232$ MeV in the $\pi N$ system. The $\Delta$ pole diagram gets 
``dressed'' when it is iterated with other graphs upon insertion in an integral equation. Also, it appears that by using only $u$-channel and $t$-channel forces it is impossible to describe the experimental $\pi N$ phases above resonance in the $P_{33}$-wave.
From the viewpoint of the quark-model this is natural, because here the 
$\Delta$ resonance is, at least partly, a genuine three-quark state, and 
should not be described as a pure $\pi N$ resonance, but should be treated at the same footing as the nucleons. We take the same attitude 
to the other meson-baryon resonances as the Roper, $S_{11}(1535)$, 
etc.
The resonance diagrams split nicely into a pole part, 
having a $(\sqrt{s}-M_0+i\epsilon)^{-1}$-factor, and a non-pole part having a $(\sqrt{s}+M_0-i\epsilon)^{-1}$-factor. Here, $M_0$ is the so-called 
``bare'' mass.
The pole-position will move to $\sqrt{s}= M_R$, where $M_R$ is the physical mass of the resonance. This determines the bare mass $M_0$.

\begin{figure}[t]
\begin{center}
\resizebox{8.25cm}{2.26cm}{\includegraphics*[1.4cm,22.6cm][10.5cm,25.cm]{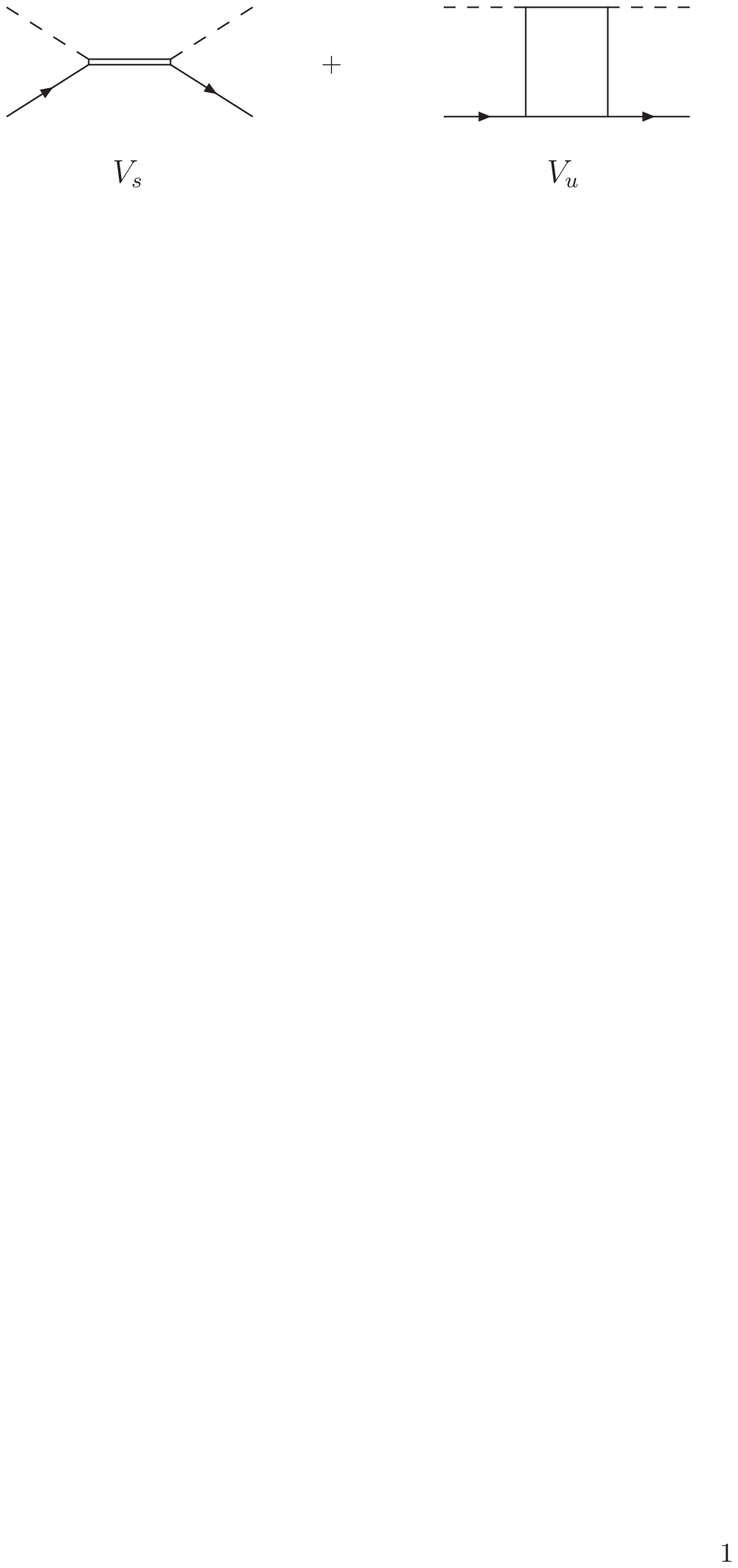}}
\end{center}
\caption{The pole potential $V_s$ contains $s$-channel diagrams, the non-pole potential $V_u$ contains $t$- and $u$-channel diagrams.}
\label{fig:4.1}
\end{figure}

\begin{figure}[h]
\begin{center}
\resizebox{8.25cm}{7.07cm}{\includegraphics*[1.cm,10.9cm][16.cm,25.cm]{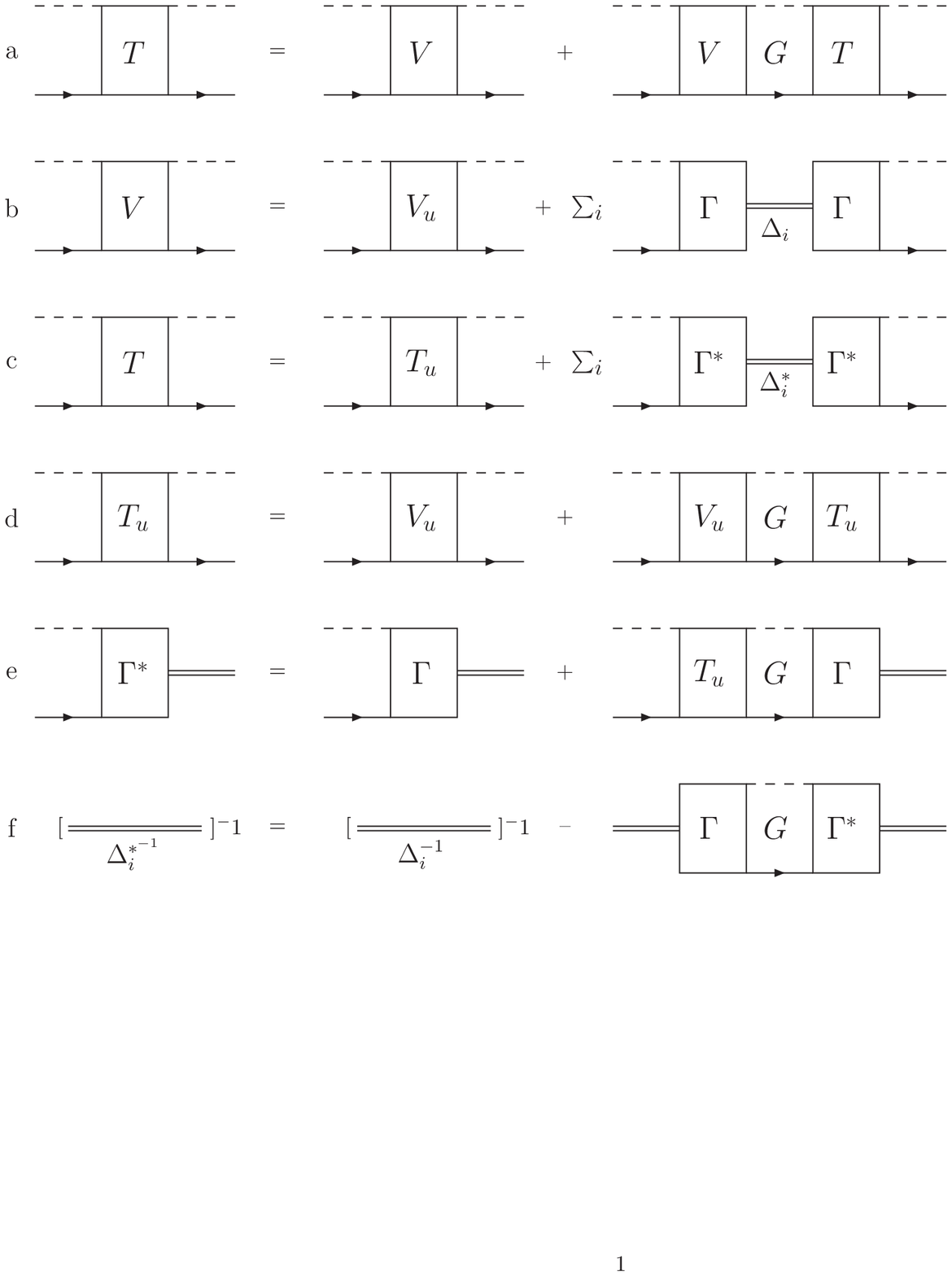}}
\end{center}
\caption{The integral equation for the amplitude in case of a non-pole and pole potential, a. integral equation for the total amplitude 
, b. the potential in terms of the non-pole and pole potential, c. the amplitude in terms of the non-pole and pole amplitude Eq. (\ref{eq:19.9a}), d. integral equation for the non-pole amplitude Eq. (\ref{eq:19.3}), e. equation for the dressed vertex Eq. (\ref{eq:19.9b}), f. equation for the dressed propagator Eq. (\ref{eq:19.10}).}
\label{fig:4.0}
\end{figure}

To implement these ideas, we follow Haymaker \cite{Hay69}. We write the total potential $V$ as a sum of a potential containing poles and a potential not containing poles 
$V({\bf p}',{\bf p}) = V_s({\bf p}',{\bf p}) + V_u({\bf p}',{\bf p})$
\footnote{
Notice that in \cite{Hay69} the $V$- and $T$-matrices differ a (-)-sign with those used here.}, see Figure \ref{fig:4.1}, 
where
\begin{equation}
 V_s({\bf p}',{\bf p}) = \sum_i \Gamma_i({\bf p}')\ \Delta_i(P)\ \Gamma_i({\bf p})
\label{eq:19.1}\end{equation}
is the pole part of the $s$-channel baryon exchanges. In Eq. (\ref{eq:19.1}) the right hand
side is written in terms of the so-called 
``bare'' couplings and masses.
We have $ \Delta_i(P) = (\sqrt{s}-M_0+i\epsilon)^{-1}$, where in the CM system $P=(\sqrt{s},{\bf 0})$. 
The other part and the 
$t$-channel and $u$-channel exchanges are contained in $V_u({\bf p}',{\bf p})$. 
In the following,
we treat explicitly the cases when there is only one $s$-channel bound state or resonance
present. It is easy to generalize this to the case with more $s$-channel poles.
Following \cite{Hay69} we define two $T$-matrices $T_j, j=1,2$ by
\begin{equation}
 T_j = V_j + V_j\ G\ T\ ,\ T = T_1 + T_2\ ,
\label{eq:19.2}\end{equation}
where $V_1=V_s$ and $V_2=V_u$. The amplitude $T_j$ is the sum of all graphs in the iteration of $T$ in which the potential $V_j$ 
``acts last''. Defining $T_u$ as the $T$-matrix for the $V_u$
interaction alone, i.e.
\begin{equation}
 T_u = V_u + V_u\ G\ T_u\ ,
\label{eq:19.3}\end{equation}
it is shown in \cite{Hay69} that 
\begin{equation}
 T_1 = T_u + T_u\ G\ T_2\ ,\  T_2 = T_s + T_s\ G\ T_u\ ,
\label{eq:19.4}\end{equation}
with 
\begin{equation}
 T_s = V_s + V_s\ H_1\ T_s\ ,\  H_1 = G + G\ T_u\ G\ .        
\label{eq:19.5}\end{equation}
Taking together these results one obtains for the total $T$-matrix the expression
\begin{equation}
 T = T_u + T_s + T_u\ G\ T_s + T_s\ G\ T_u + T_u\ G\ T_s\ G\ T_u \ .
\label{eq:19.6}\end{equation}
Since $V_s$ is a separable potential, the solution for $T_s$ in the case of one pole can be written as      
\begin{equation}
 T_s({\bf p}',{\bf p}) = \frac{\Gamma({\bf p}')\ \Gamma({\bf p})}
 {\Delta(P)^{-1}- \Sigma(P)}
 \equiv \Gamma({\bf p}')\ \Delta^*(P)\ \Gamma({\bf p})\ ,
\label{eq:19.7}\end{equation}
where we introduced the shorthand $\Delta=\Delta_i$, and defined the self-energy $\Sigma$ and the dressed propagator $\Delta^*$ by
\begin{eqnarray}
 \Sigma(P) &=& \int\widetilde{dq'}\int\widetilde{dq^{\prime\prime}}\
 \Gamma({\bf q}')\ H_1({\bf q}',{\bf q}^{\prime\prime};P)\ 
 \Gamma({\bf q}^{\prime\prime})\ , \nonumber\\ 
 \Delta^*(P) &=& \frac{\Delta(P)}{1-\Delta(P)\ \Sigma(P)}
\nonumber \\ &=&
 \Delta(P)+ \Delta(P)\ \Sigma(P)\ \Delta^*(P)\ ,
\label{eq:19.8}\end{eqnarray}
where $\widetilde{dq'}=d^3q'/(2\pi)^3$ etc..     
Inserting Eqs. (\ref{eq:19.7}) and (\ref{eq:19.8}) in Eq. (\ref{eq:19.6}), 
and exploiting time-reversal and parity invariance, which gives 
$T_u({\bf p}',{\bf p})=T_u({\bf p},{\bf p}')$, one finds the expressions for the total amplitude, dressed vertex and self-energy
\begin{eqnarray}
 T({\bf p}',{\bf p}) &=& T_u({\bf p}',{\bf p}) + 
 \Gamma^*({\bf p}')\ \Delta^*(P)\ \Gamma^*({\bf p})\ , \label{eq:19.9a}\\
 \nonumber\\
 \Gamma^*({\bf p}) &=& \Gamma({\bf p}) +\!\! \int\! \widetilde{dq}\ \Gamma({\bf q}) G({\bf q},P) T_u({\bf q},{\bf p})\ , \label{eq:19.9b}\\ \nonumber\\ 
 \Sigma(P) &=& \int\widetilde{dq}\ \Gamma({\bf q})\ G({\bf q},P)\ \Gamma^*({\bf q})\ ,
\label{eq:19.9}\end{eqnarray}
where the dressed propagator $\Delta^*(P)$ is given by
\begin{equation}
 \Delta^{*}(P)^{-1} = \Delta(P)^{-1} - \Sigma(P)\ . 
\label{eq:19.10}\end{equation}
The equations above show that the complete $T$-matrix can be computed in a 
straightforward manner, using the full-off-shell $T$-matrix $T_u({\bf p}',{\bf p})$,
defined in Eq. (\ref{eq:19.3}).  The renormalized pole position $\sqrt{s}=M_R$ is determined by the condition
\begin{eqnarray}
 0&=& \Delta^{*}(\sqrt{s}=M_R)^{-1}\nonumber \\
& =& \Delta(\sqrt{s}=M_R)^{-1} - \Sigma(\sqrt{s}=M_R)\ .    
\label{eq:19.11}\end{eqnarray}
A diagrammatic representation of the previous derived equations for the meson-baryon amplitude, potential, dressed vertex and dressed propagator is given in Figure \ref{fig:4.0}.
\subsection{Partial wave analysis}     
\label{sec:19b}
The partial wave expansion for the vertex function $\Gamma$ reads
\begin{eqnarray}
\Gamma({\bf p}) &=& \sqrt{4\pi}\sum_{L,M} \Gamma_L(p)\ Y^L_{M}(\hat{\bf p})\ ,
\label{eq:19.51}\end{eqnarray}
and similar for $\Gamma ^*$. The partial wave expansions for the amplitude $T$ reads 
\begin{equation}
T({\bf q},{\bf p}) = 4\pi \sum_{L,M} T_{L}(q,p)\ 
 Y^L_{M}(\hat{\bf q})^*\  Y^L_{M}(\hat{\bf p})\ .  
\label{eq:19.52}\end{equation}
Then, the partial wave projection of the integrals in Eqs. (\ref{eq:19.9b}) and (\ref{eq:19.9}) become
\begin{eqnarray}
 \Gamma^*_L(p) &=& \Gamma_L(p) +\! \frac{1}{2\pi^2}\int q^2dq\ \Gamma_L(q)\ 
 G(q,P)\ T_{u,L}(q,p)\ , \nonumber\\ 
 \Sigma_L(P) &=& \frac{1}{2\pi^2}\int q^2dq\ \Gamma_L(q)\ 
 G(q,P)\ \Gamma^*_{L}(q)\ .                  
\label{eq:19.53}\end{eqnarray}
In the following subsections it is understood that we deal with the partial wave quantities.
We suppress the angular momenta labels for notational convenience.

\subsection{Multiplicative renormalization parameters}
\label{sec:19c}
To start, in Eq. (\ref{eq:19.9a}) the second part on the right hand side we consider 
to be given in terms of the bare resonance mass $M_0$ and the bare 
resonance coupling $g_0$. 
We consider only the wave function and 
vertex renormalization
for the resonance, and use the multiplicative renormalization method. Then, since 
the total Lagrangian is unchanged and hermitian, unitarity is preserved. The 
Z-transformation for the resonance field reads $\Psi_{0,\mu}= \sqrt{Z_2} \Psi_{r,\mu}$, 
and for the 
resonance coupling $g_0= Z_g\ g_r$, where
the subscripts $r$ and $0$ refer to respectively the ''renormalized'', and ''bare'' field.
Applied to the $\Delta N\pi$ interaction  this gives
\begin{eqnarray}
 {\cal L}_I \sim g_0 \bar{\Psi}_{0,\mu} \psi \partial^\mu\phi =
 Z_g\sqrt{Z_2}\
 g_r \bar{\Psi}_{r,\mu} \psi \partial^\mu\phi\ , 
\label{eq:19.12}\end{eqnarray}
where $g_r = f_{\Delta N \pi}/m_{\pi^+}$ is the renormalized, i.e. the physical,
and $g_0$ the unrenormalized, i.e. the bare coupling. Introducing the renormalization constant $Z_1= Z_g \sqrt{Z_2}$, we have
\begin{eqnarray}
 {\cal L}_I &\sim & Z_1 g_r \bar{\Psi}_{r,\mu} \psi \partial^\mu\phi 
\nonumber \\
  &=& g_r \bar{\Psi}_{r,\mu} \psi \partial^\mu\phi +
 (Z_1-1) g_r \bar{\Psi}_{r,\mu} \psi \partial^\mu\phi \ .
\label{eq:19.13}\end{eqnarray}From the form of Eq. (\ref{eq:19.9b}) it is useful at this stage to distinguish functions with the bare and physical couplings $g_0$ and $g_r$. Therefore, we introduce the  vertex functions
\begin{equation}
 \Gamma^*_{u,r}(p) = \Gamma_{u,r}(p) + \int\widetilde{dq}\ \Gamma_{u,r}(q)\ 
 G(q)\ T_u(q,p)\ ,  
\label{eq:19.14}\end{equation}
with the definitions
\begin{equation}
 \Gamma_{u,r} (p) = g_{0,r}\ \bar{\Gamma}(p)\ ,\ 
 \Gamma_{u,r}^* (p) = g_{0,r}\ \bar{\Gamma}^*(p)\ ,  
\label{eq:19.15}\end{equation}
implying the relations
\begin{eqnarray}
&& \Gamma_u (p) = Z_g\ \Gamma_r(p)\ ,\ 
 \Gamma^*_u (p) = Z_g\ \Gamma^*_r(p)\ ,\ 
\nonumber \\ &&
 \Sigma_u (P) = Z_g^2\ \Sigma_r(P)\ .\ 
\label{eq:19.16}\end{eqnarray}

\subsubsection{Resonance renormalization}
Working out this renormalization scheme for the baryon resonances, we
start, in Eq. (\ref{eq:19.9a}) with the second part on the right hand side as given    
in terms of the bare resonance mass $M_0$ and bare 
resonance coupling $g_0$. 
We write this part of the amplitude as
\begin{equation}
 T_{res}(p',p) = \Gamma^*_u(p')\frac{1}{\sqrt{s}-M_0-\Sigma_u(\sqrt{s})}\ \Gamma^*_u(p)\ .    
\label{eq:19.20}\end{equation}
Next, we develop the denominator around the renormalized, i.e. the physical,
resonance mass $M_R$ and rearrange terms. We get
\begin{eqnarray}
 T_{res}(p',p) &=& \Gamma^*_u(p')
\left[\vphantom{\frac{\partial\Sigma_u}{\partial\sqrt{s}}}
\sqrt{s}-M_0-\Sigma_u(M_R)-
\right. \nonumber \\ &&\left.\hspace{1.2cm}
\left(\sqrt{s}-M_R\right)\frac{\partial\Sigma_u}{\partial\sqrt{s}}\ 
 \ldots \right]^{-1}\Gamma^*_u(p) \nonumber\\
 &=& \Gamma^*_u(p')\left[\vphantom{\frac{\partial\Sigma_u}{\partial\sqrt{s}}} (\sqrt{s}-M_R)-
\right. \nonumber \\ &&\left.\hspace{1.2cm}
\left(\sqrt{s}-M_R\right)\frac{\partial\Sigma_u}{\partial\sqrt{s}}\ 
 \ldots \right]^{-1}\Gamma^*_u(p) \nonumber\\
 &=& \Gamma^*_u(p') Z_2 \Gamma^*_u(p)\left[\vphantom{\frac{\partial^2\Sigma_u}{\left(\partial\sqrt{s}\right)^2}}       
\left(\sqrt{s}-M_R\right)-
\right. \nonumber \\ && \left.
\frac{1}{2}\left(\sqrt{s}-M_R\right)^2 Z_2 
 \frac{\partial^2\Sigma_u}{(\partial\sqrt{s})^2}\ \ldots \right]^{-1}\ .
\label{eq:19.21}\end{eqnarray}
Here, we have introduced the renormalization constant $Z_2$ defined by 
\begin{eqnarray}
  Z_2 &\equiv &
\left( 1- \left.\frac{\partial\Sigma_u}{\partial\sqrt{s}}\right|_{\sqrt{s}=M_R}
 \right)^{-1}
\nonumber \\
&=&\left( 1- Z_g^2\frac{\partial\Sigma_r}{\partial\sqrt{s}}\right)^{-1} 
= 1+ Z_1^2
\frac{\partial\Sigma_r}{\partial\sqrt{s}}
\  .
\label{eq:19.22}\end{eqnarray}
The derivatives in Eq. (\ref{eq:19.21}) w.r.t. $\sqrt{s}$ are
evaluated at the point $\sqrt{s}=M_R$, as is indicated in Eq. (\ref{eq:19.22}).

Now we require that the residue at the resonance pole is given in terms of the
physical coupling, i.e. $g_r$.
In terms of the renormalized quantities the amplitude $T_{res}$ of Eq. (\ref{eq:19.20}) reads
\begin{equation}
 T_{res}(p',p) = \Gamma^*_{ren}(p')\frac{1}{\sqrt{s}-M_R-\Sigma_{ren}^{(2)}(\sqrt{s})}\ 
 \Gamma^*_{ren}(p)\ .    
\label{eq:19.27}\end{equation}
Here we have defined the renormalized self-energy and the renormalized dressed vertex
\begin{equation}
\Sigma_{ren}^{(2)}(\sqrt{s})\equiv Z_2\Sigma_u^{(2)}(\sqrt{s})\ ,\Gamma_{ren}^*(p)\equiv \sqrt{Z_2}\Gamma_u^*(p)\ .
\label{eq:19.27b}
\end{equation}
The renormalized self-energy in the last expression in Eq. (\ref{eq:19.21}) and its first derivative are defined to be zero at the resonance position $\sqrt{s}=M_R$ and is given by 
\begin{eqnarray}
 \Sigma_{ren}^{(2)}(\sqrt{s}) &\equiv&\frac{1}{2}\left(\sqrt{s}-M_R\right)^2\left.\frac{\partial^2\Sigma_{ren}}{(\partial\sqrt{s})^2}\right|_{\sqrt{s}=M_R} +\ldots  \nonumber \\
&=&\Sigma_{ren}(\sqrt{s})-\Sigma_{ren}(M_R) 
\nonumber \\ &&
- \left(\sqrt{s}-M_R\right)\left.\frac{\partial\Sigma_{ren}}{\partial\sqrt{s}}\right|_{\sqrt{s}=M_R} \ .
\label{eq:19.26}\end{eqnarray}
We notice that the imaginary part of the self-energy is not changed by the
wave function renormalization. It is straightforward to include $\Im\Sigma(\sqrt{s})$in the resonance mass $M_R$ as well as in $\Sigma_{ren}(\sqrt{s})$.

The computation of the amplitude $T_{res}(p',p)$, Eq. (\ref{eq:19.27}), using renormalized quantities only runs as follows. From Eqs. (\ref{eq:19.16}) and (\ref{eq:19.27b}) and the definition $Z_1=Z_g\sqrt{Z_2}$ the renormalized vertex is given by
\begin{equation}
\Gamma^*_{ren}(p)=Z_1\Gamma^*_r(p)\ .
\end{equation}
{\it Notice that $\Gamma ^*_r (p_R)=\left|\Gamma^*_r(p_R)\right|exp(i\varphi ^*_r(p_R))$, and that this phase can be ignored when defining the effective decay Lagrangian in Eq. (\ref{eq:19.13})}. The renormalization condition for the vertex is that at the pole position ($\sqrt{s}=M_R$) the renormalized vertex is given in terms of the physical coupling constant
\begin{equation}
 \left| \Gamma^*_{ren} (p=p_R)\right| = Z_1\ \left|g_r \bar{\Gamma}^*(p=p_R)\right| = g_r \frac{p_R}{\sqrt{3}} \sqrt{E_R+M}\ ,
\label{eq:19.18}\end{equation}
which determines $Z_1$ and, by Eq. (\ref{eq:19.22}), $Z_2$ and $Z_g$, now the renormalized self-energy and the renormalized dressed vertex are known from Eq. (\ref{eq:19.27b}). In passing we note that the coupling $g_r = f_{\Delta N\pi}/m_\pi$, and the other factors in the second expression of Eq. (\ref{eq:19.18}) are specific for a $P_{33}$-wave resonance.

{\it As is clear from this section one can either express all quantities in terms of the bare parameters $(M_0,g_0)$ or in terms of the renormalized parameters $(M_R,g_r)$}.

For the second part of this statement we now express the bare quantities in terms of the renormalized ones. From Eqs. (\ref{eq:19.22}) and (\ref{eq:19.18}) we know $Z_g$, thus
\begin{equation}
g_0^2=Z_g^2g_r^2\ .
\end{equation}
In the following, we denote the real part of the resonance mass by $M_R$. Also, we want to renormalize at a point which is experimentally accessible. Therefore, we choose for the renormalization point the real part of the resonance position, $\sqrt{s}=M_R$. So actually we consider the real part of the self-energy, $\Re\Sigma$, in the previous derivations and from Eq. (\ref{eq:19.21}) we have
\begin{equation}
 M_R=M_0 + g_0^2\ \Re\bar{\Sigma}(M_R)\ ,     
\label{eq:19.29}\end{equation}
giving the bare mass in terms of the renormalized quantities
\begin{equation}
 M_0 = M_R -Z_g^2g_r^2\ \Re\bar{\Sigma}(M_R)\ .
\label{eq:19.29a}\end{equation}

{\it This concludes the demonstration that one may start with the physical parameters and compute the bare parameters $(g_0,M_0)$. Of course, in exploiting $M_0$ in order to force the pole position at the chosen $\sqrt{s}=M_R$ to be reasonable one must have $M_0 > 0$}.

\subsubsection{Nucleon pole renormalization}
The renormalization of the nucleon pole is completely analogous to the resonance renormalization, except for the renormalization point, which is now the nucleon mass and thus below the $\pi N$ threshold. Here the Green's function has no pole and is real. This implies that $\Re\Sigma (M_N)=\Sigma (M_N)$, in contrast to the resonance case.
 All quantities in the expression for the self-energy, Eq. (\ref{eq:19.9}), are real at the nucleon pole.

The renormalization condition for the vertex, analogous to Eq. (\ref{eq:19.18}), is that at the nucleon pole position ($\sqrt{s}=M_N$) the renormalized vertex is given in terms of the physical coupling constant
\begin{eqnarray}
 \left|\Gamma^*_{ren} (p=ip_N)\right| &=& Z_1 \left|g_r\bar{\Gamma}^*(p=ip_N)\right|
\nonumber \\&=&
 \left|\frac{f_r}{m_{\pi}} \frac{\sqrt{3}\ i\;p_N}{ \sqrt{E_N+M}}(\sqrt{s}+M)\right| \ ,
\label{eq:19.18y}\end{eqnarray}
in case of pv-coupling. This determines the renormalization constant $Z_1$. In passing we note that the factor in the second expression of Eq. (\ref{eq:19.18y}) is specific for a $P_{11}$-wave nucleon pole. Since the nucleon pole position lies below the $\pi N$ threshold, $\Gamma^*(ip_N)$ and in Eq. (\ref{eq:19.9b}) $\Gamma(ip_N)$ and $T_u(q,ip_N)$ are imaginary.

\subsection{Generalization to the multi-pole case}
\label{sec:19d}
In case of multiple pole contributions we have the generalized expression for the pole potential Eq. (\ref{eq:19.1})
\begin{equation}
 V_s({\bf p}',{\bf p}) = \sum_i \Gamma_i({\bf p}')\ \Delta_i(P)\ \Gamma_i({\bf p})\ .
\label{eq:19.35}\end{equation}
From Eq. (\ref{eq:19.5}) one finds, using Eq. (\ref{eq:19.35}) that the pole amplitude $T_s$ can be written as
\begin{equation}
 T_s({\bf p}',{\bf p}) = \sum_i \Gamma_i({\bf p}')\ \Delta_i(P)\ A_i({\bf p})\ .
\label{eq:19.36}\end{equation}
Substituting this again in Eq. (\ref{eq:19.5}) one finds
\begin{eqnarray}
\Gamma_i({\bf p})&=&
 \left[\Delta_i^{-1}(P) \delta_{ij}- 
\int\int \Gamma_i({\bf p}^{\prime\prime})\
\times \right. \nonumber \\ &&\left.
 H_1({\bf p}^{\prime\prime},{\bf p}';P)\ \Gamma_j({\bf p}')\vphantom{\frac{A}{A}}\right]\Delta_j(P) A_j({\bf p})\ ,
\label{eq:19.37}\end{eqnarray}
which can be solved, and leads to the separable $T_s$-matrix
\begin{eqnarray}
 T_s({\bf p}',{\bf p}) &=& \sum_{ij} \Gamma_i({\bf p}')\ 
 \left[\Delta^{-1}(P) - \int\int \Gamma({\bf p}^{\prime\prime})\
\times \right. \nonumber \\ &&\left.
 H_1({\bf p}^{\prime\prime},{\bf p}';P)\ \Gamma({\bf p}')
 \vphantom{\frac{A}{A}}\right]^{-1}_{ij} \Gamma_j({\bf p}) \nonumber\\
  &\equiv& \sum_{ij} \Gamma_i({\bf p}')\ \left[\Delta^{-1}(P) - \Sigma(P)
 \vphantom{\frac{A}{A}}\right]^{-1}_{ij} \Gamma_j({\bf p})\ , \nonumber \\
\label{eq:19.38}\end{eqnarray}
which obviously is a generalization of Eq. (\ref{eq:19.7}).  In Eq. (\ref{eq:19.38}) the quantities $\Delta^{-1}(P)$, $\Gamma({\bf p})$, and $H_1({\bf p}'',{\bf p}';P)$ stand 
respectively for a diagonal matrix, a vector, and a constant in resonance-space.
Above, we have introduced the generalized self-energy in resonance-space as
\begin{equation}
 \Sigma_{ij}(P) =  \int\int \Gamma_i({\bf p}^{\prime\prime})\
 H_1({\bf p}^{\prime\prime},{\bf p}';P)\ \Gamma_j({\bf p}')\ .
\label{eq:19.39}\end{equation}

\subsection{Baryon mixing}
\label{sec:19e}
In this paragraph we consider the case of two different nucleon states, called $N_1$ and $N_2$.
Apart from their masses they have identical quantum numbers. In particular, this applies to the 
$(I=\frac{1}{2},J^P=\frac{1}{2}^+)$-states 
$N$ and the Roper resonance, i.e. the $P_{11}$-wave. Obviously, the 
resonance-space is two-dimensional. Starting with the bare states $N_1$ and $N_2$, these 
states will communicate with each other through the transition to the $\pi N$-states, and will 
themselves not be eigenstates of the strong Hamiltonian. The 
eigenstates of the strong Hamiltonian are identified with the physical states $N$ and
the Roper, which are mixtures of $N_1$ and $N_2$.
To perform the renormalization similarly to the case with only one resonance, we have 
in order to define the physical couplings at the physical states to diagonalize the 
propagator. This can be achieved using a complex orthogonal $2\times 2$-matrix ${\cal O}$, 
${\cal O}\widetilde{\cal O} = \widetilde{\cal O}{\cal O} = 1$. 
We can write, similar to Pascalutsa and Tjon \cite{Pas00},
\begin{equation}
 {\cal O} = \left( \begin{array}{cc}
 \cos\chi & \sin\chi \\ -\sin\chi & \cos\chi \end{array} \right)\ ,
\label{eq:19.40}\end{equation}
where $\chi$ is the complex $(N_1,N_2)$-mixing angle.
Now, since $N_1$ and $N_2$ have the same quantum numbers, apart from their couplings and masses,
their $\pi N$-vertices are isomorphic. This implies that the self-energy matrix in Eq. (\ref{eq:19.39})
can be written as
\footnote{Notice that we distinguish the nucleon in the $\pi N$-state from $N_{1,2}$-states.}
\begin{eqnarray}
\lefteqn{
 \left(\begin{array}{cc} 
 \Sigma_{11}(P) & \Sigma_{12}(P) \\ \Sigma_{21}(P) & \Sigma_{22}(P) \end{array}\right)_u =
}\nonumber \\&&
 \left(\begin{array}{cc} 
 g_{N_1 N\pi}^2 & g_{N_1 N\pi} g_{N_2 N\pi} \\
 g_{N_1 N\pi} g_{N_2 N\pi} & g_{N_2 N\pi}^2  \end{array}\right)_u \bar{\Sigma}(P)\ ,
\label{eq:19.41}\end{eqnarray}
while for the vertices we have
\begin{equation}
 \left(\begin{array}{c} 
 \Gamma_{N_1} \\ \Gamma_{N_2} \end{array}\right)_u =
 \left(\begin{array}{c} 
 g_{N_1 N\pi} \\ g_{N_2 N\pi} \end{array}\right)_u
 \bar{\Gamma}\ .
\label{eq:19.42}\end{equation}
The propagator in Eq. (\ref{eq:19.38}) is diagonalized by the angle
\begin{eqnarray}
\chi(P) &=& \frac{1}{2} \arctan\left[ 2\left(\vphantom{\left(\frac{M_{N_2}}{g_{N_2}\Sigma(P)}\right)^{-1}_u} \frac{g_{N_1N\pi}}{g_{N_2N\pi}} 
 - \frac{g_{N_2N\pi}}{g_{N_1N\pi}}
\right.\right. \nonumber \\ &&\left. \left.
 -\frac{M_{N_2}-M_{N_1}}{g_{N_1N\pi}g_{N_2N\pi}\bar{\Sigma}(P)}\right)_u^{-1}\right]\ .     
\label{eq:19.43}\end{eqnarray}
We write $\Sigma=\Sigma_u$ in the following for notational convenience. The corresponding eigenvalues are 
\begin{eqnarray}
 \Delta^*(P)^{-1}(\pm) &=& \sqrt{s}- \frac{1}{2}\left(M_{0,1}+M_{0,2}\right) - \Sigma(\pm,P)\ ,\nonumber \\
 \Sigma(\pm,P) &=&\left[\vphantom{\left[\Sigma_{22}(P)^2\right]^{1/2}} \left(\Sigma_{11}(P)+\Sigma_{22}(P)\right) \pm 
 \left[\vphantom{\left(\Sigma_{22}(P)\right)^2}
\left(M_{0,2}-M_{0,1}
\right. \right. \right. \nonumber \\ &&\left. 
+\Sigma_{22}(P)-\Sigma_{11}(P)\right)^2 
\nonumber \\ && \left. \left.
+ 4 \Sigma_{12}(P)^2\right]^{1/2}\right]/2\ . 
\label{eq:19.44}
\end{eqnarray}
Here, we denoted the unrenormalized masses by $M_{0,1}=M_{N_1}$ for the nucleon, and by $M_{0,2}=M_{N_2}$ 
for the Roper resonance. Likewise, the unrenormalized couplings are denoted as 
$g_{0,1} \equiv g_{N_1N\pi,u}$ and $g_{0,2} \equiv g_{N_2N\pi,u}$. Then, for example 
$\Sigma_{ij}(P) = g_{0,i} g_{0,j} \bar{\Sigma}(P)$. 
The resonance amplitude $T_{res}$ is a generalization of the second term in Eq. (\ref{eq:19.9a}) and can be rewritten as follows
\begin{eqnarray}
 T_{res}(p',p) &=& 
  \sum_{ij} \Gamma^*_i(p')\ \Delta^{*}_{ij}(P) \Gamma^*_j(p) \nonumber\\ &=&
  \sum_{i} \left(\Gamma^*(p'){\cal O}\right)_i \left(\widetilde{\cal O}\Delta^{*}(P){\cal O}\right)_{ij}
 \left(\widetilde{\cal O}\Gamma^*(p)\right)_j \nonumber\\  &=&
  \sum_{\alpha=\pm} \left(\Gamma^*(p'){\cal O}\right)_\alpha d_\alpha^{-1}(P) \left(\widetilde{\cal O}\Gamma^*(p)\right)_\alpha ,                 
\label{eq:19.45}\end{eqnarray}
where the diagonalized propagator is
\begin{eqnarray}
  d_\alpha(P) &=& \sqrt{s}- \frac{1}{2}\left(M_{0,1}+M_{0,2}\right) - \Sigma(\alpha,P) .                 
\label{eq:19.46}\end{eqnarray}
Unlike in \cite{Pas98} we renormalize the eigenstate $\alpha=(-)$ at the nucleon pole, and the 
eigenstate $\alpha=(+)$ at the Roper resonance position. That is the reason why we formulate the 
procedure in terms of the bare or unrenormalized parameters and not directly in terms of the 
physical parameters. This way we can utilize Eqs. (\ref{eq:19.41}) and (\ref{eq:19.42}). As we will see,
we get four equations from the renormalization conditions on the masses and couplings, with the set of 
four unknowns $\left\{ M_{0,1}, M_{0,2}, g_{0,1}, g_{0,2}\right\}$. 

For both $\alpha$-solutions we have, using $M_0= (M_{0,1}+M_{0,2})/2$, that the resonance amplitude is
\begin{eqnarray}
 T_{res}(\alpha) &=& \Gamma^*_u(\alpha,p')\frac{1}{\sqrt{s}-M_0-\Sigma(\alpha,\sqrt{s})}\ \Gamma^*_u(\alpha,p)  \nonumber\\
&=& \Gamma^*_u(\alpha,p')\left[\vphantom{\frac{\partial \Sigma(\alpha)}{\partial \sqrt{s}}}
\sqrt{s}-M_0-\Sigma(\alpha,M_R(\alpha))-
 \right.\nonumber \\ &&\left.
\left(\sqrt{s}-M_R(\alpha)\right)\frac{\partial\Sigma(\alpha)}{\partial\sqrt{s}}
 \ldots \right]^{-1}\Gamma^*_u(\alpha,p) \nonumber\\
 &=& \Gamma^*_u(\alpha,p')\left[\vphantom{\frac{\partial \Sigma(\alpha)}{\partial \sqrt{s}}}
\left(\sqrt{s}-M_R(\alpha)\right)-
 \right.\nonumber \\ &&\left.
\left(\sqrt{s}-M_R(\alpha)\right)\frac{\partial\Sigma(\alpha)}{\partial\sqrt{s}}\ 
 \ldots \right]^{-1}\Gamma^*_u(\alpha,p) \nonumber\\
 &=&\Gamma^*_u(\alpha,p')\ Z(\alpha)\ \Gamma^*_u(\alpha,p)\left[\vphantom{\frac{\partial^2 \Sigma(\alpha)}{\left(\partial \sqrt{s}\right)^2}}      
(\sqrt{s}-M_R(\alpha) -
\right.\nonumber \\ &&\left.
\frac{1}{2}(\sqrt{s}-M_R(\alpha))^2 Z(\alpha)
 \frac{\partial^2\Sigma(\alpha)}{(\partial\sqrt{s})^2}\ \ldots \right]\ ,
\label{eq:19.47}\end{eqnarray}
here we introduced the renormalization constants $Z(\alpha)$ defined by 
\begin{equation}
  Z(\alpha) \equiv 
\left( 1- \left.\frac{\partial\Sigma(\alpha)}{\partial\sqrt{s}}\right|_{\sqrt{s}=M_R(\alpha)}
 \right)^{-1}\ .
\label{eq:19.48}\end{equation}
Also we can define $\Sigma_{ren}(\alpha,\sqrt{s}) \equiv Z(\alpha)\Sigma(\alpha,\sqrt{s})$ similar to Eq. (\ref{eq:19.27b}).
Analogous to Eq. (\ref{eq:19.26}) we introduce the renormalized self-energy by 
\begin{eqnarray}
 \Sigma_{ren}^{(2)}(\alpha,\sqrt{s}) &=& 
\frac{1}{2}\left(\sqrt{s}-M_R(\alpha)\right)^2
  \frac{\partial^2\Sigma_{ren}(\alpha)}{(\partial\sqrt{s})^2} +
 \ldots \nonumber \\&=&
 \Sigma_{ren}(\alpha,\sqrt{s})-\Sigma_{ren}(\alpha,M_R(\alpha)) -
 \nonumber \\ &&
\left(\sqrt{s}-M_R(\alpha)\right)
  \frac{\partial\Sigma_{ren}(\alpha)}{\partial\sqrt{s}} \ .
\label{eq:19.49}\end{eqnarray}
where the derivatives are evaluated at the point $\sqrt{s}=M_R(\alpha)$.
The resonance amplitude $T_{res}(\alpha)$ in Eq. (\ref{eq:19.47}) in terms of the renormalized quantities reads
\begin{eqnarray}
 T_{res}(\alpha) &=& 
 \Gamma^*_{ren}(\alpha,p')\left[\sqrt{s}-M_{R}(\alpha)-\Sigma_{ren}^{(2)}(\alpha,\sqrt{s})\right]^{-1}
\nonumber \\ && \times 
 \Gamma^*_{ren}(\alpha,p)\ ,                   
\label{eq:19.50}\end{eqnarray}
where the renormalized vertex is 
\begin{equation}
 \Gamma^*_{ren}(\alpha,p) \equiv \sqrt{Z(\alpha)}\ \Gamma^*_{u}(\alpha,p)\ .
\label{eq:19.51a}\end{equation}
In the previous we have suppressed the momentum dependence of $T_{res}(\alpha)$ for notational convenience. The renormalization is now performed by application of the following renormalization conditions:                       
\begin{itemize}
\item[(i)] Mass-renormalization: The physical masses $M_R(\alpha)$ are given implicitly by 
\begin{equation}
 M_R(\alpha) = M_0 + \Sigma\left(\alpha,M_R(\alpha)\right)\ .
\label{eq:19.52a}\end{equation}
\item[(ii)] Coupling-renormalization: The physical coupling constants $g_r(\alpha)$ are given by 
\begin{eqnarray}
\lefteqn{
\left| \lim_{\sqrt{s} \rightarrow M_R(\alpha)} \left(\sqrt{s}-M_R(\alpha)\right)\ T_{res}(\alpha)\right| 
}\nonumber \\&=&
 \left|\Gamma^*_{ren}(\alpha,p_R)\right|^2 \nonumber \\ 
 &=& Z(\alpha) \left|\Gamma^*_{u}(\alpha,p_R)\right|^2 
\ .
\label{eq:19.53a}
\end{eqnarray}
\end{itemize}
Eqs. (\ref{eq:19.52a}) and (\ref{eq:19.53a}) constitute four equations. These can be solved for the four bare parameters 
$\left\{ M_{0,1}, M_{0,2}, g_{0,1}, g_{0,2}\right\}$ using as input the physical masses and coupling constants. We get 
\begin{eqnarray}
 g_{0,1} &=& g_{0,1}\left[g_r(+),g_r(-); M_R(+), M_R(-)\right]\ , \nonumber\\
 g_{0,2} &=& g_{0,2}\left[g_r(+),g_r(-); M_R(+), M_R(-)\right]\ , \nonumber\\
 M_{0,1} &=& M_{0,1}\left[g_r(+),g_r(-); M_R(+), M_R(-)\right]\ , \nonumber\\
 M_{0,2} &=& M_{0,2}\left[g_r(+),g_r(-); M_R(+), M_R(-)\right] . 
\label{eq:19.54}\end{eqnarray}
From these we obtain the renormalization constants:
\begin{equation}
 Z_g(-) \equiv g_{0,1}/g_r(-)\ \ ,\ \ Z_g(+) \equiv g_{0,2}/g_r(+)\ .
\label{eq:19.55}\end{equation}
Notice that after the diagonalization of the propagator we have two uncoupled systems $\alpha=\pm$. Therefore, it is natural to define, in analogy with the single resonance case, the $Z_1(\alpha)$-factors by
\begin{eqnarray}
 \Gamma^*_{ren}(\alpha,p) &=& \sqrt{Z_2(\alpha)}\ \Gamma^*_u(\alpha,p) 
\nonumber \\
&\equiv& Z_1(\alpha)\ Z_g^{-1}(\alpha)\ \Gamma^*_u(\alpha,p)
\nonumber \\
& \equiv&
 Z_1(\alpha)\ \Gamma^*_r(\alpha,p)\ ,     
\label{eq:19.56}\end{eqnarray}
where $Z_2(\alpha) \equiv Z(\alpha)$. 
Rotating back to the basis $(N_1,N_2)$ we find the $Z$-transformation
on the original basis before the diagonalization of the propagator. 
This $Z$-transformation on the unmixed fields is a $2\times 2$-matrix.
Note, that in Eqs. (\ref{eq:19.55}) and (\ref{eq:19.56}) we have defined several $Z$-factors suggestively. In order to find out how these constants are related to the $Z$-matrices alluded to above, we would have to work out this $Z$-transformation in detail. This we do not attempt, since it 
is not really necessary here.

From the input of the four physical parameters $\left\{M_R(\alpha), g_r(\alpha)\right\}$ 
one computes the bare parameters. Using the latter one computes $\Sigma_{ren}(\alpha,\sqrt{s})$ and 
$\Gamma^*_{ren}(\alpha,p)$. This defines the resonance part of the amplitudes unambiguously.

\section{The $\pi N$ interaction}
\label{chap:6}
In this section we show the results of the fit of the NSC $\pi N$-model to the most recent energy-dependent phase shift analysis of Arndt et al.\cite{Arn95} (SM95)
. We find a good agreement between the calculated and empirical phase shifts, up to $T_{{\rm lab}}=600$ MeV for the lower partial waves. The results of the fit  to the Arndt phase shifts are shown in Figure \ref{fig:6.1}. The calculated phase shifts are also compared with the Karlsruhe-Helsinki phase shift analysis
 \cite{Koch80} (KH80) in Figure \ref{fig:6.2}. The parameters of the NSC $\pi N$-model are given in Tables \ref{tab:6.2} and \ref{tab:6.3}.

Some results of the renormalization procedure for the $s$-channel diagrams, discussed in Sec. \ref{chap:4}, are given. The bare coupling constants and masses are listed in Table \ref{tab:6.3}, and the energy dependence of the renormalized self-energy of the nucleon and $\Delta$ are shown in Figure \ref{fig:6.4}.

\subsection{The NSC $\pi N$-model}
\begin{figure}[t]
\begin{center}
\resizebox{8.25cm}{1.65cm}{\includegraphics*[2cm,23.0cm][19cm,26cm]{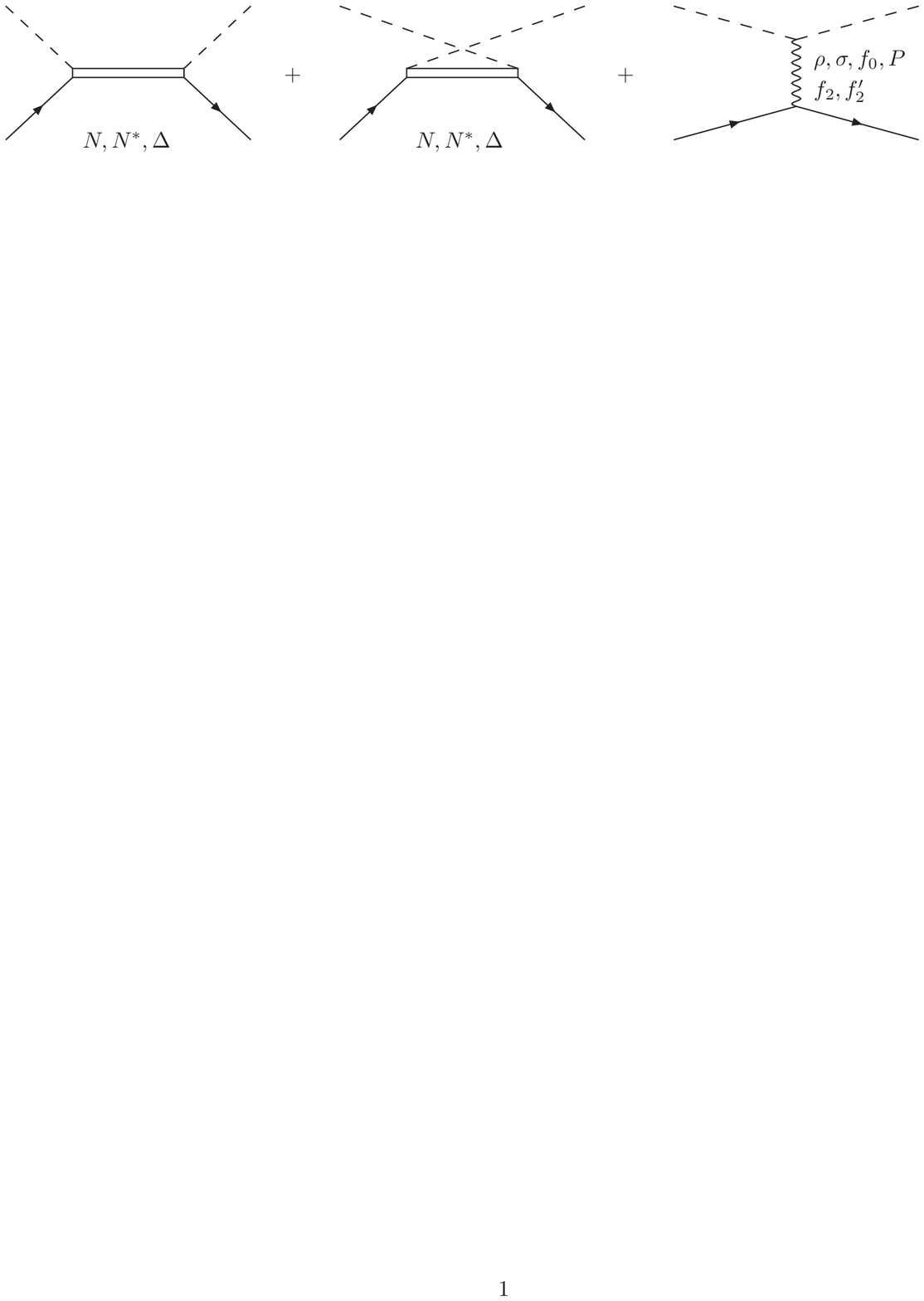}}
\end{center}
\caption{Contributions to the $\pi N$ potential from the $s$-, $u$- and $t$-channel Feynman diagrams. The external dashed and solid lines are always the $\pi$ and $N$ respectively.}
\label{fig:6.0}
\end{figure}

The potential for the $\pi N$-interactions consists of the one-meson-exchange and one-baryon-exchange Feynman diagrams, derived from effective meson-baryon interaction Hamiltonians, see paper I and Sec. \ref{chap:5}. The diagrams contributing to the $\pi N$ potential are given in Figure \ref{fig:6.0}. The partial wave potentials together with the $\pi N$ Green's function constitute the kernel of the integral equation for the partial wave $T$-matrix 
which is solved numerically to find the observable quantities or the phase shifts. We solve the partial wave $T$-matrix by matrix inversion and we use the method introduced by Haftel and Tabakin \cite{Haf70} to deal numerically with singularities in the physical region in the Green's function.

The interaction Hamiltonians from which the Feynman diagrams are derived, are explicitly given below for the $\pi N$ system. We use the pseudovector coupling for the $NN\pi$ vertex
\begin{eqnarray}
{\cal H}_{NN\pi}&=&\frac{f_{NN\pi}}{m_{\pi^+}}\left(\bar{N}\gamma_5\gamma_{\mu}\mbox{\boldmath $\tau$}N\right)\cdot\partial^{\mu}\mbox{\boldmath $\pi$}\ ,
\label{eq:6.1}
\end{eqnarray}
the same structure is used for the Roper, and for the $S_{11}(1535)$ we use a similar coupling where the $\gamma_5$ is omitted. The $NN\pi$ coupling constant is quite well determined and is fixed in the fitting procedure. For the $N\Delta\pi$ vertex we use the conventional coupling
\begin{eqnarray}
{\cal H}_{N\Delta\pi}&=&\frac{f_{N\Delta\pi}}{m_{\pi^+}}\left(\bar{\Delta}_\mu\mbox{\boldmath $T$}N\right)\cdot\partial^{\mu}\mbox{\boldmath $\pi$} + H.c.\ ,
\label{eq:6.2}
\end{eqnarray}
where $\mbox{\boldmath $T$}$ is the transition operator between isospin-$\frac{1}{2}$ isospin-$\frac{3}{2}$ states \cite{Car71}. The only vector-meson exchanged in $\pi N$ scattering is the $\rho$. The $NN\rho$ and $\pi\pi\rho$ couplings we use are
\begin{eqnarray}
{\cal H}_{NN\rho}&=&g_{NN\rho}\left(\bar{N}\gamma_{\mu}\mbox{\boldmath $\tau$}N\right)\cdot\mbox{\boldmath $\rho$}^{\mu}
\nonumber \\ &&
+\frac{f_{NN\rho}}{4 {\cal M}}\left(\bar{N}\sigma_{\mu\nu}\mbox{\boldmath $\tau$}N\right)\cdot\left(\partial^{\mu}\mbox{\boldmath $\rho$}^{\nu}-\partial^{\nu}\mbox{\boldmath $\rho$}^{\mu}\right)\ , \nonumber \\
{\cal H}_{\pi\pi\rho}&=&\frac{g_{\pi\pi\rho}}{2}\mbox{\boldmath $\rho$}_{\mu}\cdot\mbox{\boldmath $\pi$}\times \stackrel{\leftrightarrow}\partial^{\mu}\mbox{\boldmath $\pi$}\ ,
\end{eqnarray}
we remark that the vector-meson dominance model predicts the ratio of the tensor and vector coupling to be $\kappa_{\rho}=f_{NN\rho}/g_{NN\rho}=3.7$, but in $\pi N$ models it appears to be considerably lower \cite{Pea90,Sat96,Lah99,Pas00}. We also find a lower value for $\kappa_{\rho}$, see Table \ref{tab:6.2}. The scalar-meson couplings have the simple structure
\begin{eqnarray}
{\cal H}_{NN\sigma}&=&g_{NN\sigma}\bar{N}N\sigma\ ,
\end{eqnarray}
\begin{eqnarray}
{\cal H}_{\pi\pi\sigma}&=&\frac{g_{\pi\pi\sigma}}{2}m_{\pi^+}\sigma\mbox{\boldmath $\pi$}\cdot\mbox{\boldmath $\pi$}\ .
\end{eqnarray}

\begin{figure}[t]
\begin{center}
\resizebox{8.25cm}{9.18cm}{\includegraphics*[2cm,2.0cm][20cm,27cm]{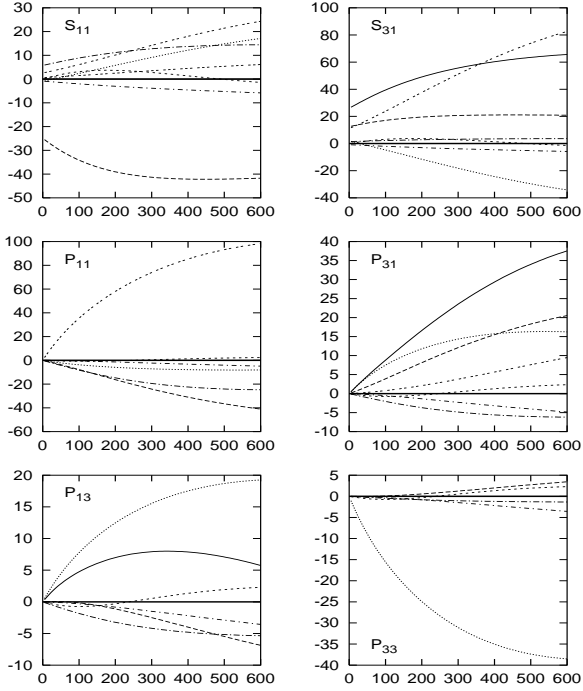}}
\end{center}
\caption{The total $\pi N$ partial wave potentials $V_L$ as a function of $T_{{\rm lab}}$ (MeV) are given by the solid line. For the $S_{11}$-, $P_{11}$- and $P_{33}$-wave the resonance pole and total contributions are omitted. The various contributions are a. the long dashed line: $\rho$, b. short dashed line: scalar-mesons and Pomeron, c. the dotted line: nucleon-exchange, d. the long dash-dotted line: $\Delta$-exchange, e. the short dash-dotted line: tensor-mesons, f. the double dashed line: nucleon or $\Delta$ pole, g. the triple dashed line: Roper pole.}
\label{fig:6.3}
\end{figure}

In contrast with other $\pi N$ models, we consider the scalar-mesons as genuine $SU_f(3)$ octet particles. Therefore not only the $\sigma$ is exchanged but also the $f_0(975)$ having the same structure for the coupling, both giving an attractive contribution. The contribution of $\sigma$-exchange is, however, much larger than the contribution of $f_0$-exchange. A repulsive contribution is obtained from Pomeron-exchange, also having the same structure for the coupling. The contributions of the Pomeron and the scalar-mesons cancel each other almost completely, as can be seen in the figures for the partial wave potentials, Figure \ref{fig:6.3}. This cancellation is important in order to comply with the soft-pion theorems for low-energy $\pi N$ scattering \cite{Wei66,Adl65,Swa89}.
The $\sigma$ and the $\rho$ are treated as broad mesons, for details about the treatment we refer to \cite{NRS75}. The $\sigma$ is not considered as an $SU_f(3)$ particle in other $\pi N$ models, but e.g. as an effective representation of correlated two-pion-exchange \cite{Sch94,Sch95,Pas00}, in that case its contribution may be repulsive in some partial waves.

We consider the exchange of the two isoscalar tensor-mesons $f_2$ and $f_2'$, the structure of the couplings we use is 
\begin{eqnarray}
 {\cal H}_{NNf_2} &=& \left[\frac{iF_{1NNf_2}}{4}\bar{N}\left(
 \gamma_\mu \stackrel{\leftrightarrow}{\partial_\nu} +
 \gamma_\nu \stackrel{\leftrightarrow}{\partial_\mu} 
 \right) N 
\right. \nonumber \\ && \left.
-\frac{F_{2NNf_2}}{4}\bar{N}\
 \stackrel{\leftrightarrow}{\partial^\mu} 
 \stackrel{\leftrightarrow}{\partial^\nu} 
 N \right]f_2^{\mu\nu}\ , \nonumber \\
{\cal H}_{\pi\pi f_2}&=&\frac{g_{\pi\pi f_2}}{m_{\pi^+}}\ f_2^{\mu\nu}\left(\partial_{\mu}\mbox{\boldmath $\pi$} \cdot \partial_{\nu}\mbox{\boldmath $\pi$}\right)\ ,
\end{eqnarray}
and the coupling of $f_2'$ is similar to the $f_2$ coupling. Similar as for the scalar-mesons $f_0$ and $\sigma$, the $f_2'$ contribution is very small compared to the $f_2$ contribution.

The isospin structure results in the isospin factors 
listed in Table \ref{tab:6.0}, see also Appendix \ref{app:AA}.
 The spin-space amplitudes in paper I need to be multiplied by these isospin factors to find the complete $\pi N$ amplitude.

\begin{table}[t]
\caption{The isospin factors for the various exchanges for a given total isospin $I$ of the $\pi N$ system, see Appendix \ref{app:AA}.}
\centering
\begin{ruledtabular}
\begin{tabular}{crr}
Exchange & $I=\frac{1}{2}$& $I=\frac{3}{2}$ \\
\hline
$\sigma ,f_0,f_2,f_2'            $ &  1&  1  \\
$\rho                   $ & -2&  1  \\
$N (s-{\rm channel})     $ &  3&  -  \\
$N ({u-\rm channel})     $ & -1&  2  \\
$\Delta (s-{\rm channel})$ &  -&  1  \\
$\Delta (u-{\rm channel})$ &  $\frac{4}{3}$&  $\frac{1}{3}$  \\
\end{tabular}
\end{ruledtabular}
\label{tab:6.0}
\end{table}

Summarizing we consider in the $t$-channel the exchanges of the scalar-mesons $\sigma$, $f_0$, the Pomeron, the vector-meson $\rho$ and the tensor-mesons $f_2$ and $f_2'$, and in the $u$- and $s$-channel the exchanges of the baryons $N$, $\Delta$, Roper and $S_{11}$.

The latter two resonances were included in the NSC $\pi N$-model to give a good description of the $P_{11}$- and $S_{11}$-wave  phase shifts at higher energies, their contribution at lower energies is small.
These resonances were also included in the model of Pascalutsa and Tjon \cite{Pas00}.

It is instructive to examine the relative strength of the contributions of the various exchanges for each partial wave. The on-shell partial wave potentials are given for each partial wave in Figure \ref{fig:6.3}. The pole contributions for the $\Delta$, Roper and $S_{11}$ are omitted from the $P_{33}$-, $P_{11}$- and $S_{11}$-wave respectively to show the other contributions more clearly.

We remark that for the $s$-channel diagrams only the positive-energy intermediate state develops a pole and is nonzero only in the partial wave having the same quantum numbers as the considered particle. The negative-energy intermediate state (background contribution), which is also included in a Feynman diagram,
 does not have a pole and may contribute to other waves having the same isospin. These background contributions from the nucleon and $\Delta$ pole to the $S_{11}$- and $S_{31}$-wave respectively are not small.


The Pomeron-$\sigma$ cancellation is clearly seen in all partial waves. The nucleon-exchange is quite strong in the $P$-waves, except for the $P_{11}$-wave where the nucleon pole is quite strong and gives a repulsive contribution, which causes the negative phase shifts at low energies in this wave. The change of sign of the phase shift in the $P_{11}$-wave is caused by the attractive $\rho$ and $\Delta$-exchange.

The $\Delta$ pole dominates the $P_{33}$ -wave, but also a large contribution is present in the $S_{31}$-wave and a small contribution in the $P_{31}$-wave is seen. This contribution results from the spin-1/2 component of the Rarita-Schwinger propagator. The $\Delta$-exchange is present in all partial waves. A significant contribution of $\rho$-exchange is seen in all partial waves, except the $P_{33}$-wave, which is dominated by nucleon-exchange and of course the $\Delta$ pole. A modest contribution from the tensor-mesons is seen in all partial waves.

\begin{figure}[t]
\begin{center}
\resizebox{8.25cm}{11.42cm}{\includegraphics*[2cm,2.0cm][20cm,27cm]{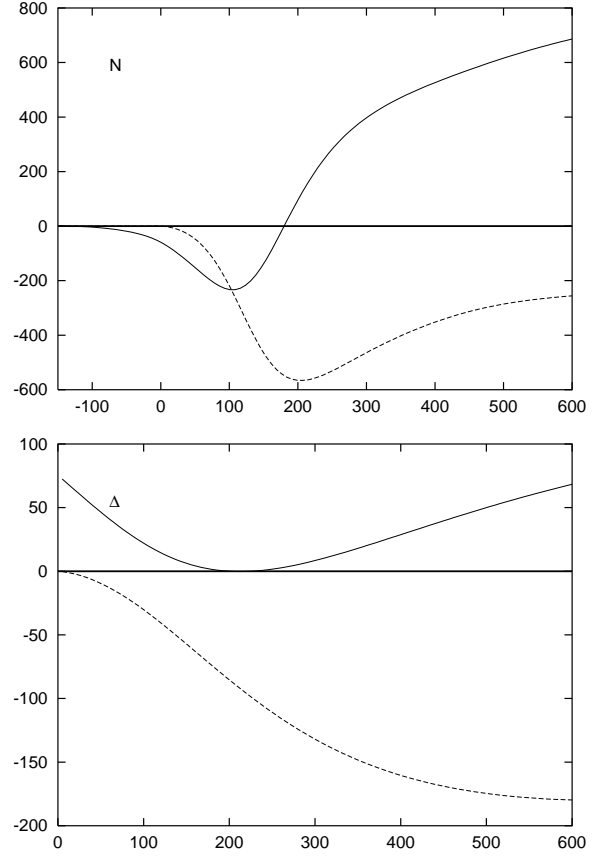}}
\end{center}
\caption{The renormalized self-energy $\Sigma^{(2)}_{ren}$ (MeV) of Eqs. (\ref{eq:19.49}) and (\ref{eq:19.26}) for the nucleon and the $\Delta$ as a function of $T_{{\rm lab}}$ (MeV). The real part is given by the solid line and the imaginary part is given by the dashed line.}
\label{fig:6.4}
\end{figure}

When solving the integral equation for the $T$-matrix, the propagator and vertices of the $s$-channel diagrams get dressed. The renormalization procedure, described in Sec. \ref{chap:4}, determines the bare masses and coupling constants in terms of the physical parameters. The physical parameters and bare parameters obtained from the fitting procedure are given in Tables \ref{tab:6.2} and \ref{tab:6.3} respectively.
%
The self-energy of the baryons in the $s$-channel is renormalized, ensuring a pole at the physical mass of the baryons. For the nucleon and the $\Delta$ we show the energy dependence of 
the renormalized self-energy in Figure \ref{fig:6.4}. This figure clearly shows that the real part of the renormalized self-energy of the $\Delta$ and its derivative vanish at the $\Delta$ pole, by definition. This is of course also the case for the nucleon renormalized self-energy, however, the nucleon pole lies below the $\pi N$ threshold.

\subsubsection{Decay coupling constants}
\label{subsec:decay}
The physical coupling constants of the resonances included in the NSC model can be estimated by relating the width of the resonance to the $T$-matrix element of its decay into two particles, in this case $\pi N$. This relation for the two-particle decay is derived in \cite{Pil67}, the two-particle width is
\begin{eqnarray}
\Gamma(p)&=&\frac{p}{4M^2}\int\frac{d\cos\theta}{4\pi}\sum_{\sigma}\left|T\right|^2,
\label{eq:res1}
\end{eqnarray}
where $M$ is the resonance mass and the absolute square of the $T$-matrix is summed over the nucleon spin. The decay processes $\Delta\rightarrow \pi N$, $N^*\rightarrow \pi N$ and $S_{11}\rightarrow \pi N$ are considered in order to find an estimate for the coupling constants $f_{N\Delta\pi}$, $f_{NN^*\pi}$ and $f_{NS_{11}\pi}$ respectively. The $T$-matrix elements of the various decays in lowest order can be calculated using the interaction Hamiltonians defined in Sec. \ref{chap:5} and paper I, 
Eq. (\ref{eq:res1})
gives us the estimates for the coupling constants
\begin{eqnarray}
\frac{f^2_{N\Delta\pi}}{4\pi}&=&3\frac{M_{\Delta}}{E+M}\frac{m_{\pi^+}^2\Gamma}{p^3}\approx 0.39\ , \nonumber \\
\frac{f^2_{NN^*\pi}}{4\pi}&=&\frac{1}{3}\frac{m_{\pi^+}^2}{\left(M_{N^*}+M\right)^2}\frac{(E+M)M_{N^*}\Gamma}{p^3}\approx 0.012\ , \nonumber \\
\frac{f^2_{NS_{11}\pi}}{4\pi}&=&\frac{1}{3}\frac{m_{\pi^+}^2}{\left(M_{S_{11}}-M\right)^2}\frac{M_{S_{11}}}{E+M}\frac{\Gamma}{p}\approx 0.002\ .
\label{eq:res3}
\end{eqnarray}
The numerical values are obtained by using the Breit-Wigner masses and widths from the Particle Data Group.

The coupling constants for the decay of the $\rho$, $\sigma$ and $f_2$ into two pions can be estimated in the same way
\begin{eqnarray}
\frac{g_{\pi\pi\rho}}{\sqrt{4\pi}}&=&\sqrt{\frac{3}{2}m_{\rho}^2\frac{\Gamma}{p^3}}\approx 1.70\ , \nonumber \\
\frac{g_{\pi\pi\sigma}}{\sqrt{4\pi}}&=&\sqrt{2\frac{m_{\sigma}^2}{m_{\pi^+}^2}\frac{\Gamma}{p}}\approx 10.6\ ,\nonumber \\
\frac{g_{\pi\pi f_2}}{\sqrt{4\pi}}&=&\sqrt{\frac{15}{16}m_{f_2}^2m_{\pi^+}^2\frac{\Gamma}{p^5}}\approx 0.224\ .
\label{eq:dec1}
\end{eqnarray}

\subsection{Results and discussion for $\pi N$ scattering}

\begin{figure}[t]
\begin{center}
\resizebox{8.25cm}{9.35cm}{\includegraphics*[2cm,2.0cm][21cm,27cm]{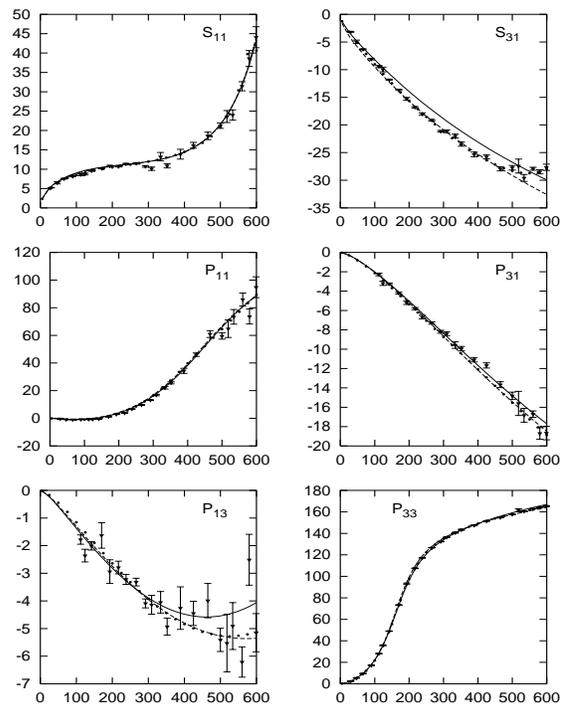}}
\end{center}
\caption{The $S$-wave and $P$-wave $\pi N$ phase shifts $\delta$ (degrees) as a function of $T_{{\rm lab}}$ (MeV). The empirical phases are from SM95 \cite{Arn95}, the dots are the multi-energy phases and the triangles with error bars are the single-energy phases. The NSC $\pi N$-model is given by the solid lines, the dashed line is the model without tensor-mesons.}
\label{fig:6.1}
\end{figure}

\begin{figure}[t]
\begin{center}
\resizebox{8.25cm}{9.35cm}{\includegraphics*[2cm,2.0cm][21cm,27cm]{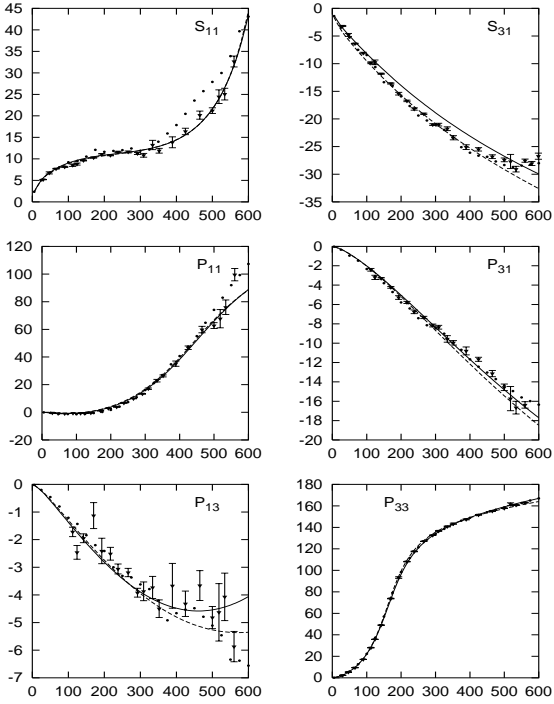}}
\end{center}
\caption{The $S$-wave and $P$-wave $\pi N$ phase shifts $\delta$ (degrees) as a function of $T_{{\rm lab}}$ (MeV). The empirical phases are from KH80 \cite{Koch80}, the dots are the multi-energy phases and the triangles with error bars are the single-energy phases. The NSC $\pi N$-model is given by the solid lines, the dashed line is the model without tensor-mesons.}
\label{fig:6.2}
\end{figure}

We have fitted the NSC $\pi N$-model to the energy-dependent SM95 partial wave analysis up to pion kinetic laboratory energy $T_{{\rm lab}}=600$ MeV. The results are shown in Figures \ref{fig:6.1} and \ref{fig:6.2}, showing the calculated and empirical phase shift for the SM95 and KH80 phase shift analyses respectively. The calculated and empirical scattering lengths for the $S$- and $P$-waves 
are listed in Table \ref{tab:6.1}.




\begin{table}[t]
\caption{The calculated and empirical $\pi N$ $S$-wave and $P$-wave scattering lengths in units of $m_{\pi}^{-1}$ and $m_{\pi}^{-3}$.}
\begin{ruledtabular}
\centering
\begin{tabular}{crrr}
Scat. length & Model& SM95 \cite{Arn95}& KH80 \cite{Koch80} \\
\hline
$S_{11}$ &  0.171&  0.172&  0.173$\pm$0.003 \\
$S_{31}$ & -0.096& -0.097& -0.101$\pm$0.004 \\
$P_{11}$ & -0.060& -0.068& -0.081$\pm$0.002 \\
$P_{31}$ & -0.037& -0.040& -0.045$\pm$0.002 \\
$P_{13}$ & -0.031& -0.021& -0.030$\pm$0.002 \\
$P_{33}$ &  0.213&  0.209&  0.214$\pm$0.002 \\
\end{tabular}
\end{ruledtabular}
\label{tab:6.1}
\end{table}

A good agreement between the NSC $\pi N$-model and the empirical phase shifts is found, but at higher energies some deviations are observed in some partial waves. These deviations may be caused by inelasticities, which become important at higher energies and have not been considered in this model.
The scattering lengths have been reproduced quite well, except for the $I=\frac{1}{2}$ $P$-waves, here the NSC $\pi N$-model scattering lengths deviate a little from \cite{Arn95}.


First we attempted to generate the $\Delta$ resonance 
dynamically, however, it was not possible to find the correct energy behavior for the $P_{33}$ phase shift. Then we considered the $\Delta$ resonance, at least partially, as a three-quark state and included it explicitly in the potential, as is done in the modern $\pi N$ literature, and immediately found the correct energy behavior for the $P_{33}$ phase shift. The other resonances have been treated in the same way.

We use six different cutoff masses, which are free parameters in the fitting procedure. For the nucleon and the Roper we use the same cutoff mass, for the two scalar-mesons we use the same cutoff mass and also for the two tensor-mesons the same cutoff mass is used. The masses of the mesons and the nucleon have been fixed in the fitting procedure, but the masses of the resonances are free parameters. 

Table \ref{tab:6.2} shows that the pole positions of these resonances are not necessarily exactly the same as the resonance positions, due to the non-resonance part of the amplitude, see Eq. (\ref{eq:19.9a}). The $\Delta$ and Roper resonate at respectively $\sqrt{s}=1232$ MeV and $\sqrt{s}=1440$ MeV while the poles are located at $\sqrt{s}=1254$ MeV and $\sqrt{s}=1440$ MeV respectively.

In order to obtain a good fit, we had to introduce an off-mass-shell damping for the $u$-channel $\Delta$-exchange, we used the factor
$
{\rm exp}\left[\left(u-M_{\Delta}^2\right) \gamma^2/M_{\Delta}^2\right]\ ,
$
where $\gamma=1.18$ was a free parameter in the fitting procedure.

Only the product of two coupling constants are determined in the fitting procedure. Therefore the triple-meson coupling constants are fixed at the value calculated from their decay width, see subsection \ref{subsec:decay}, and the baryon-baryon-meson coupling constant is a free parameter in the fitting procedure. The resonance coupling constants are first calculated from their decay width, see subsection \ref{subsec:decay}, but are also treated as free parameters. The fitted and calculated values deviate only a little.


\begin{table}[t]
\caption{NSC $\pi N$-model parameters: coupling constants, masses (MeV) and cutoff masses (MeV). Numbers with an asterisk were fixed in the fitting procedure.}
\begin{ruledtabular}
\centering
\begin{tabular}{crcrrcrrr}
Exch. &\multicolumn{6}{c}{Coupling Constants}  &Mass &$\Lambda$  \\
\hline
$\rho$ &$\frac{g_{NN\rho}g_{\pi\pi\rho}}{4\pi}$&$\! \!$=&$\! \!$1.333 & $\frac{f_{NN\rho}}{g_{NN\rho}}$&$\! \!$=&$\! \!$$2.121$ &$770^*$ & 838 \\
$\sigma$ &$\frac{g_{NN\sigma}g_{\pi\pi\sigma}}{4\pi}$&$\! \!$=&$\! \!$$26.196^*$& && &$760^*$ &1126  \\
$f_0$ &$\frac{g_{NNf_0}g_{\pi\pi f_0}}{4\pi}$&$\! \!$=&$\! \!$$-1.997^*$& && &$975^*$ &1126  \\
$f_2$ &$\frac{g_{NNf_2}g_{\pi\pi f_2}}{4\pi}$&$\! \!$=&$\! \!$$0.157^*$ &$\frac{f_{NNf_2}}{g_{NNf_2}}$&$\! \!$=&$\! \!$$0.382^*$ &$1270^*$ &412  \\
$f_2'$ &$\frac{g_{NNf_2'}g_{\pi\pi f_2'}}{4\pi}$&$\! \!$=&$\! \!$$0.003^*$& $\frac{f_{NNf_2'}}{g_{NNf_2'}}$&$\! \!$=&$\! \!$$3.393^*$ &$1525^*$ &412  \\
Pom. &$\frac{g_{NN P}g_{\pi\pi P}}{4\pi}$&$\! \!$=&$\! \!$4.135& && &315 &  \\
$N$ &$\frac{f_{NN\pi}^2}{4\pi}$&$\! \!$=&$\! \!$$0.075^*$& && &$938.3^*$ &665  \\
$\Delta$ &$\frac{f_{N\Delta\pi}^2}{4\pi}$&$\! \!$=&$\! \!$0.478& && &1254 &603  \\
$N^*$ &$\frac{f_{NN^*\pi}^2}{4\pi}$&$\! \!$=&$\! \!$0.023& && &1440 &665  \\
$S_{11}$ &$\frac{f_{NS_{11}\pi}^2}{4\pi}$&$\! \!$=&$\! \!$0.003& && &1567 &653  \\
\end{tabular}
\end{ruledtabular}
\label{tab:6.2}
\end{table}
\begin{table}[t]
\caption{Renormalization parameters: bare masses (MeV) and coupling constants. The renormalization conditions determine the bare parameters in terms of the model parameters in Table \ref{tab:6.2} .}
\begin{ruledtabular}
\centering
\begin{tabular}{crr}
Exch. &Bare Coupling Constants &Bare Mass \\
\hline
$N$ &$\frac{f_{0NN\pi}^2}{4\pi}=0.013$ &1187 \\
$\Delta$ &$\frac{f_{0N\Delta\pi}^2}{4\pi}=0.167$ &1399 \\
$N^*$ &$\frac{f_{0NN^*\pi}^2}{4\pi}=0.015$ &1831 \\
$S_{11}$ &$\frac{f_{0NS_{11}\pi}^2}{4\pi}=0.018$ &1774 \\
\end{tabular}
\end{ruledtabular}
\label{tab:6.3}
\end{table}
The NSC $\pi N$-model has 17 free physical fit parameters; 3 meson and Pomeron coupling constants, 6 cutoff masses, 4 masses, 3 decay coupling constants and $\gamma$.
The values of the coupling constants, listed in Table \ref{tab:6.2}, are in good agreement with the literature; $g_{NN\rho}=0.78$ and $g_{NN\sigma}=2.47$. However, the tensor coupling constant for the $\rho$, $f_{NN\rho}/g_{NN\rho}=2.12$ is small compared with values obtained in $NN$ models and the vector dominance value of 3.7. Other $\pi N$ models, \cite{Pas00,Pea90}, also suffer from this problem. The $NN\pi$ coupling constant, which is quite well determined in the $NN$ interaction, has been fixed in the NSC $\pi N$-model. We notice that for the tensor-mesons we used the coupling constants $g_T={\mathcal M}F_1+{\mathcal M}^2F_2$ and $f_T=-{\mathcal M}^2F_2$ in Table \ref{tab:6.2}.

The two conditions in the renormalization procedure for the pole contributions result in the two renormalization constants, i.e. the bare coupling constant and mass, listed in Table \ref{tab:6.3}.
We found the bare coupling constants to be smaller than the physical coupling constants except for the $S_{11}$ resonance. The bare masses are larger than the physical masses for each type of exchange, the interaction shifts the bare mass down to the physical mass. Pascalutsa and Tjon \cite{Pas00} find a larger physical mass than bare mass for the Roper. This is probably caused by the choice of the renormalization point. They renormalize the Roper contribution at the nucleon pole, we think it is more natural to perform the renormalization at the Roper pole.

Besides the discussed NSC $\pi N$-model, we also considered a model that does not contain tensor-mesons. We fitted this model to the empirical phase shifts and the results of the fit are given by the dashed lines in Figures \ref{fig:6.1} and \ref{fig:6.2}. We notice that in two partial waves a noticeable difference can be seen between the two models, the $S_{11}$ partial wave is described better by this model than by the NSC $\pi N$-model. It is hard to say which model works better for the $P_{13}$ partial wave, since the single-energy phase shifts have large error bars and both models are in agreement with the $P_{13}$ phase shifts. The tensor-mesons are important for a good description of the $K^+ N$ data, this is shown in the next section. The $\pi N$ scattering lengths are approximately the same for both models.

The parameters belonging to this model are listed in Table \ref{tab:6.4}, and the bare masses and coupling constants are given in Table \ref{tab:6.5}. The values of these parameters are essentially the same as the NSC $\pi N$-model parameters.

\begin{table}[t]
\caption{Parameters of the NSC $\pi N$-model without tensor-mesons: coupling constants, masses (MeV) and cutoff masses (MeV). Numbers with an asterisk were fixed in the fitting procedure.}
\begin{ruledtabular}
\centering
\begin{tabular}{crcrrcrrr}
Exch. &\multicolumn{6}{c}{Coupling Constants} &Mass &$\Lambda$  \\
\hline
$\rho$ &$\frac{g_{NN\rho}g_{\pi\pi\rho}}{4\pi}$&$\! \!$=&$\! \!$1.282 & $\frac{f_{NN\rho}}{g_{NN\rho}}$&$\! \!$=&$\! \!$$1.730$ &$770^*$ & 717 \\
$\sigma$ &$\frac{g_{NN\sigma}g_{\pi\pi\sigma}}{4\pi}$&$\! \!$=&$\! \!$$26.196^*$& && &$760^*$ &864  \\
$f_0$ &$\frac{g_{NNf_0}g_{\pi\pi f_0}}{4\pi}$&$\! \!$=&$\! \!$$-1.997^*$& && &$975^*$ &864  \\
Pom. &$\frac{g_{NN P}g_{\pi\pi P}}{4\pi}$&$\! \!$=&$\! \!$4.453& && &296 &  \\
$N$ &$\frac{f_{NN\pi}^2}{4\pi}$&$\! \!$=&$\! \!$$0.075^*$& && &$938.3^*$ &728  \\
$\Delta$ &$\frac{f_{N\Delta\pi}^2}{4\pi}$&$\! \!$=&$\! \!$0.470& && &1249 &659  \\
$N^*$ &$\frac{f_{NN^*\pi}^2}{4\pi}$&$\! \!$=&$\! \!$0.021& && &1441 &728  \\
$S_{11}$ &$\frac{f_{NS_{11}\pi}^2}{4\pi}$&$\! \!$=&$\! \!$0.003& && &1557 &482  \\
\end{tabular}
\end{ruledtabular}
\label{tab:6.4}
\end{table}
\begin{table}[h]
\caption{Renormalization parameters of the NSC $\pi N$-model without tensor-mesons: bare masses (MeV) and coupling constants. The renormalization conditions determine the bare parameters in terms of the model parameters in Table \ref{tab:6.4}.}
\begin{ruledtabular}
\centering
\begin{tabular}{crr}
Exch. &Bare Coupling Constants &Bare Mass \\
\hline
$N$ &$\frac{f_{0NN\pi}^2}{4\pi}=0.011$ &1203 \\
$\Delta$ &$\frac{f_{0N\Delta\pi}^2}{4\pi}=0.159$ &1417 \\
$N^*$ &$\frac{f_{0NN^*\pi}^2}{4\pi}=0.022$ &1944 \\
$S_{11}$ &$\frac{f_{0NS_{11}\pi}^2}{4\pi}=0.016$ &1602 \\
\end{tabular}
\end{ruledtabular}
\label{tab:6.5}
\end{table}

Since the $S$-wave scattering lengths are reproduced well, the soft-pion theorems for low-energy $\pi N$ scattering \cite{Wei66,Adl65} are satisfied in the NSC $\pi N$-model, without the need for a derivative coupling for the $\pi\pi\sigma$-vertex
. In view of chiral perturbation theory inspired models, the chiral $c_1$-, $c_3$- and $c_4$-terms are described implicitly by the NSC $\pi N$-model, since this model gives a good description of the empirical phase shifts.

\section{The $K ^+ N$ interaction}
\label{chap:7}
In this section we present the NSC $K^+ N$-model
and show the results of the fit to the energy-dependent phase shift analysis of Hyslop et al. \cite{Hys92} (SP92)
. The NSC $K^+ N$-model phase shifts are also compared with the single-energy phase shift analyses of Hashimoto \cite{Has83} and Watts et al. \cite{Wat80}. We find a fair agreement between the calculated and empirical phase shifts, up to $T_{{\rm lab}}=600$ MeV for the lower partial waves. The results of the fit are shown in Figures \ref{fig:7.1} and \ref{fig:7.2} and the parameters of the NSC $K^+ N$-model are given in Table \ref{tab:7.1}.

Since the various phase shift analyses \cite{Hys92, Has83, Wat80} are not always consistent and have quite large error bars, we also give a comparison between the experimental observables 
and the NSC model prediction. The total elastic cross sections 
up to $T_{{\rm lab}}$=600 MeV are shown in Figure \ref{fig:7.3}. The differential cross sections for the elastic processes $K^+p\rightarrow K^+p$ and $K^+n\rightarrow K^+n$ at various values of $T_{{\rm lab}}$ are shown in Figures \ref{fig:7.4} and \ref{fig:7.5}. For the same elastic processes, the polarizations at various values of $T_{{\rm lab}}$ are shown in Figure \ref{fig:7.6}.

\subsection{The NSC $K^+ N$-model}
\begin{figure}[t]
\begin{center}
\resizebox{8.25cm}{1.65cm}{\includegraphics*[2cm,23.0cm][18cm,26cm]{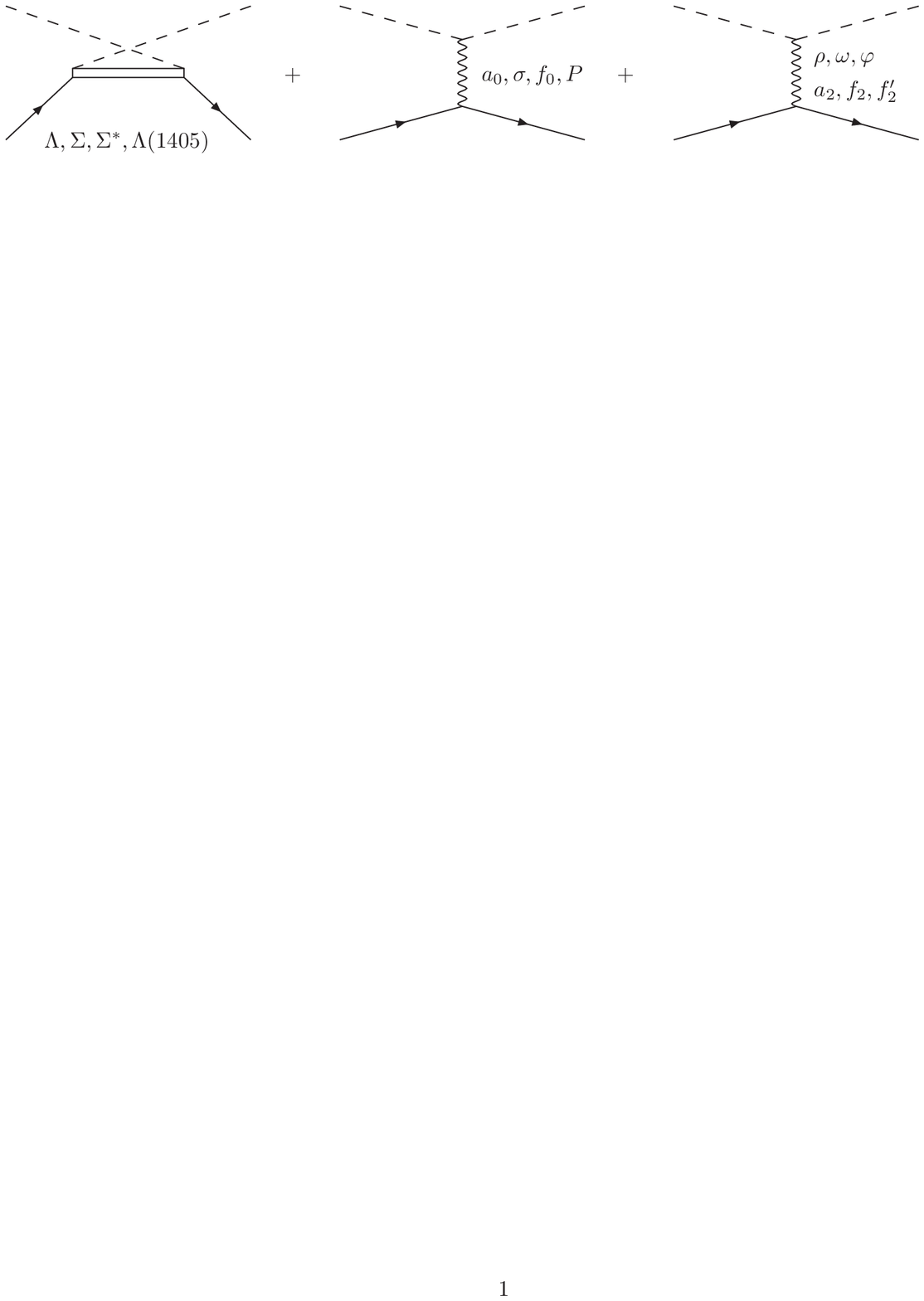}}
\end{center}
\caption{Contributions to the $K^+ N$ potential from the $u$- and $t$-channel Feynman diagrams. The external dashed and solid lines are always the $K^+$ and $N$ respectively.}
\label{fig:7.0}
\end{figure}

The NSC $K^+ N$-model is an $SU_f(3)$ extension of the NSC $\pi N$-model and consists analogously of the one-meson-exchange and one-baryon-exchange Feynman diagrams.
The various diagrams contributing to the $K^+ N$ potential are given in Figure \ref{fig:7.0}. 
The interaction Hamiltonians from which the Feynman diagrams for the $K^+ N$ system are derived, are explicitly given below. We use the pseudovector coupling for the $N\Lambda K$ and $N\Sigma K$ vertex
\begin{eqnarray}
{\cal H}_{N\Lambda K}&=&\frac{f_{N\Lambda K}}{m_{\pi^+}}\left(\bar{N}\gamma_5\gamma_{\mu}\partial^{\mu}K\right)\Lambda + H.c.\ , \nonumber \\
{\cal H}_{N\Sigma K}&=&\frac{f_{N\Sigma K}}{m_{\pi^+}}\left(\bar{N}\gamma_5\gamma_{\mu}\mbox{\boldmath $\tau$}\partial^{\mu}K\right)\cdot\mathbf{ \Sigma} + H. c.\ ,
\end{eqnarray}
 the coupling constants are determined by the  $NN\pi$ coupling constant and the $F/(F+D)$-ratio, $\alpha_{P}$. For the $\Lambda(1405)$ we use a similar coupling where the $\gamma_5$ is omitted. For the $N\Sigma^* K$ vertex we use, just as for the $N\Delta\pi$ vertex, the conventional coupling
\begin{eqnarray}
{\cal H}_{N\Sigma^* K}&=&\frac{f_{N\Sigma^* K}}{m_{\pi^+}}\left(\bar{N}\mbox{\boldmath $\tau$}\partial^{\mu}K\right)\cdot\mathbf{ \Sigma}^*_{\mu} + H.c.\ .
\end{eqnarray}
Since the $SU_f(3)$ decuplet occurs only once in the direct product of two octets, there is no mixing parameter $\alpha$ for this coupling. The $N\Sigma^* K$ coupling is determined by $SU_f(3)$, $f^2_{N\Sigma^* K}=f^2_{N\Delta\pi}/3$. Besides the $\rho$ also the isoscalar vector-mesons $\omega$ and $\varphi$ are exchanged.
 The following vector-meson couplings are used
\begin{eqnarray}
{\cal H}_{NN\rho}&=&g_{NN\rho}\left(\bar{N}\gamma_{\mu}\mbox{\boldmath $\tau$}N\right)\cdot\mbox{\boldmath $\rho$}^{\mu}
\nonumber \\ &&
+\frac{f_{NN\rho}}{4 {\cal M}}\left(\bar{N}\sigma_{\mu\nu}\mbox{\boldmath $\tau$}N\right)\cdot\left(\partial^{\mu}\mbox{\boldmath $\rho$}^{\nu}-\partial^{\nu}\mbox{\boldmath $\rho$}^{\mu}\right)\ , \nonumber \\
{\cal H}_{NN\omega}&=&g_{NN\omega}\bar{N}\gamma_{\mu}N\omega^{\mu}
\nonumber \\ &&
+\frac{f_{NN\omega}}{4 {\cal M}}\bar{N}\sigma_{\mu\nu}N\left(\partial^{\mu}\omega^{\nu}-\partial^{\nu}\omega^{\mu}\right)\ , 
\label{eq:7.3}
\end{eqnarray}
\begin{eqnarray}
{\cal H}_{KK\rho}&=&g_{KK\rho}\mbox{\boldmath $\rho$}_{\mu}\cdot\left(iK^\dagger\mbox{\boldmath $\tau$} \stackrel{\leftrightarrow}\partial^{\mu}K\right)\ , \nonumber \\
{\cal H}_{KK\omega}&=&g_{KK\omega}\omega_{\mu}\left(iK^\dagger \stackrel{\leftrightarrow}\partial^{\mu}K\right)\ .
\label{eq:7.4}
\end{eqnarray}
The coupling of $\varphi$ is similar to the $\omega$ coupling. Although we include $\varphi$-exchange its contribution is negligible compared to $\omega$-exchange. The coupling constants $g_{KK\omega}$ and $g_{KK\varphi}$ are fixed by $SU_f(3)$ in terms of $g_{\pi\pi\rho}$ and $\theta _V$. The $NN\omega$ coupling constant is a free parameter and the $NN\varphi$ coupling constant depends on $\theta _V$, $\alpha_V$ and the other two coupling constants. In addition to $\sigma$- and $f_0$-exchange, also the isovector scalar-meson $a_0$ is exchanged
, the following scalar-meson couplings are used
\begin{eqnarray}
{\cal H}_{NNa_0}&=&g_{NNa_0}\left(\bar{N}\mbox{\boldmath $\tau$}N\right)\cdot\mbox{\boldmath $a_0$}\ , \nonumber \\
{\cal H}_{NN\sigma}&=&g_{NN\sigma}\bar{N}N\sigma \ ,
\end{eqnarray}
\begin{eqnarray}
{\cal H}_{KKa_0}&=&g_{KKa_0}m_{\pi^+}\mbox{\boldmath $a_0$}\cdot\left(K^\dagger\mbox{\boldmath $\tau$}K\right)\ , \nonumber \\
{\cal H}_{KK\sigma}&=&g_{\pi\pi\sigma}m_{\pi^+}\sigma K^\dagger K\ .
\end{eqnarray}
The $f_0$ coupling is similar to the $\sigma$ coupling. 
Besides the exchange of the $f_2$ and the $f_2'$ also the isovector tensor-meson $a_2$ is exchanged.
 The following tensor-meson couplings are used
\begin{eqnarray}
 {\cal H}_{NNa_2} &=& \left[\frac{iF_{1NNa_2}}{4}\bar{N}\left(
 \gamma_\mu \stackrel{\leftrightarrow}{\partial_\nu} +
 \gamma_\nu \stackrel{\leftrightarrow}{\partial_\mu} 
 \right)\mbox{\boldmath $\tau$} N 
\right. \nonumber \\ && \left.
-\frac{F_{2NNa_2}}{4}\bar{N}\
 \stackrel{\leftrightarrow}{\partial^\mu} 
 \stackrel{\leftrightarrow}{\partial^\nu} \mbox{\boldmath $\tau$}
 N \right]\cdot \mbox{\boldmath $a$}_2^{\mu\nu}\ , \nonumber \\
 {\cal H}_{NNf_2} &=& \left[\frac{iF_{1NNf_2}}{4}\bar{N}\left(
 \gamma_\mu \stackrel{\leftrightarrow}{\partial_\nu} +
 \gamma_\nu \stackrel{\leftrightarrow}{\partial_\mu} 
 \right) N 
\right. \nonumber \\ && \left.
-\frac{F_{2NNf_2}}{4}\bar{N}\
 \stackrel{\leftrightarrow}{\partial^\mu} 
 \stackrel{\leftrightarrow}{\partial^\nu} 
 N \right]f_2^{\mu\nu}\ ,
\label{eq:7.3a}
\end{eqnarray}
\begin{eqnarray}
{\cal H}_{KKa_2}&=&\frac{g_{KKa_2}}{m_{\pi^+}}\ \mbox{\boldmath $a$}_2^{\mu\nu}\cdot\left(\partial_{\mu}K^\dagger\mbox{\boldmath $\tau$} \partial_{\nu}K\right)\ , \nonumber \\
{\cal H}_{KKf_2}&=&\frac{g_{KKf_2}}{m_{\pi^+}}\ f_2^{\mu\nu}\left(\partial_{\mu}K^\dagger \partial_{\nu}K\right)\ .
\label{eq:7.4a}
\end{eqnarray}
The coupling of $f_2'$ is similar to the $f_2$ coupling.
A repulsive contribution is obtained from Pomeron-exchange which is assumed to couple as a singlet and the value of its coupling constant is determined in the $\pi N$ system.

The isospin structure gives the isospin factors 
listed in Table \ref{tab:7.0}, see also Appendix \ref{app:AA}. The spin-space amplitudes in paper I need to be multiplied by these isospin factors to find the complete $K^+ N$ amplitude.

\begin{table}[t]
\caption{The isospin factors for the various exchanges for a given total isospin $I$ of the $K^+ N$ system, see Appendix \ref{app:AA}.}
\begin{ruledtabular}
\centering
\begin{tabular}{crr}
Exchange & $I=0$& $I=1$ \\
\hline
$\sigma ,f_0,\omega ,\varphi ,f_2,f_2'         $ &  1&  1  \\
$a_0 ,\rho ,a_2                     $ &-3&  1  \\
$\Lambda               $ & -1&  1  \\
$\Sigma                 $  &  3&  1  \\
\end{tabular}
\end{ruledtabular}
\label{tab:7.0}
\end{table}

Summarizing we consider in the $t$-channel the exchanges of the scalar-mesons $\sigma$, $f_0$ and $a_0$, the Pomeron, the vector-mesons $\omega$, $\varphi$ and $\rho$ and the tensor-mesons $a_2$, $f_2$ and $f_2'$, and in the $u$-channel the exchanges of the baryons $\Lambda$, $\Sigma$, $\Sigma (1385)$ ($\Sigma^*$) and $\Lambda (1405)$ ($\Lambda^*$).


The Coulomb interaction is neglected in the NSC model. Its contribution to the partial wave phase shifts is in principle relevant at very low energies. However, for the $K^+ N$ interaction we will not only investigate the phase shifts, but also some scattering observables. The differential cross section and polarization in the $K^+p\rightarrow K^+p$ channel as a function of the scattering angle clearly show the effect of the Coulomb peak at forward angles, the differential cross sections blow up and the polarizations go to zero. For the description of these scattering observables we correct for the Coulomb interaction by replacing the spin-nonflip and spin-flip scattering amplitudes $\tilde{f}$ and $\tilde{g}$ in paper I
 by \cite{Bra73,Has83}
\begin{eqnarray}
\tilde{f}&=&\sum_{L}\left[\left(L+1\right)F_{L+\frac{1}{2},L}+LF_{L-\frac{1}{2},L}\right]e^{2i\phi_L}P_L(\cos\theta)
\nonumber \\ && 
+f_C\ , \nonumber \\
\tilde{g}&=&\sum_{L}\left[F_{L+\frac{1}{2},L}-F_{L-\frac{1}{2},L}\right]e^{2i\phi_L}\sin\theta \ \frac{dP_L(\cos\theta)}{d\cos\theta}\ .
\end{eqnarray}

Here $f_C$ is the Coulomb amplitude and $\phi_L$ are the Coulomb phase shifts, defined respectively as
\begin{eqnarray}
f_C&=&-\frac{\alpha}{2kv\sin^2(\theta/2)}e^{-i\frac{\alpha}{v}\ln\left(\sin^2(\theta/2)\right)}\ , \nonumber \\
\phi_L&=&\sum_{n=1}^L\arctan \left(\frac{\alpha}{nv}\right)\ ,
\end{eqnarray}
where $k$ is the CM momentum, $v$ is the relative velocity of the particles in the CM system, $\theta$ is the CM scattering angle and $\alpha$ is the fine structure constant.

\begin{figure}[t]
\begin{center}
\resizebox{8.25cm}{8.8cm}{\includegraphics*[2cm,10.5cm][19.5cm,27cm]{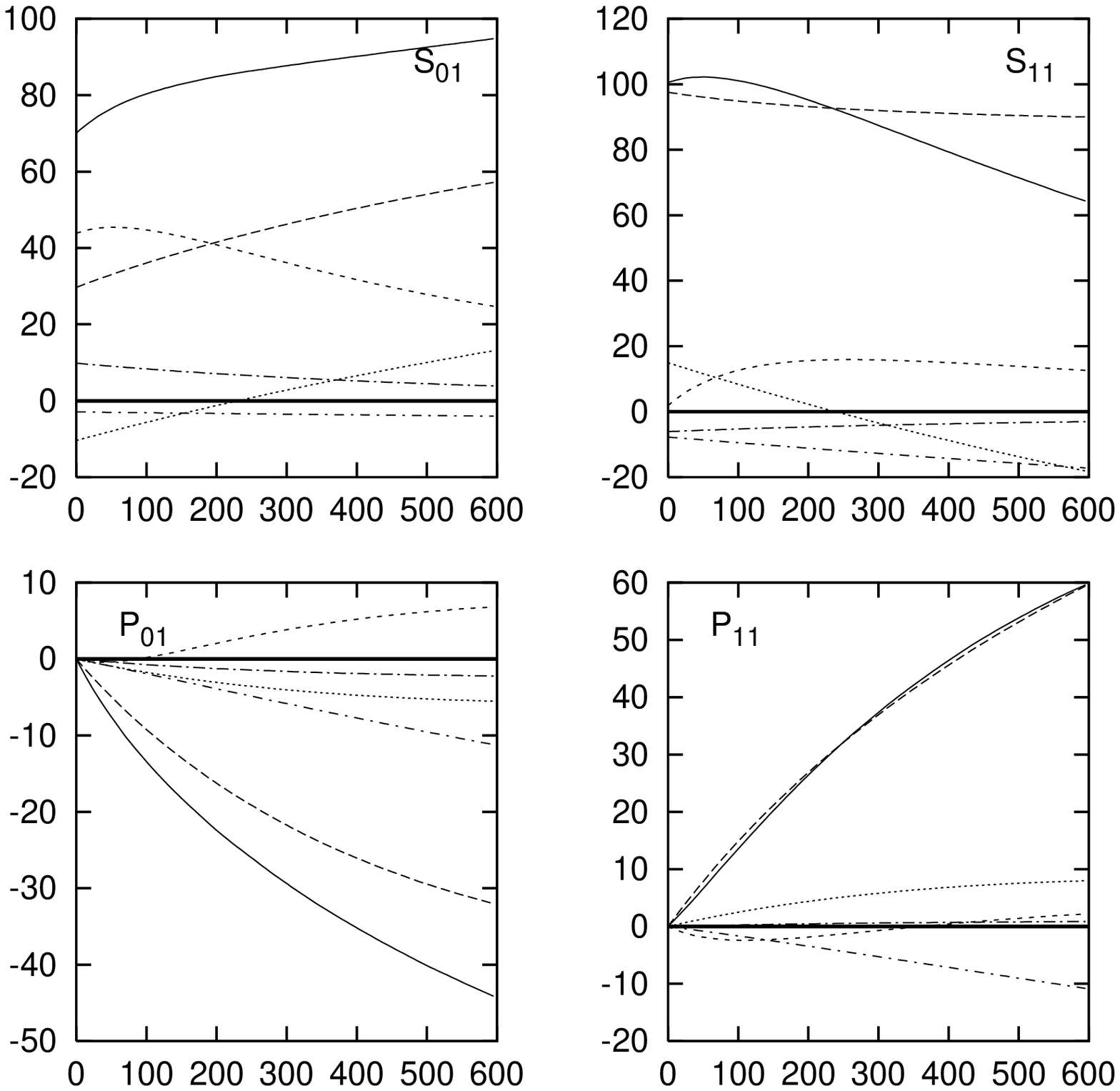}}
\end{center}
\caption{The total $K^+ N$ partial wave potentials $V_L$ as a function of $T_{{\rm lab}}$ (MeV) are given by the solid line. The various contributions are a. the long dashed line: vector-mesons, b. short dashed line: scalar-mesons and Pomeron, c. the dotted line: $\Lambda$ and $\Sigma$, d. the long dash-dotted line: $\Sigma^*$ and $\Lambda^*$, e. the short dash-dotted line: tensor-mesons.}
\label{fig:7.1a}
\end{figure}

\begin{figure}[t]
\begin{center}
\resizebox{8.25cm}{8.8cm}{\includegraphics*[2cm,10.5cm][19.5cm,27cm]{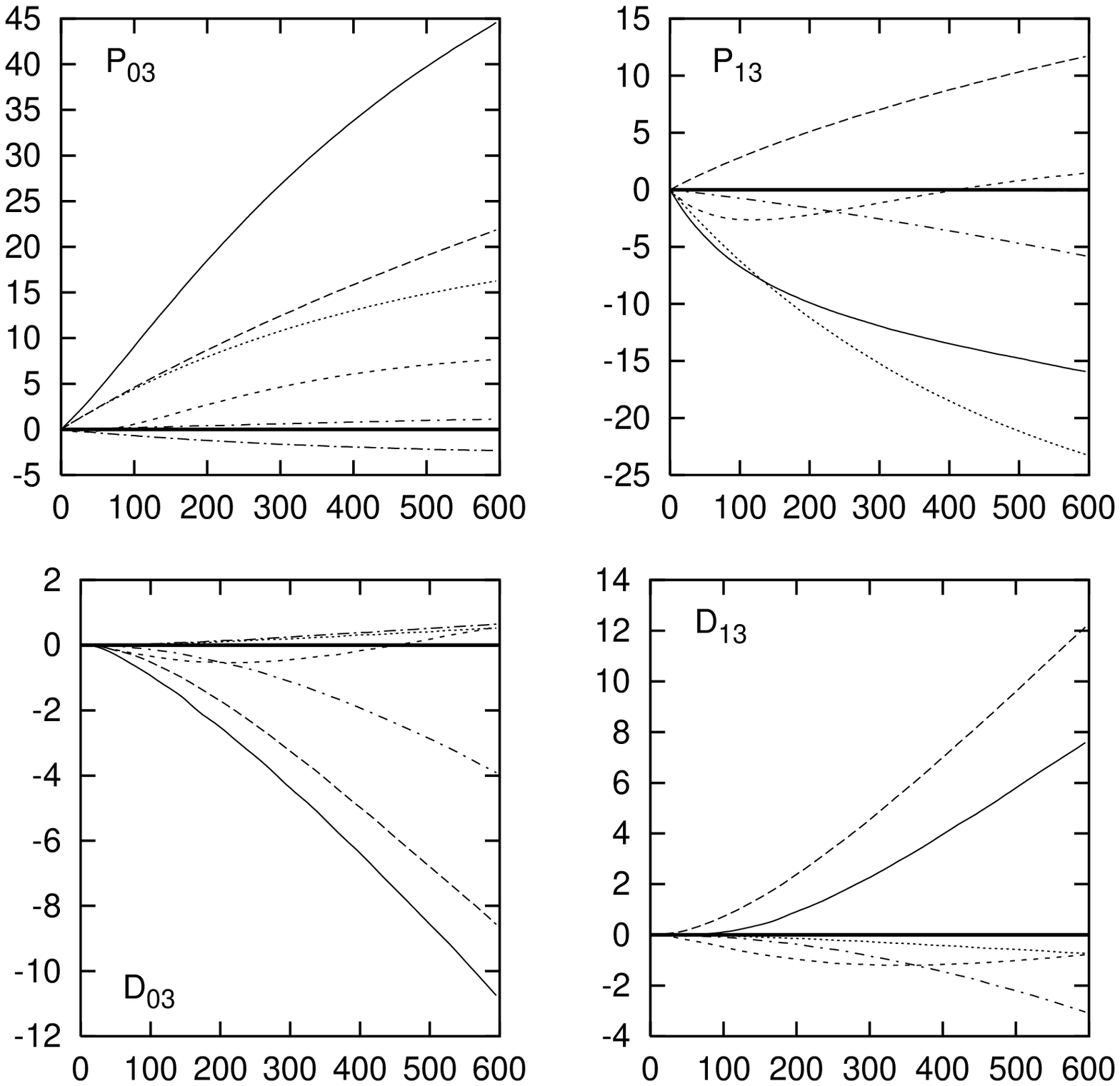}}
\end{center}
\caption{The total $K^+ N$ partial wave potentials $V_L$ as a function of $T_{{\rm lab}}$ (MeV) are given by the solid line. The various contributions are a. the long dashed line: vector-mesons, b. short dashed line: scalar-mesons and Pomeron, c. the dotted line: $\Lambda$ and $\Sigma$, d. the long dash-dotted line: $\Sigma^*$ and $\Lambda^*$, e. the short dash-dotted line: tensor-mesons.
}
\label{fig:7.1b}
\end{figure}

It is instructive to examine the relative strength of the different exchanges
 that contribute to the partial wave $K^+ N$ potentials. The on-shell potentials are given in Figures \ref{fig:7.1a} and \ref{fig:7.1b} for each partial wave.
The largest contribution 
 comes from vector-meson-exchange, $\omega$-exchange gives the largest contribution and the isospin splitting of the vector-mesons is caused by $\rho$-exchange. Especially the $S_{11}$, $P_{01}$ and $P_{11}$ partial waves are dominated by vector-meson-exchange. 

The cancellation between the scalar-mesons and the Pomeron in the $K^+ N$ interaction is less than in the $\pi N$ interaction, so the scalar-mesons and the Pomeron give a relevant contribution. Specifically a large repulsive contribution is seen in the $S$-waves.

The contribution from $\Lambda$- and $\Sigma$-exchange is large in the $J=\frac{3}{2}$ $P$-waves, and small in the other partial waves. This exchange plays in particular an important role in describing the rise of the $P_{13}$ phase shift. The contribution of the strange resonances $\Sigma^*$ and $\Lambda^*$ is practically negligible over the whole energy range in all partial waves.

The tensor-mesons give a relevant contribution in most partial waves, especially at higher energies. The inclusion of tensor-meson-exchange in the $K^+ N$ potential improved the description of the phase shifts at higher energies. 

\subsection{Results and discussion for $K^+ N$ scattering}
\begin{figure}[t]
\begin{center}
\resizebox{8.25cm}{7.15cm}{\includegraphics*[2cm,14.8cm][19.5cm,27cm]{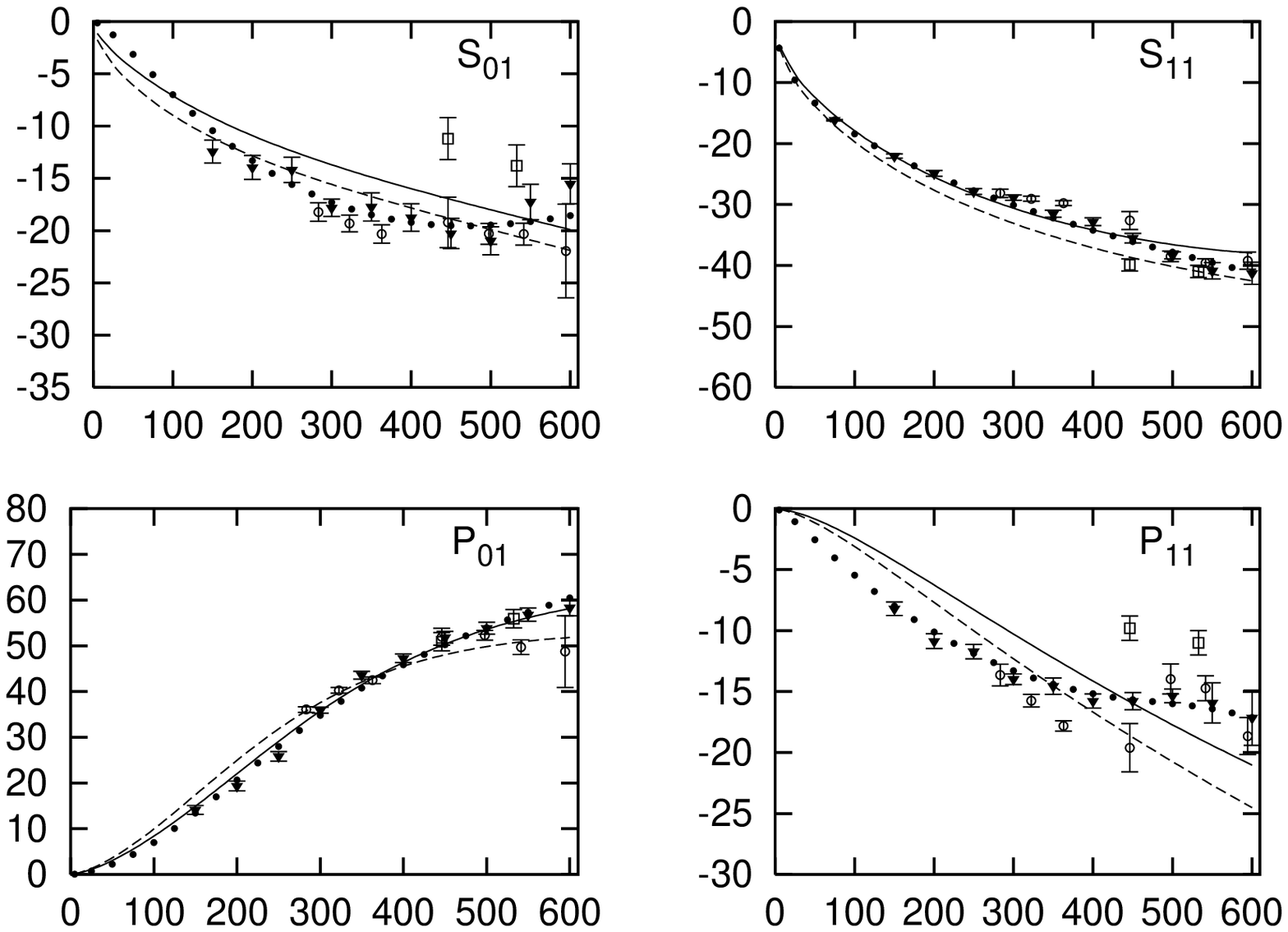}}
\end{center}
\caption{The $S$-wave and $P$-wave $K^+ N$ phase shifts $\delta$ (degrees) as a function of $T_{{\rm lab}}$ (MeV). The empirical phases are from SP92 \cite{Hys92}: multi-energy phases (dots) and single-energy phases (filled triangles), \cite{Has83} single-energy phases (open circles), \cite{Wat80} single-energy phases (open squares). The NSC $K^+ N$-model is given by the solid line, the dashed line is the model without tensor-mesons.}
\label{fig:7.1}
\end{figure}

\begin{figure}[t]
\begin{center}
\resizebox{8.25cm}{7.15cm}{\includegraphics*[2cm,14.8cm][19.5cm,27cm]{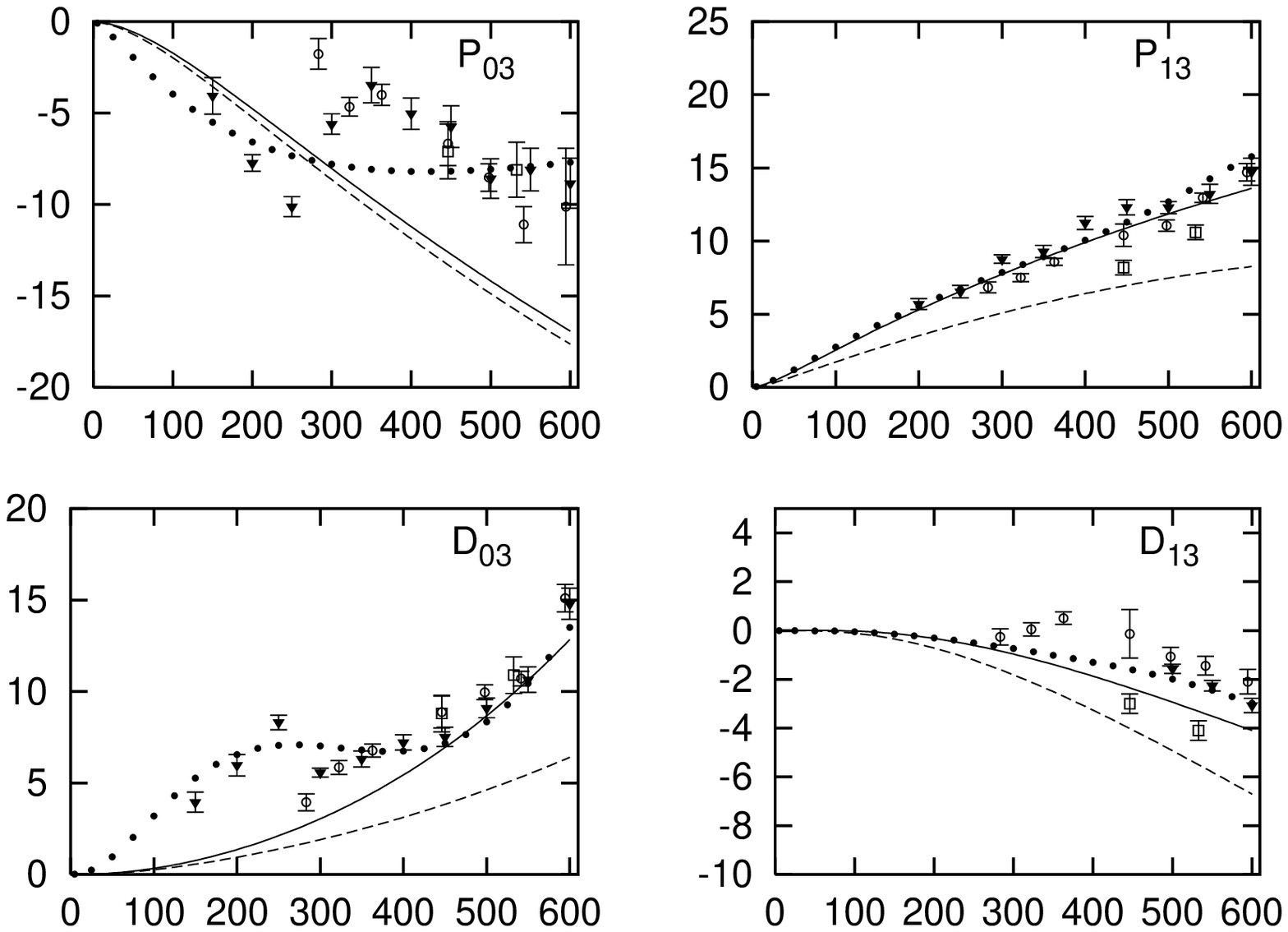}}
\end{center}
\caption{The $P$-wave and $D$-wave $K^+ N$ phase shifts $\delta$ (degrees) as a function of $T_{{\rm lab}}$ (MeV). The empirical phases are from SP92 \cite{Hys92}: multi-energy phases (dots) and single-energy phases (filled triangles), \cite{Has83} single-energy phases (open circles), \cite{Wat80} single-energy phases (open squares). The NSC $K^+ N$-model is given by the solid line, the dashed line is the model without tensor-mesons.}
\label{fig:7.2}
\end{figure}

We have fitted the NSC $K^+ N$-model
 to the energy-dependent SP92 partial wave analysis up to kaon kinetic laboratory energy $T_{{\rm lab}}=600$ MeV. The results of the fit are shown in Figures \ref{fig:7.1} and \ref{fig:7.2}. 
Table \ref{tab:7.2} shows the calculated and empirical $S$- and $P$-wave scattering lengths.

\begin{table}[t]
\caption{The calculated and empirical $K^+ N$ $S$-wave and $P$-wave scattering lengths in units of $fm$ and $fm^3$.}
\begin{ruledtabular}
\centering
\begin{tabular}{crrrr}
Scat. length & Model& SP92 \cite{Hys92}& \cite{Mar75} &\cite{Dov82} \\
\hline
$S_{01}$ & -0.09 & 0.00  & -0.04 & 0.03$\pm$ 0.15\\
$S_{11}$ & -0.28 & -0.33 & -0.32 &-0.30$\pm$ 0.03\\
$P_{01}$ & 0.137 & 0.08  & 0.086 &\\
$P_{11}$ & -0.035& -0.16 & -0.032&\\
$P_{03}$ & -0.020& -0.13 & -0.019&\\
$P_{13}$ & 0.059 & 0.07  & 0.021 & \\
\end{tabular}
\end{ruledtabular}
\label{tab:7.2}
\end{table}


\begin{figure}[t]
\begin{center}
\resizebox{8.25cm}{9.9cm}{\includegraphics*[2cm,2.0cm][20cm,27cm]{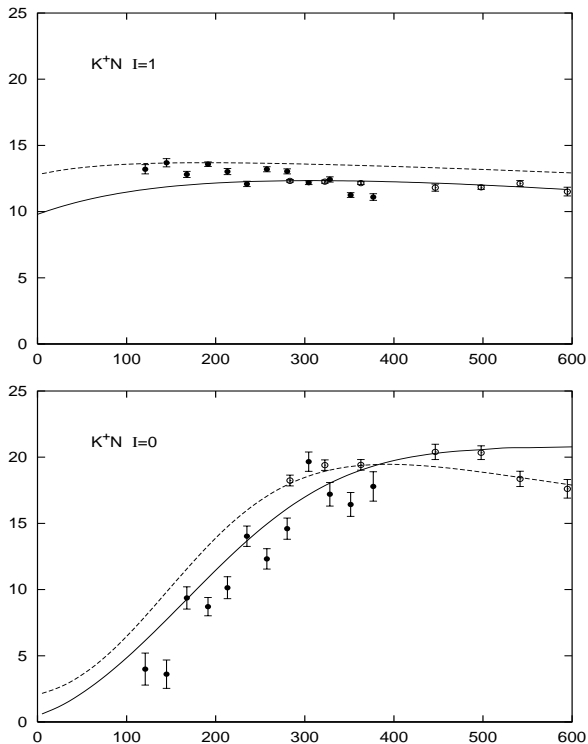}}
\end{center}
\caption{The total elastic $K^+ N$ cross section $\sigma$ (mb) as a function of $T_{{\rm lab}}$ (MeV) for both isospin channels. The experimental cross sections are from \cite{Bow70} (full circles) and \cite{Has83} (empty circles). The NSC $K^+ N$-model is given by the solid line, the dashed line is the model without tensor-mesons.}
\label{fig:7.3}
\end{figure}

\begin{figure}[t]
\begin{center}
\resizebox{8.25cm}{6.9cm}{\includegraphics*[2cm,14.8cm][19.5cm,27cm]{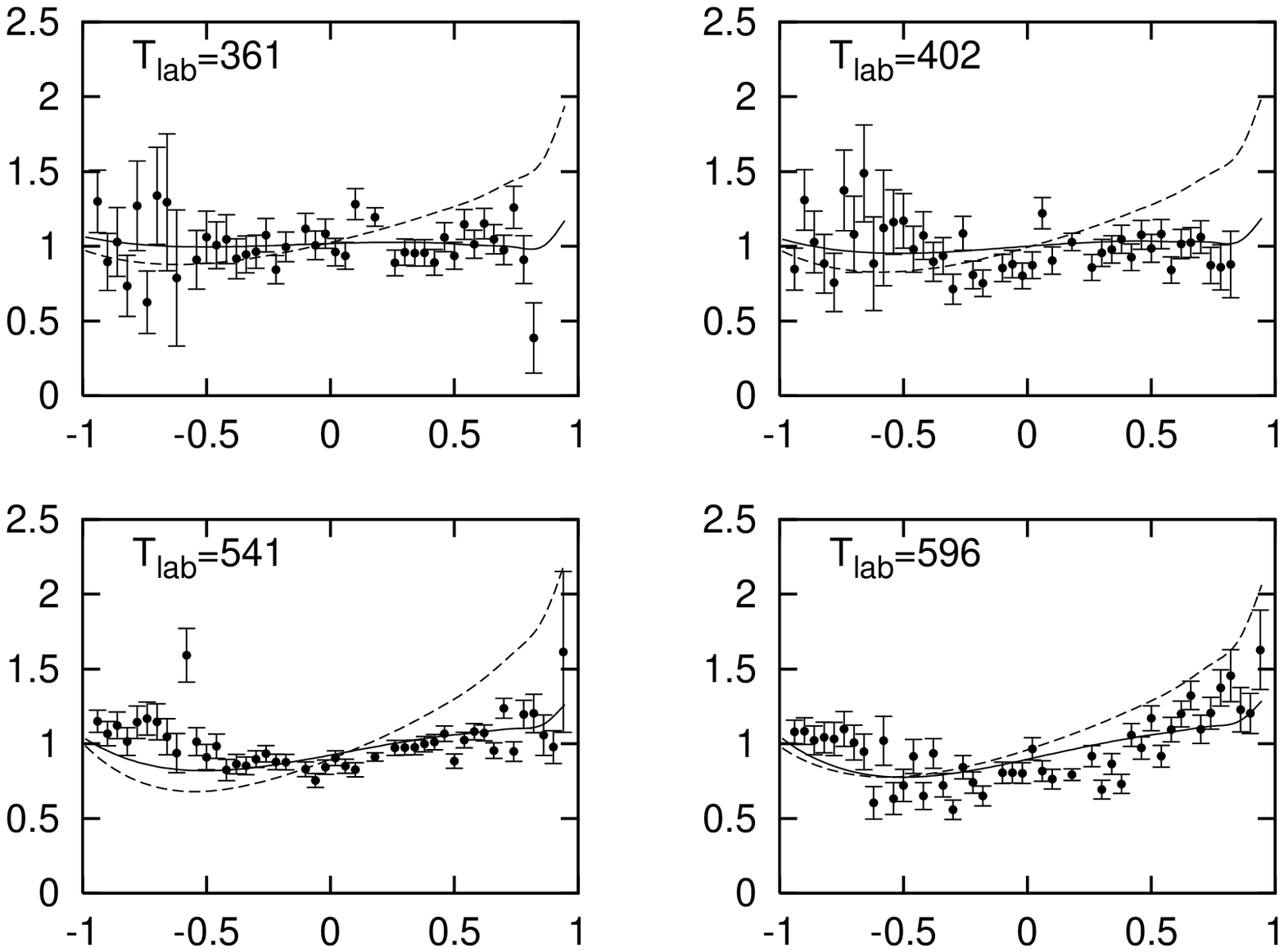}}
\end{center}
\caption{The  $K^+ p \rightarrow K^+p$ differential cross section $d \sigma / d\Omega$ (mb/sr) as a function of $\cos \theta$, where $\theta$ is the CM scattering angle. The experimental differential cross sections are from \cite{Cha77}. The NSC $K^+ N$-model is given by the solid line, the dashed line is the model without tensor-mesons.}
\label{fig:7.4}
\end{figure}

\begin{figure}[t]
\begin{center}
\resizebox{8.25cm}{9.56cm}{\includegraphics*[2cm,8.4cm][19.5cm,27cm]{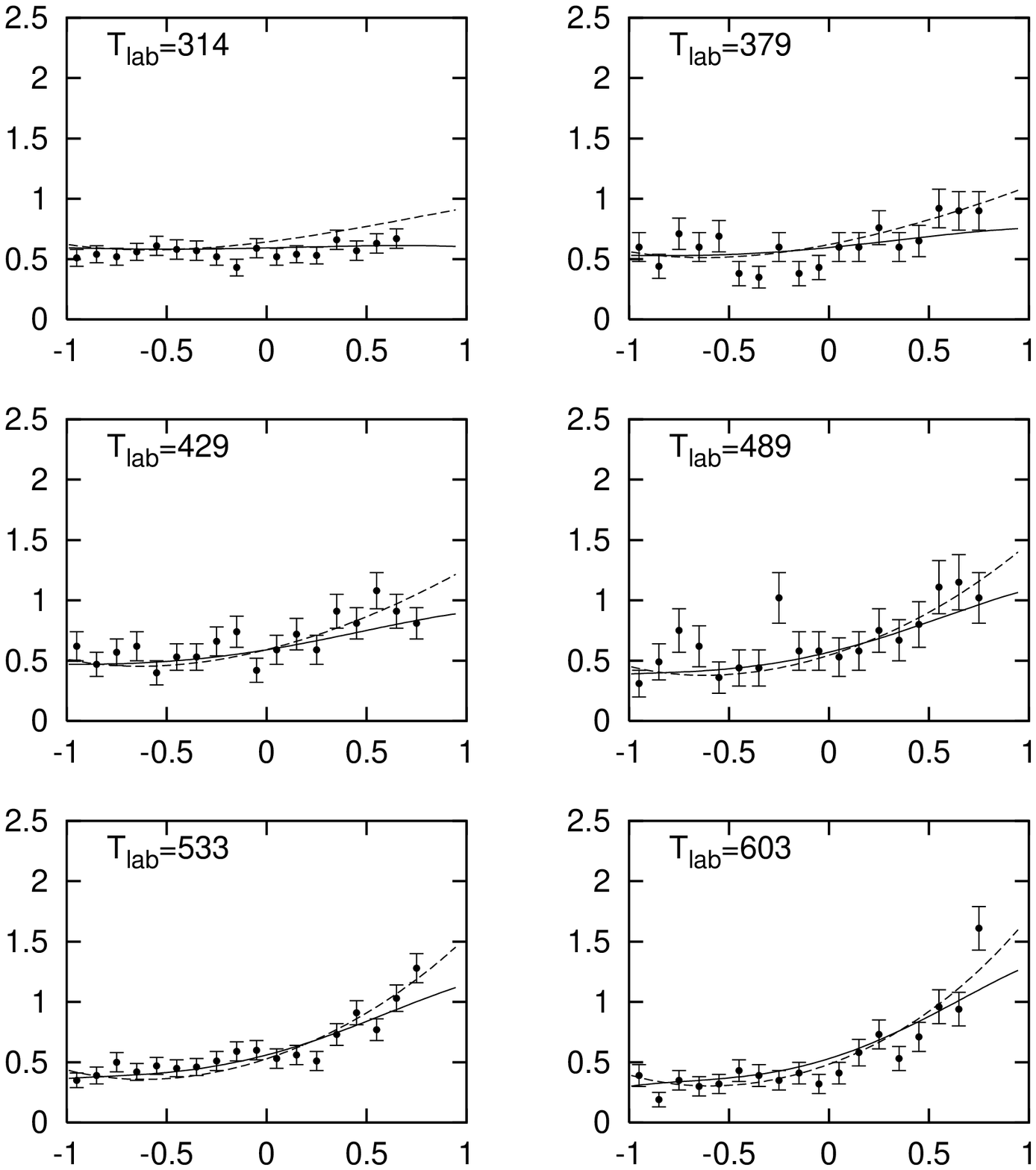}}
\end{center}
\caption{The  $K^+ n \rightarrow K^+n$ differential cross section $d \sigma / d\Omega$ (mb/sr) as a function of $\cos \theta$, where $\theta$ is the CM scattering angle. The experimental differential cross sections are from \cite{Gia73}. The NSC $K^+ N$-model is given by the solid line, the dashed line is the model without tensor-mesons.}
\label{fig:7.5}
\end{figure}

\begin{figure}[t]
\begin{center}
\resizebox{8.25cm}{9.56cm}{\includegraphics*[2cm,8.4cm][19.5cm,27cm]{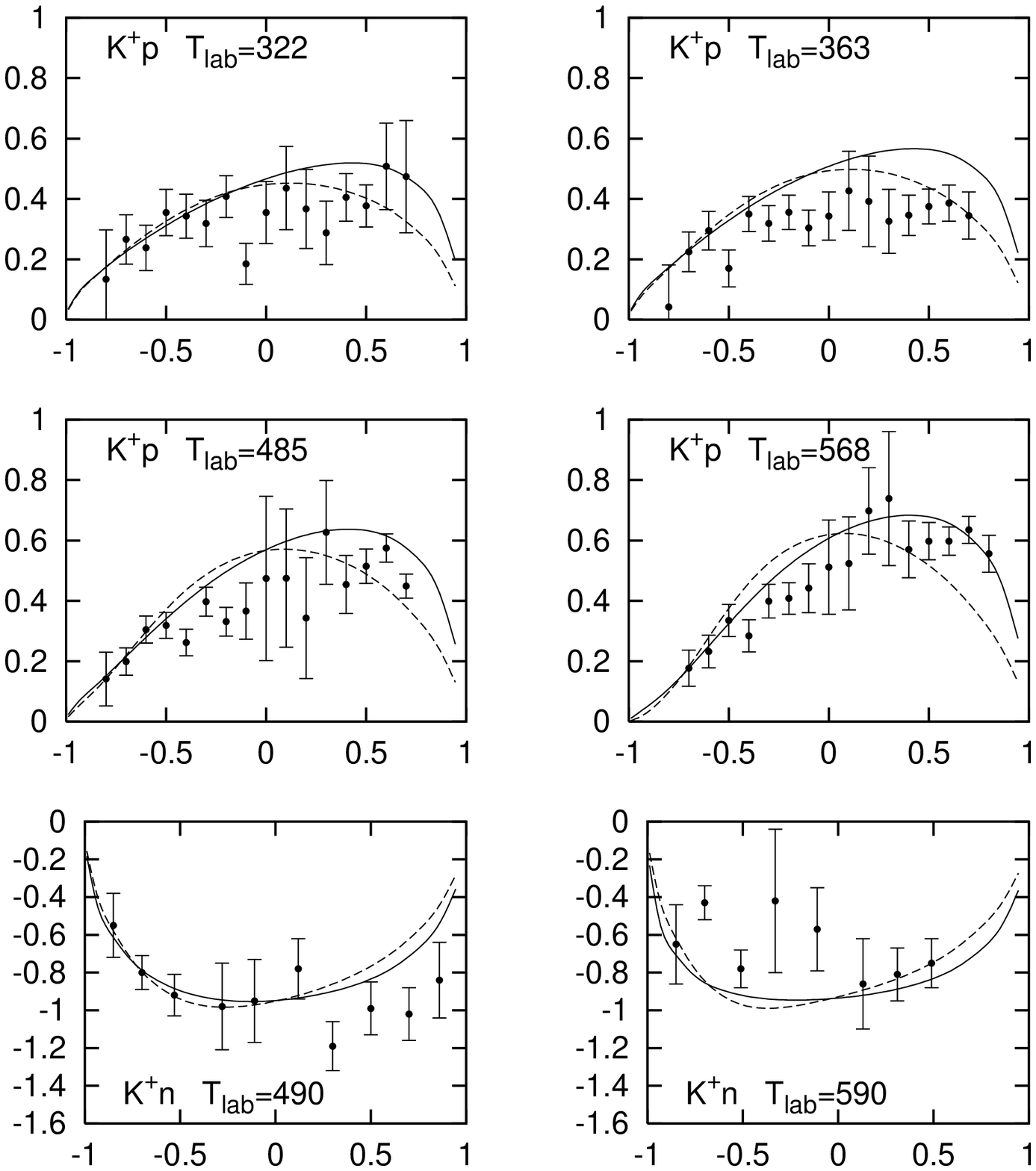}}
\end{center}
\caption{The $K^+ p \rightarrow K^+p$ and $K^+ n \rightarrow K^+n$ polarizations $P$  as a function of $\cos \theta$, where $\theta$ is the CM scattering angle. The experimental polarizations are from \cite{Lov81, Rob80}. The NSC $K^+ N$-model is given by the solid line, the dashed line is the model without tensor-mesons.}
\label{fig:7.6}
\end{figure}

A reasonable agreement between the NSC $K^+ N$-model and the empirical phases up to $T_{{\rm lab}}=600$ MeV is obtained, but the energy behavior of the empirical multi-energy phases in the $P_{11}$, $P_{03}$ and $D_{03}$ partial waves is not reproduced well by the NSC $K^+ N$-model. This, however, is also the case for the J\"{u}lich $K^+ N$ models \cite{HDH95,Had02}. The various phase shift analyses are not very consistent in these partial waves, in particular the behavior of the SP92 multi-energy $P_{03}$ and $D_{03}$ phases deviates much from the different single-energy phases. The low-energy structure of the multi-energy $D_{03}$ phase is not expected. One should wonder if this strange structure causes problems for other partial waves in the phase shift analysis.

The $S$-wave scattering lengths 
listed in Table \ref{tab:7.2}, are reproduced well. For the $P$-waves the situation is less clear, the empirical $P$-wave scattering lengths found in the two partial wave analyses \cite{Hys92} and \cite{Mar75} are contradictory. The model $P_{13}$ partial wave scattering length is in reasonable in agreement with \cite{Hys92}. The $P_{11}$ and $P_{03}$ scattering lengths are in agreement with \cite{Mar75}.

Since the various phase shift analyses do not always give consistent results and one should wonder how well the multi-energy SP92 phase shifts represent the experimental data, we also compared the NSC $K^+ N$-model with the experimental scattering observables directly. The total elastic cross sections 
as a function of $T_{{\rm lab}}$ are shown in Figure \ref{fig:7.3}. The experimental isospin one ($K^+p$) total elastic cross section is known quite accurately, the isospin zero total elastic cross section is known to less accuracy. The NSC $K^+ N$-model reproduces both total elastic cross sections quite well. The differential cross sections for the channels $K^+p\rightarrow K^+p$ and $K^+n\rightarrow K^+n$, having quite large error bars, are shown in Figures \ref{fig:7.4} and \ref{fig:7.5} as a function of the scattering angle. 
They are described well by the NSC $K^+ N$-model. Finally the polarizations, also having large error bars, are given in Figure \ref{fig:7.6} for the same channels as a function of the scattering angle. 
Again a good agreement between the model and the experimental values is seen.

Although the empirical phase shifts are not in all partial waves described very well by the NSC $K^+ N$-model, the scattering observables as well as the $S$-wave scattering lengths 
are.
We remark that the description of the experimental scattering data and the phase shifts by the NSC $K^+ N$-model, containing only one-particle-exchange processes, is as least as good as that of the J\"{u}lich models \cite{HDH95,Had02}. These models, however, used two-particle-exchanges to describe the experimental data.

The parameters of the NSC $K^+ N$-model 
searched and fixed in the fitting procedure are listed in Table \ref{tab:7.1}. The NSC $K^+ N$-model has six different cutoff masses, which are free parameters in the fitting procedure. For the three scalar-mesons we use the same cutoff mass, for the vector-mesons we use the same cutoff mass for the $\rho$ and $\varphi$, but allow for a different value for the $\omega$ in order to find a better description of the $S_{11}$ and $P_{01}$ partial waves at higher energies. For the three tensor-mesons, necessary to fit the $S_{11}$, $P_{01}$ and $P_{13}$ partial waves simultaneously, we use the same cutoff mass. For the Pomeron mass we take the value found for the NSC $\pi N$-model, the meson and baryon masses have been fixed in the fitting procedure.

\begin{table}[t]
\caption{NSC $K^+ N$-model parameters: coupling constants, masses and cutoff masses (MeV). Coupling constants with an asterisk were not searched in the fitting procedure, but constrained via $SU_f(3)$ or simply put to some value used in previous work. An $SU_f(3)$-breaking factor $\lambda_V=0.764$ for the vector and $\lambda_S=0.899$ for the scalar-mesons was found.}
\begin{ruledtabular}
\centering
\begin{tabular}{crcrrcrrr}
Exch. &\multicolumn{6}{c}{Coupling Constants} &Mass &$\Lambda$  \\
\hline
$\rho$ &$\frac{g_{NN\rho}g_{KK\rho}}{4\pi}$&$\! \!$=&$\! \!$$0.667^*$&  $\frac{f_{NN\rho}}{g_{NN\rho}}$&$\! \!$=&$\! \!$$5.285$ &770 & 1563 \\
$\omega$ &$\frac{g_{NN\omega}g_{KK\omega}}{4\pi}$&$\! \!$=&$\! \!$2.572&  $\frac{f_{NN\omega}}{g_{NN\omega}}$&$\! \!$=&$\! \!$$0.345$ &783 & 1805 \\
$\varphi$ &$\frac{g_{NN\varphi}g_{KK\varphi}}{4\pi}$&$\! \!$=&$\! \!$$-0.573^*$&  $\frac{f_{NN\varphi}}{g_{NN\varphi}}$&$\! \!$=&$\! \!$$0.932^*$ &1020 & 1563 \\
$a_0$ &$\frac{g_{NNa_0}g_{KKa_0}}{4\pi}$&$\! \!$=&$\! \!$3.461& && &980 &712  \\
$\sigma$ &$\frac{g_{NN\sigma}g_{KK\sigma}}{4\pi}$&$\! \!$=&$\! \!$$20.676^*$& && &760 &712  \\
$f_0$ &$\frac{g_{NNf_0}g_{KKf_0}}{4\pi}$&$\! \!$=&$\! \!$$4.203^*$& && &975 &712  \\
$a_2$ &$\frac{g_{NNa_2}g_{KKa_2}}{4\pi}$&$\! \!$=&$\! \!$$0.019$&  $\frac{f_{NNa_2}}{g_{NNa_2}}$&$\! \!$=&$\! \!$$-3.161$ &1320 & 854 \\
$f_2$ &$\frac{g_{NNf_2}g_{KKf_2}}{4\pi}$&$\! \!$=&$\! \!$0.080&  $\frac{f_{NNf_2}}{g_{NNf_2}}$&$\! \!$=&$\! \!$$0.382$ &1270 & 854 \\
$f_2'$ &$\frac{g_{NNf_2'}g_{KKf_2'}}{4\pi}$&$\! \!$=&$\! \!$$0.022^*$&  $\frac{f_{NNf_2'}}{g_{NNf_2'}}$&$\! \!$=&$\! \!$$3.393^*$ &1525 & 854 \\
Pom. &$\frac{g_{NN P}g_{KKP}}{4\pi}$&$\! \!$=&$\! \!$$4.135^*$& && &$315^*$ &  \\
$\Lambda$ &$\frac{f_{\Lambda NK}^2}{4\pi}$&$\! \!$=&$\! \!$$0.074^*$& && &1116 &1029  \\
$\Sigma$ &$\frac{f_{\Sigma NK}^2}{4\pi}$&$\! \!$=&$\! \!$$0.006^*$& && &1189 &1029  \\
$\Sigma^*$ &$\frac{f_{\Sigma^*NK}^2}{4\pi}$&$\! \!$=&$\! \!$$0.147^*$& && &1385 &1052  \\
$\Lambda^*$ &$\frac{f_{\Lambda^*NK}^2}{4\pi}$&$\! \!$=&$\! \!$$0.710^*$& && &1405 &1052  \\
\end{tabular}
\end{ruledtabular}
\label{tab:7.1}
\end{table}

Ideal mixing is assumed for the vector-mesons, so $\theta_V=35,26^{\circ}$, the $F/(F+D)$-ratios are fixed to the values in \cite{MRS89}, $\alpha_V^e=1.0$ and $\alpha_V^m=0.275$. This fixes the PPV coupling constants in terms of the empirical determined $f_{\pi\pi\rho}$ and leaves $g_{NN\omega}$ and $f_{NN\omega}$ as fit parameters, the fitted values are in agreement with the literature. The tensor coupling $f_{NN\rho}$ is in principle determined in the NSC $\pi N$-model, but since its value was determined to be very low we also fit this parameter in the NSC $K^+N$-model and we found a larger value than in the NSC $\pi N$-model. We remark that the exchange of the vector-meson $\varphi$ is considered for consistency, but its contribution is negligible.
For the scalar-mesons $g_{NN\sigma}$ and $g_{NNf_0}$ are determined in the NSC $\pi N$-model, we use $g_{NNa_0}$ and $\theta_S$ as fit parameters, all scalar-meson coupling constants are then determined.
For the tensor-mesons we use the $F/(F+D)$-ratios $\alpha_T^e=1.0$ and $\alpha_T^m=0.4$ and an almost ideal mixing angle $\theta_T=37.50$. This fixes the $PPT$ coupling constants in terms of $f_{\pi\pi f_2}$. We notice that the tensor-meson coupling constants $g_T={\mathcal M}F_1+{\mathcal M}^2F_2$ and $f_T=-{\mathcal M}^2F_2$ are used in Table \ref{tab:7.1}.

The $\Lambda NK$ and $\Sigma NK$ coupling constants are determined  by $f_{NN\pi}$ and fixing $\alpha_P$ at the value in \cite{MRS89} $\alpha_P=0.355$. The Pomeron is considered as an $SU_f(3)$-singlet and its coupling to the $K^+ N$ system is determined in the NSC $\pi N$-model. 
For the $\Lambda^*$ coupling constant we take an average value from \cite{NRS79}. In the fitting procedure we found that it was desirable to allow for an $SU_f(3)$-breaking for the scalar- and vector-meson couplings. The breaking factors we found are $\lambda_S=0.899$ and $\lambda_V=0.764$.

The NSC $K^+ N$-model has 17 free physical parameters; 8 coupling constants, 1 mixing angle, 6 cutoff masses and 2 $SU_f(3)$ breaking parameters. From the $\pi N$ fit we have $g_{NN\rho}=0.78$ and $g_{NN\sigma}=2.47$, from the $K^+ N$ fit we have $g_{NN\omega}=3.03$ and $g_{NNa_0}=0.78$.

Besides the discussed NSC $K^+ N$-model, we also considered a model that does not contain tensor-mesons. We fitted this model to the empirical phase shifts and the results of the fit are given by the dashed lines in Figures \ref{fig:7.1} and \ref{fig:7.2}. The parameters of this model are listed in Table \ref{tab:7.2a}. We remark that in the $P_{13}$ and $D_{03}$ partial waves a noticeable difference can be seen between the two models. These partial waves as well as the $S_{11}$ and $P_{01}$ partial waves are described better by the NSC $K^+ N$-model, i. e. the model including the tensor-mesons. The total cross sections and $K^+p\rightarrow K^+p$ differential cross sections are described better by the NSC $K^+ N$-model, while the $K^+n\rightarrow K^+n$ differential cross sections and the polarizations are described equally well.

Summarizing, the NSC $K^+ N$-model gives a reasonable description of the empirical partial wave phase shifts and also the $S$-wave scattering lengths are reproduced well. The scattering observables, investigated because the various phase shift analyses are not always consistent, are described satisfactory by this model.

\begin{table}[t]
\caption{Parameters of the NSC $K^+ N$-model without tensor-mesons: coupling constants, masses and cutoff masses (MeV). Coupling constants with an asterisk were not searched in the fitting procedure, but constrained via $SU_f(3)$ or simply put to some value used in previous work. An $SU_f(3)$-breaking factor $\lambda_V=0.918$  for the vector and $\lambda_S=0.900$ for the scalar-mesons was found.}
\begin{ruledtabular}
\centering
\begin{tabular}{crcrrcrrr}
Exch. &\multicolumn{6}{c}{Coupling Constants} &Mass &$\Lambda$  \\
\hline
$\rho$ &$\frac{g_{NN\rho}g_{KK\rho}}{4\pi}$&$\! \!$=&$\! \!$$0.641^*$&  $\frac{f_{NN\rho}}{g_{NN\rho}}$&$\! \!$=&$\! \!$$5.443$ &770 & 1547 \\
$\omega$ &$\frac{g_{NN\omega}g_{KK\omega}}{4\pi}$&$\! \!$=&$\! \!$2.215&  $\frac{f_{NN\omega}}{g_{NN\omega}}$&=&$0.345$ &783 & 1704 \\
$\varphi$ &$\frac{g_{NN\varphi}g_{KK\varphi}}{4\pi}$&$\! \!$=&$\! \!$$-0.243^*$&  $\frac{f_{NN\varphi}}{g_{NN\varphi}}$&$\! \!$=&$\! \!$$1.842^*$ &1020 & 1547 \\
$a_0$ &$\frac{g_{NNa_0}g_{KKa_0}}{4\pi}$&$\! \!$=&$\! \!$3.806& && &980 &909  \\
$\sigma$ &$\frac{g_{NN\sigma}g_{KK\sigma}}{4\pi}$&$\! \!$=&$\! \!$$26.068^*$& && &760 &909  \\
$f_0$ &$\frac{g_{NNf_0}g_{KKf_0}}{4\pi}$&$\! \!$=&$\! \!$$1.168^*$& && &975 &909  \\
Pom. &$\frac{g_{NN P}g_{KKP}}{4\pi}$&$\! \!$=&$\! \!$$4.453^*$& && &$296^*$ &  \\
$\Lambda$ &$\frac{f_{\Lambda NK}^2}{4\pi}$&$\! \!$=&$\! \!$$0.074^*$& && &1116 &1041  \\
$\Sigma$ &$\frac{f_{\Sigma NK}^2}{4\pi}$&$\! \!$=&$\! \!$$0.006^*$& && &1189 &1041  \\
$\Sigma^*$ &$\frac{f_{\Sigma^*NK}^2}{4\pi}$&$\! \!$=&$\! \!$$0.147^*$& && &1385 &629  \\
$\Lambda^*$ &$\frac{f_{\Lambda^*NK}^2}{4\pi}$&$\! \!$=&$\! \!$$0.710^*$& && &1405 &629  \\
\end{tabular}
\end{ruledtabular}
\label{tab:7.2a}
\end{table}

\subsection{Exotic resonances}
Evidence for the existence of a resonance structure in the isospin zero $K^+ N$ system at low energies has recently been found in various measurements from SPring-8, ITEP, Jefferson Lab and ELSA \cite{Nak03, Bar03, Ste03, Bart03}. The exotic resonance, a $qqqq\bar{q}$-state, was called $Z^*$ but is now renamed as $\Theta^+$. The experimental values for its mass and decay width are $\sqrt{s}\simeq1540$ MeV and $\Gamma_{\Theta^+}\leq25$ MeV. This is in good agreement with the theoretical predictions of Diakonov et al. \cite{Dia97} based on the chiral quark-soliton model, giving $\sqrt{s}\simeq1530$ MeV and $\Gamma_{\Theta^+}\simeq15$ MeV, isospin $I=0$ and spin-parity $J^P=\frac{1}{2}^+$.

The present $K^+ N$ scattering data does not explicitly show this resonance structure, but some fluctuations in the isospin zero scattering data around $\sqrt{s}=1540$ MeV are present, however the decay width of the $\Theta^+$ is expected to be quite small. Arndt et al. \cite{Arn03} have reanalyzed the $K^+ N$ scattering database and investigated the possibility of a resonance structure in their $K^+ N$ phase shift analysis. Since their last phase shift analysis \cite{Hys92} no new scattering data has become available. Arndt et al. concluded that the $\Theta^+$ decay width must indeed be quite small in view of the present scattering data. They concluded that $\Gamma_{\Theta^+}$ is not much larger than a few MeV.

In this subsection the NSC $K^+ N$-model, describing the experimental data well far beyond the $\Theta^+$ resonance region, is used to examine the influence of including this resonance explicitly on the total elastic isospin zero $K^+ N$ cross section. This has also been done by the J{\"u}lich group \cite{Hai03,Sib04,Sib05}. The $\Theta^+$ resonance is assumed to be present in the $P_{01}$ partial wave. The procedure for including the $\Theta^+$ resonance explicitly in the $K^+ N$ system is completely the same as for the $\Delta$ in the $\pi N$ system. This renormalization procedure, giving a good description of the $\pi N$ $P_{33}$ partial wave, is described in detail in Sec. \ref{chap:4}.

A pole diagram for the $\Theta^+$ resonance with bare mass and coupling constant $M_0$ and $g_0$ is added to the $K^+ N$ potential, iteration in the integral equation dresses the vertex and self-energy. The renormalization procedure ensures a pole at the physical $\Theta^+$ mass and the vanishing of the self-energy and its first derivative at the pole position. The bare mass and coupling constant are in the renormalization procedure determined in terms of the physical parameters. The physical $KN\Theta^+$ coupling constant is calculated using the decay width and Eq. (\ref{eq:res3}). We mention that we did not fit the model which includes the $\Theta^+$ resonance to the scattering data, but simply used the NSC $K^+ N$-model, added the $\Theta^+$ pole diagram and observed the change in the cross section.

\begin{figure}[t]
\begin{center}
\resizebox{8.25cm}{4.79cm}{\includegraphics*[2cm,2.0cm][20cm,14.1cm]{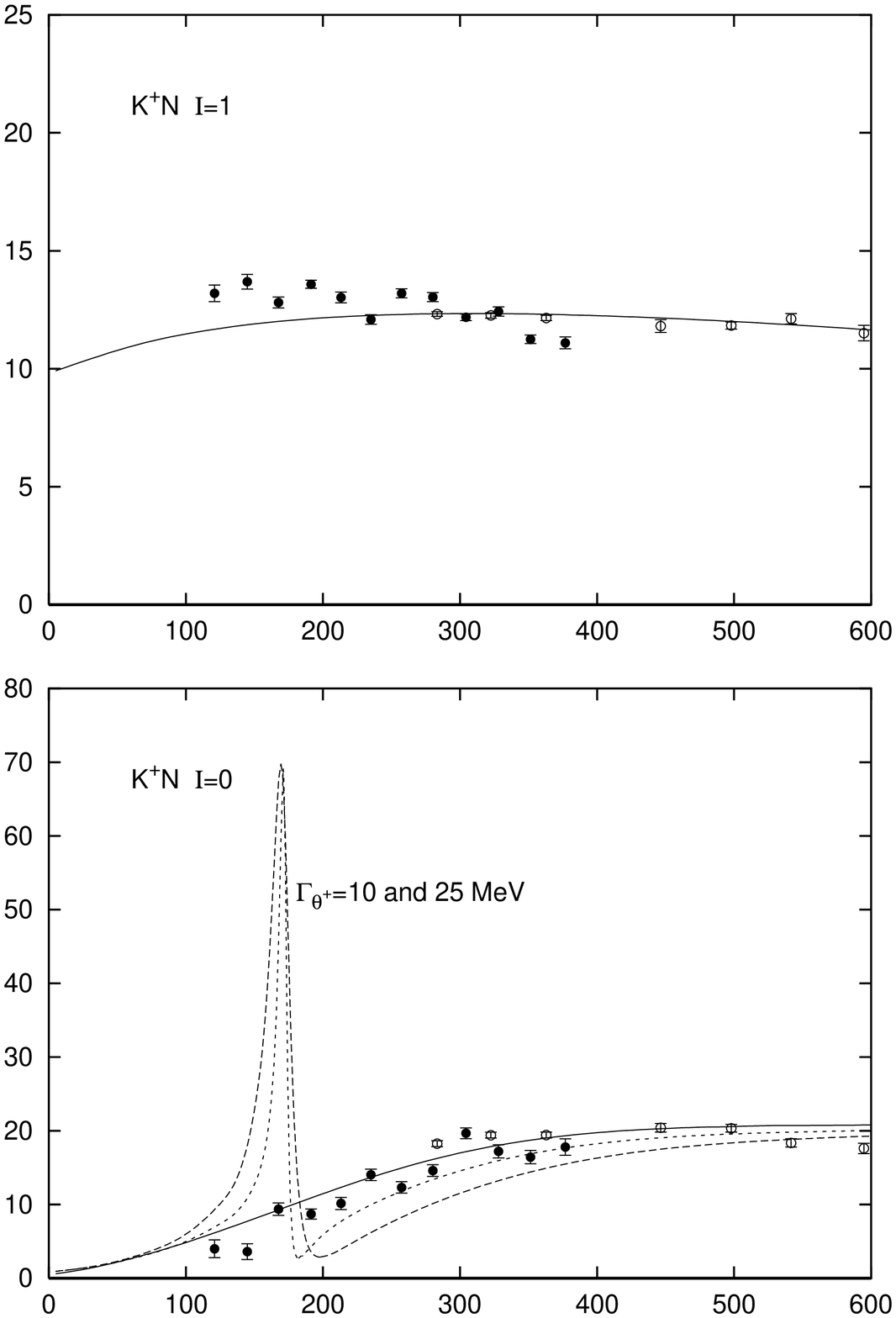}}
\end{center}
\caption{The $\Theta ^+$ resonance included in the NSC $K^+ N$-model. The total elastic $K^+ N$ cross section $\sigma$ (mb) is given as a function of $T_{{\rm lab}}$ (MeV). The experimental cross sections are from \cite{Bow70} (full circles) and \cite{Has83} (empty circles). The NSC $K^+N$-model is given by the solid line.}
\label{fig:7.7}
\end{figure}

The total elastic cross section in the isospin zero channel, predicted by the NSC $K^+ N$-model, is given in Figure \ref{fig:7.7} by the solid line. Inclusion of the $\Theta^+$ resonance results in a peak in the isospin zero cross section around $\sqrt{s}=1540$ MeV or $T_{{\rm lab}}=171$ MeV. We calculated the influence of the $\Theta^+$ resonance on the isospin zero cross section for two values of its decay width, $\Gamma_{\Theta^+}=$ 10 and 25 MeV, corresponding to the short and long dashed curves.

Far away from the resonance position
the dashed curves coincide with the solid NSC $K^+ N$ curve. It is clear that the smaller the $\Theta^+$ decay width the narrower the peak and the more the dashed curve coincides with the solid NSC $K^+ N$ curve. It is hard to reconcile the present isospin zero $K^+ N$ scattering data with a $\Theta^+$ resonance decay width larger than 10 MeV, unless the $\Theta^+$ resonance lies much closer to threshold, where no scattering data is available. In both cases new and accurate scattering experiments, especially at low energies and around $\sqrt{s}=1540$ MeV, would be desirable.

\section{Summary and outlook}

In paper I the NSC model was derived. Its application to the $\pi N$ interaction presented in this paper shows that the soft-core approach of the Nijmegen group not only gives a good description of the $NN$ and $YN$ data, but also the $\pi N$ data are described well in this approach. The NSC $\pi N$-model serves as a solid basis for the NSC $K^+ N$-model, assumed to be connected via $SU_f(3)$-symmetry. 

In the $\pi N$ cross section some resonances are present at low and intermediate energies, e.g. the $\Delta$ and the Roper. It turned out that these resonances can not be described by using only a $\pi N$ potential, i.e. they could not be generated dynamically. This confirms the quark-model picture. We consider these resonances as, at least partially, genuine three-quark states and we treat them in the same way as the nucleon. Therefore we have included $s$-channel diagrams for these resonances in the NSC $\pi N$-model. However, this is done carefully in a renormalized procedure, i.e. a procedure in which physical coupling constants and masses are used.

The NSC $\pi N$-model contains the exchanges of the baryons $N$, $\Delta$, Roper and $S_{11}$ and the scalar-mesons $\sigma$ and $f_0$, vector-meson $\rho$ and tensor-mesons $f_2$ and $f_2'$. An excellent fit to the empirical $S$- and $P$- wave phase shifts up to pion laboratory energy 600 MeV is given in Sec. \ref{chap:6}. We found normal values for the coupling constants and cutoff masses, except for a low value of $f_{NN\rho}/g_{NN\rho}$, which is also a problem in other $\pi N$ models. The scattering lengths have been reproduced well.
The soft-pion theorems for low-energy $\pi N$ scattering are satisfied, since the $S$-wave scattering lengths are described well. The $c_1$-, $c_2$-, $c_3$- and $c_4$-terms in chiral perturbation theory are described implicitly by the NSC $\pi N$-model, higher derivative terms in chiral perturbation theory are effectively described by the propagators and Gaussian form factors in the NSC $\pi N$-model.

The NSC $K^+ N$-model
 and the fit to the experimental data are presented in Sec. \ref{chap:7}. The model contains the exchanges of the baryons $\Lambda$, $\Sigma$, $\Sigma^*$ and $\Lambda^*$, the scalar-mesons $a_0$, $\sigma$ and $f_0$, the vector-mesons $\rho$, $\omega$ and $\varphi$ and the tensor-mesons $a_2$, $f_2$ and $f_2'$. The quality of the fit to the empirical phase shifts up to kaon laboratory energy 600 MeV is not as good as for the NSC $\pi N$-model, but the NSC $K^+ N$-model certainly reflects the present state of the art.
The scattering observables
and the $S$-wave scattering lengths are reproduced well.

Low energy (exotic) resonances have never been seen in the present $K^+ N$ scattering data, however, recently indications for the existence of a narrow resonance in the isospin zero $K^+N$ system have been found in several photo-production experiments. We have included this resonance $\Theta^+(1540)$ in the NSC $K^+ N$-model, in the same way as we included resonances in the NSC $\pi N$-models, and investigated its influence as a function of its decay width on the total cross section. We concluded that, in view of the present scattering data, its decay width must be smaller than 10 MeV.

The present NSC $\pi N$- and $K^+ N$-models 
could be improved by adding two-particle-exchange processes to the $\pi N$ and $K^+ N$ potentials, similar to the extended soft-core $NN$ and $YN$ models.
Also, the Coulomb interaction, which in principle plays a role at very low energies, has not been considered here.

Finally this work provides the basis for the extension of the soft-core approach to the antikaon-nucleon ($\bar{K}N$) interaction, and to meson-baryon interactions in general. The $\bar{K}N$ system is already at threshold coupled to the $\Lambda\pi$ and $\Sigma\pi$ channels. The coupled channels treatment for this system is similar to that of the $YN$ system.

\begin{acknowledgements}
The authors would like to thank Prof. J. J. de Swart and Prof. R. G. E. Timmermans for stimulating discussions.
\end{acknowledgements}

\appendix

\section{OBE and baryon-exchange isospin factors}
\label{app:AA}

We outline the calculation of the isospin factors for the meson-baryon interactions, making use of the Wigner 6-j and 9-j symbols, 
\cite{Edm57}, this reference also gives relations for interchanging the labels of Clebsch-Gordan coefficients. An example for the $\pi N$ and for the $K^+ N$ interaction is given.\\

\begin{figure}[b]
\begin{center}
\resizebox{8.25cm}{2.15cm}{\includegraphics*[3cm,23.4cm][13cm,26.cm]{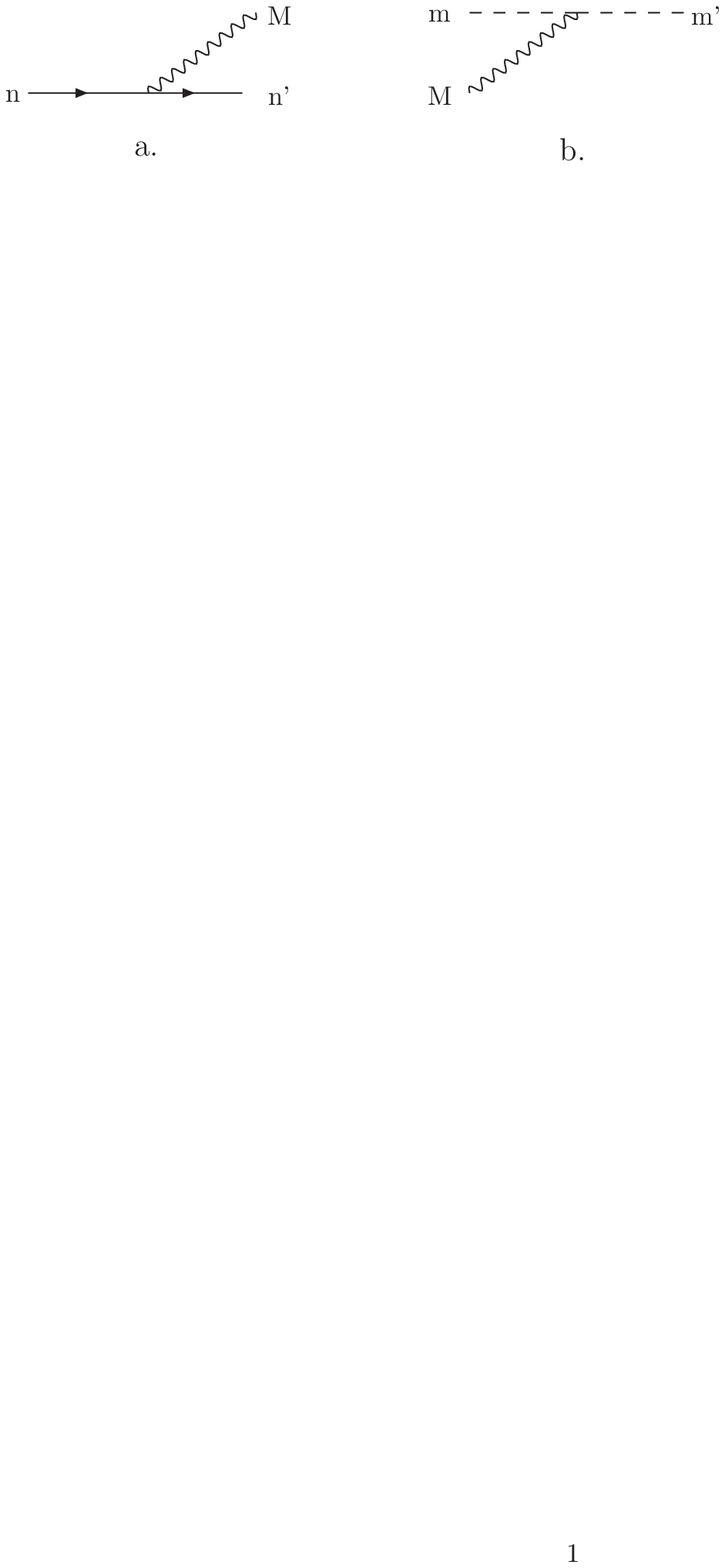}}
\end{center}
\caption{Figure a. shows the $\rho$ emission vertex and figure b. shows the $\rho$ absorption vertex.}
\label{fig:bb.1}
\end{figure}

\begin{center}
{\bf (i) Baryon-exchange in $\pi N$ interactions:}
\end{center}
\begin{figure}[b]
\begin{center}
\resizebox{8.25cm}{2.57cm}{\includegraphics*[6cm,23.2cm][15cm,26.cm]{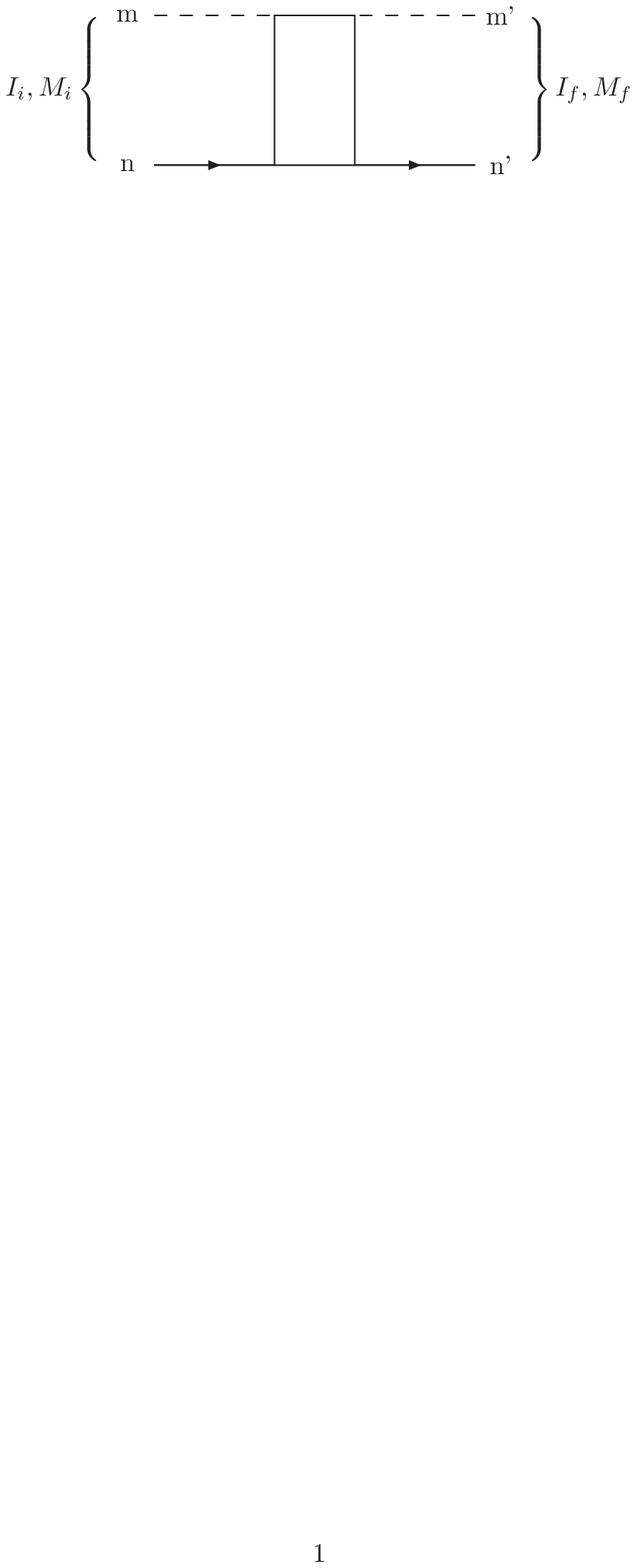}}
\end{center}
\caption{The matrix element for the total isospin, $m$ is the $z$-component of the pion isospin and $n$ is the $z$-component of the nucleon isospin.}
\label{fig:b.0}
\end{figure}
The isospin matrix element for a given total final and initial isospin in the $\pi N$ system reads
\begin{eqnarray}
 \langle I_f\ M_f| {\cal H}| I_i\ M_i\rangle &=& 
C^{1\ \ \, \frac{1}{2}\ \ I_f }_{m'\ n'\ M_f}\ C^{1\ \ \frac{1}{2}\ \, I_i }_{m\ n\ M_i}\
\nonumber \\ && \times
\langle \pi_{m'}\ N_{n'}|{\cal H}|\pi_m\ N_n\rangle\ ,           
\label{eq:b.1} \end{eqnarray}
where $I$ is the total isospin of the system and $M$ its $z$-component, $m$ is the $z$-component of the pions isospin and $n$ is the $z$-component of the nucleons isospin, see Figure \ref{fig:b.0}.
We can rewrite the first Clebsch-Gordan coefficient in Eq. (\ref{eq:b.1}), \cite{Edm57},
\begin{eqnarray}
C^{1\ \ \, \frac{1}{2}\ \ I_f }_{m'\ n'\ M_f}&=&(-)^{\frac{3}{2}-I_f}\ C^{\frac{1}{2}\ \, 1\ \ \ I_f }_{\, n'\; m'\ M_f}\ .
\label{eq:b.1a}
\end{eqnarray}
\begin{figure}[t]
\begin{center}
\resizebox{8.25cm}{2.15cm}{\includegraphics*[3cm,23.4cm][13cm,26.cm]{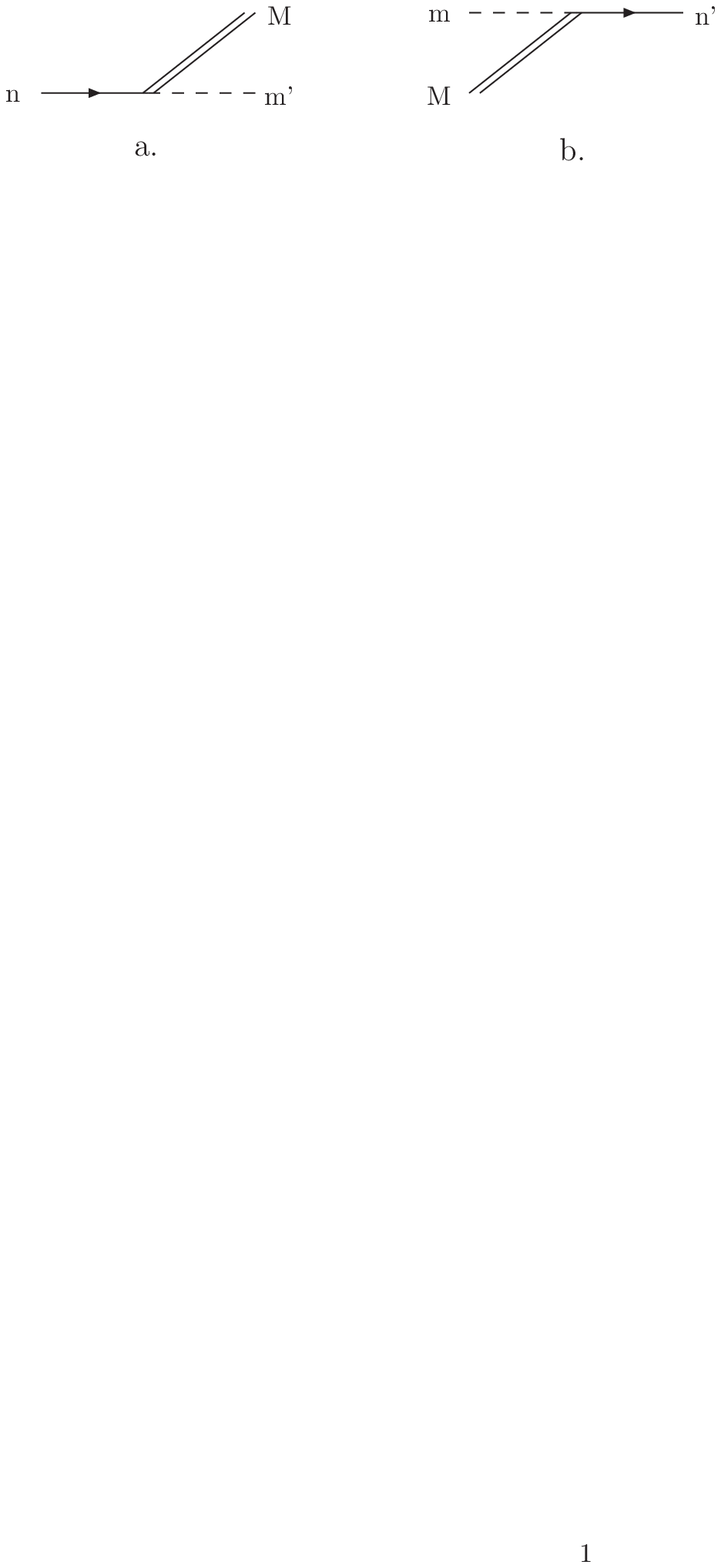}}
\end{center}
\caption{Figure a. shows the baryon emission vertex and figure b. shows the baryon absorption vertex.}
\label{fig:b.1}
\end{figure}
For baryon-exchange the isospin interaction Hamiltonian ${\cal H}$ for either the $NN\pi$ or the $N\Delta\pi$ vertex is
\begin{eqnarray}
{\cal H}&= &
\left<i\left|\left|T^\prime\right|\right|1/2\right>
C^{\frac{1}{2}\ \; 1 \ \;\; i }_{\: n\ \, m\ M}\ \left[\psi^*_MN_n\pi^*_m+ N^*_n\psi_M\pi_m\right]\ ,\nonumber \\
\label{eq:b.2}
\end{eqnarray}
where $\psi_M$ denotes either the nucleon with $i=\frac{1}{2}$ or the $\Delta$ with $i=\frac{3}{2}$ and $T^\prime$ denotes $\mbox{\boldmath $\tau $}$ or $\mbox{\boldmath $T $}$. Here $\pi_{+1}=-\left(\pi_1+i\pi_2\right)/\sqrt{2}$, $\pi_{-1}=\left(\pi_1-i\pi_2\right)/\sqrt{2}$ and $\pi_0=\pi_3$, we note that $\pi_m=(-)^m\pi^*_{-m}$. The baryon emission vertex shown in Figure \ref{fig:b.1} gives, besides the reduced matrix element, the factor
\begin{eqnarray}
(-)^{m'}\ C^{\frac{1}{2}\ \ \ \; 1\ \ \ i }_{\;n\ -m'\ M}
&=&(-)^{2i-M-\frac{1}{2}}\sqrt{\frac{2i+1}{3}}\ C^{\frac{1}{2}\ \ \: i\ \ \, 1 }_{\:n\: -M\ m'}\ . \nonumber \\
\label{eq:b.3}
\end{eqnarray}
The baryon absorption vertex shown in Figure \ref{fig:b.1} gives, besides the reduced matrix element, the factor
\begin{eqnarray}
(-)^{m}\ C^{\frac{1}{2}\ \ \ \, 1 \ \ \: i }_{\, n'\ -m \ M}
&=&-\sqrt{\frac{2i+1}{2}}\ C^{1\ \ i\ \ \frac{1}{2}}_{m\ M\ n'}\ .
\label{eq:b.4}
\end{eqnarray}
Using Eqs. (\ref{eq:b.1a}), (\ref{eq:b.3}) and (\ref{eq:b.4}) we find for the total isospin matrix element of Eq. (\ref{eq:b.1})
\begin{eqnarray}
\langle I_f\ M_f| {\cal H}| I_i\ M_i\rangle&=&
(-)^{1+3i-I_f+(i-M)}\ \frac{2i+1}{\sqrt{6}} \nonumber \\
&&\times C^{\frac{1}{2}\ \, 1\ \ \, I_f }_{n'\ m'\ M_f}\ C^{1\ \, \frac{1}{2}\ I_i }_{m\ n\ M_i}\ 
\nonumber \\ && \times
C^{\frac{1}{2}\ \ \, i\ \ \, 1 }_{n\ -M\ m'}\ C^{1\ \ i\ \ \frac{1}{2}}_{m\ M\ n'}
\nonumber \\ && \times
\left<i\left|\left|T^\prime\right|\right|1/2\right>^2\ ,
\label{eq:b.5}
\end{eqnarray}
using the identity $(-)^{i-M}=\sqrt{2i+1}\ C^{\, i\ \ \ \: \,  i\ \ 0 }_{M\ -M\ 0}$, we find
\begin{eqnarray}
\langle I_f\ M_f| {\cal H}| I_i\ M_i\rangle&=&
(-)^{1+3i-I_f}\sqrt{\frac{2i+1}{6}}\ \left(2i+1\right)
\nonumber \\ && \times
\left<i\left|\left|T^\prime\right|\right|1/2\right>^2
\left[\begin{array}{ccc}1 & i & \frac{1}{2} \\ \frac{1}{2} & i & 1 \\ I & 0 & I \end{array}\right] \nonumber \\
&=&-
(2i+1)\left<i\left|\left|T^\prime\right|\right|1/2\right>^2
\nonumber \\ &&\times
\left\{\begin{array}{ccc}\frac{1}{2} & 1 & i \\ \frac{1}{2} & 1 & I \end{array} \right\} \ ,
\label{eq:b.6}
\end{eqnarray}
here we have used the conservation of isospin $I_f=I_i=I$. 
For nucleon-exchange the reduced matrix element is $\langle \frac{1}{2} \| \mbox{\boldmath $\tau $} \| \frac{1}{2} \rangle = \sqrt{3}$ and for $\Delta$-exchange it is $\langle \frac{3}{2} \| \mbox{\boldmath $T$} \| \frac{1}{2} \rangle = 1$. 
We find the isospin factors given in Table \ref{tab:b.1}.
\begin{table}[t]
\caption{The isospin factors for nucleon- and $\Delta$-exchange for a given total isospin $I$ of the $\pi N$ system.}
\begin{ruledtabular}
\centering
\begin{tabular}{crr}
Exchange & $I=\frac{1}{2}$& $I=\frac{3}{2}$ \\
\hline
$N $ & -1& 2 \\
$\Delta$ & $\frac{4}{3}$ & $\frac{1}{3}$\\
\end{tabular}
\end{ruledtabular}
\label{tab:b.1}
\end{table}
\\
\noindent
\begin{center}
{\bf (ii) $\rho$-exchange in $K^+ N$ interactions:}
\end{center}
The isospin matrix element for a given total final and initial isospin in the $K^+ N$ system reads
\begin{eqnarray}
 \langle I_f\ M_f| {\cal H}| I_i\ M_i\rangle &=& 
C^{\frac{1}{2}\ \ \frac{1}{2}\ \ I_f }_{m'\ n'\ M_f}\ C^{\frac{1}{2}\ \, \frac{1}{2}\ \, I_i }_{m\ n\ M_i}\ 
\nonumber \\ && \times
\langle K_{m'}\ N_{n'}|{\cal H}|K_m\ N_n\rangle\ ,        
\label{eq:bb.1} \end{eqnarray}
where $I$ is the total isospin of the system and $M$ its $z$-component, $m$ is the $z$-component of the kaon isospin and $n$ is the $z$-component of the nucleon isospin.
For $\rho$-exchange the isospin interaction Hamiltonians ${\cal H}$ for the $NN\rho$ and $KK\rho$ vertex are
\begin{eqnarray}
{\cal H}_{NN\rho}&= &
\sqrt{3}\ 
C^{\frac{1}{2}\ 1\ \: \frac{1}{2} }_{n\ M\ n'}\ N^*_{n'}N_n\rho^*_M\ , \nonumber \\
{\cal H}_{KK\rho}&= &
\sqrt{3}\ 
C^{\frac{1}{2}\ \: 1\ \ \frac{1}{2} }_{m\ M\ m'}\ K^*_{m'}K_m\rho^*_M\ , 
\label{eq:bb.2}
\end{eqnarray}
we note that $\rho_m=(-)^m\rho^*_{-m}$. The $\rho$ emission vertex shown in Figure \ref{fig:bb.1} gives the factor
\begin{eqnarray}
\sqrt{3}\ (-)^{-M}\ C^{\frac{1}{2}\ \ \, 1\ \ \frac{1}{2}}_{n\ -M\ n'}\ .
\label{eq:bb.3}
\end{eqnarray}
The $\rho$ absorption vertex shown in Figure \ref{fig:bb.1} gives the factor
\begin{eqnarray}
\sqrt{3}\ C^{\frac{1}{2}\ \, 1\ \ \frac{1}{2} }_{m\ M\ m'}\ .
\label{eq:bb.4}
\end{eqnarray}
Using Eqs. (\ref{eq:bb.2}), (\ref{eq:bb.3}) and (\ref{eq:bb.4}) we find for the total isospin matrix element of Eq. (\ref{eq:bb.1})
\begin{eqnarray}
\langle I_f\ M_f| {\cal H}| I_i\ M_i\rangle&=&
(-)^{-M}\ 3\ 
C^{\frac{1}{2}\ \ \frac{1}{2}\ \ I_f }_{m'\ n'\ M_f}\ C^{\frac{1}{2}\ \frac{1}{2}\ I_i }_{m\ n\ M_i}\ 
\nonumber \\ && \times
C^{\frac{1}{2}\ \ \; 1\ \; \frac{1}{2} }_{n\ -M\ n'}\ C^{\frac{1}{2}\ \, 1\ \; \frac{1}{2}}_{m\ M\ m'}\ ,
\label{eq:bb.5}
\end{eqnarray}
applying the identity $(-)^{-M}=-\sqrt{3}\ C^{\ \ 1\ \ \, 1\ \ 0 }_{-M\ M\ 0}$, we find
\begin{eqnarray}
\langle I_f\ M_f| {\cal H}| I_i\ M_i\rangle&=&-
3\sqrt{3}\left[\begin{array}{ccc}\frac{1}{2} & 1 & \frac{1}{2} \\ \frac{1}{2} & 1 & \frac{1}{2} \\ I & 0 & I \end{array}\right] \nonumber \\
&=&
2I\left(I+1\right)-3 \ ,
\label{eq:bb.6}
\end{eqnarray}
here we have used the conservation of isospin $I_f=I_i=I$. 
For $I=0$ we find an isospin factor of $-3$ and for $I=1$ a factor $1$.
Other isospin factors can be calculated in the same way, all relevant isospin factors for the $\pi N$ and $K^+ N$ interaction are listed in Tables \ref{tab:6.0} and \ref{tab:7.0} respectively.

\end{document}